\documentclass[reqno,12pt]{amsart}

\usepackage{amsmath,amssymb,amsthm,amsfonts,amsxtra}
\usepackage{cases}
\usepackage{fullpage}
\usepackage{enumitem}
\usepackage{graphicx}
\usepackage{float}
\usepackage{caption}
\usepackage{subcaption}

\numberwithin{equation}{section} 

\newtheorem{thm}{Theorem}{\bf}{\em}
\newtheorem{cor}{Corollary}{\bf}{\em}
\newtheorem{prop}{Proposition}{\bf}{\em}
\newtheorem{lem}{Lemma}{\bf}{\em}
\newtheorem{rem}{Remark}{\bf}{\em}
{\bf}{\em}

\def\pe{\perp}
\def\pa{\parallel}

\def\mr{\mathrm}
\def\mk{\mathfrak}

\def\bpm{\begin{pmatrix}}
\def\epm{\end{pmatrix}}

\def\Rnum{\mathbb{R}}
\def\Cnum{\mathbb{C}}

\def\sgn{{\rm sgn}}

\def\const{{\rm const.}}

\def\d{\partial}

\def\T{\hat{\mr{t}}}
\def\N{\hat{\mr{n}}}

\def\sech{\mathrm{sech}}
\def\CN{\mathrm{cn}}
\def\SN{\mathrm{sn}}
\def\DN{\mathrm{dn}}
\def\AM{\mathrm{am}}

\def\K{\mathrm{K}}
\def\E{\mathrm{E}}

\def\Rop{{\mathcal R}}
\def\Hop{{\mathcal H}}
\def\Dop{{\mathcal D}}
\def\Jop{{\mathcal J}}

\def\C{{\mathcal C}}

\def\Ref#1{Ref.~\cite{#1}}
\def\Refs#1{Refs.~\cite{#1}}

\def\eg/{e.g.}
\def\ie/{i.e.}

\begin{document}

\allowdisplaybreaks[3]
\tolerance =99999

\title{Geometric curve flows in the plane\\ and mKdV loop solutions} 

\author{
Stephen C. Anco$^1$
\lowercase{and}
Jaskaran Maan$^2$
\\\\
${}^1$
D\lowercase{\scshape{epartment}} \lowercase{\scshape{of}} M\lowercase{\scshape{athematics and}} S\lowercase{\scshape{tatistics}}\\
B\lowercase{\scshape{rock}} U\lowercase{\scshape{niversity}}\\
S\lowercase{\scshape{t.}} C\lowercase{\scshape{atharines}}, C\lowercase{\scshape{anada}}
\\
${}^2$
D\lowercase{\scshape{epartment}} \lowercase{\scshape{of}} P\lowercase{\scshape{hsyics}}\\
B\lowercase{\scshape{rock}} U\lowercase{\scshape{niversity}}\\
S\lowercase{\scshape{t.}} C\lowercase{\scshape{atharines}}, C\lowercase{\scshape{anada}}
}

\thanks{$^1$sanco@brocku.ca, $^2$jm22tp@brocku.ca}

\begin{abstract}
There is a well known correspondence between geometric curve flows in the Euclidean plane
and solutions of the modified Korteweg-de Vries (mKdV) equation.
For each type of mKdV travelling wave,
the resulting geometric curve flows are derived here
through a simple quadrature formula and studied in detail. 
These curve flows can be divided into two broad types: 
travelling loops, and rotating loops.
Travelling loops are shown to arise from mKdV solitons, cnoidal (Jacobi cn) and dnoidal (Jacobi dn) waves,
the latter being periodic.
Rotating loops comprise 
asymptotically circular ones that are obtained from both
mKdV solitary waves on a non-zero background and mKdV rational waves,
as well as periodic ones that are produced by mKdV rational elliptic (cn and dn) waves.
A specialization of periodic loops, both open and closed,
is shown to yield rational cosine loops. 
An explicit description of each of these types of curve flows is used
to characterize their main features, 
including the condition under which closed loops exist.
\end{abstract}

\maketitle


\section{Introduction}
\label{sec:intro}

A close connection is known to exist between 
integrable geometric curve flows in the Euclidean plane 
and the modified Korteweg-de Vries (mKdV) equation in soliton theory. 
The simplest example is the curve flow \cite{Nak.Seg.Wad.1992,Gol.Pet.1993}
\begin{equation}\label{mkdv.curveflow.eqn}
\vec r_t = \tfrac{1}{2}\kappa^2 \T +\kappa_s \N
\end{equation}
for a vector $\vec r(s,t)$ in terms of 
the unit tangent $\T= |\vec{r}_s|^{-1} \vec{r}_s$ 
and the unit normal $\N = |\vec{r}_{ss}|^{-1} \vec{r}_{ss}$, 
where $\kappa(s,t)$ is the curvature and $s$ is the arclength. 
This equation \eqref{mkdv.curveflow.eqn} describes a flow that locally preserves arclength 
while the curvature evolves by the (focusing) mKdV soliton equation
\begin{equation}\label{mkdv.eqn}
\kappa_t = \tfrac{3}{2} \kappa^2 \kappa_s + \kappa_{sss} . 
\end{equation}
In addition, 
the winding number of the curve 
and the area enclosed by the curve are preserved,
along with an infinite number of integrals of motion $\oint C\,ds$
where each density $C$ is given in terms of $\kappa$ and $s$-derivatives of $\kappa$. 

There are several interesting physical applications of this geometric curve flow: 
modeling the evolution of the vortex patch boundaries in a thin, sheet-like layer of an incompressible fluid
\cite{Gol.Pet.1991,Gol.Pet.1992}; 
dynamics of the boundary of electron cloud densities in thin-layered materials with strong electromagnetic fields
\cite{Wex.Dor.1999a,Wex.Dor.1999b}; 
shapes of cylindrical membranes in fluid lipid bi-layers
\cite{Vas.Djo.Mla.2008}; 
solitary waves in an elastic rod 
\cite{Nis.1997}.

Every solution $\kappa(s,t)$ of the mKdV equation \eqref{mkdv.eqn} 
produces a solution $\vec r(s,t)$ to the curve flow equation \eqref{mkdv.curveflow.eqn},
which is generally called a \emph{loop solution}.
These solutions have a well-known equivalent description
as a curve flow in the complex plane via the correspondence
$z(s,t) = x(s,t) + i y(s,t)$ where $\vec{r}(s,t) = \big(x(s,t),y(s,t)\big)$
in Cartesian coordinates $(x,y)$. 

The purpose of the present paper is to study comprehensively
the loop solutions arising from all of the different types of mKdV travelling wave solutions.
In particular, the mKdV travelling waves consist of 
soliton waves, rational (heavy-tail) waves,
solitary waves on a non-zero background, 
cnoidal and dnoidal waves,
and rational-elliptic periodic waves comprising Jacobi cn and dn families,
including a special case of rational cosine waves. 

Solitons are known to produce a single travelling loop 
which asymptotically approaches a straight line.
Cnoidal and dnoidal waves are respectively given by Jacobi cn and dn functions
and yield travelling loops that are periodic.
Both solitary waves on a non-zero background and rational waves
produce a rotating loop that is wound asymptotically around a circle
whose radius is determined by the boundary condition of the waves at infinity. 
Rational-elliptic waves produce rotating periodic loops that lie in an annulus.
In general, these loops are open; they can be closed under a certain condition.
Examples of soliton loops and rational-elliptic loops have been given in
\Refs{Wex.Dor.1999a,Nak.Wad.1992}. 

The main results of the present study will be to provide a detailed explicit description of
all of the types of loop solutions and their properties.
Asymptotically circular loops and rational cosine loops have not been studied previously,
and many new loop shapes not seen in previous studies are highlighted: 
\begin{itemize}
\item
  Cnoidal open loops vary in shape from a non-intersecting loop 
  to a loop having one or more self-intersections (twists)
\item
  One cnoidal loop, shaped like a figure 8, is closed
\item
   Asymptotically circular loops contain a self-intersection (twist)
  that lies outside of the asymptotic circle in the case of rational waves
  and either outside or inside in the case of solitary waves on a respectively negative or positive background
\item
  Rational-elliptic periodic open loops lie in an annulus
\item
  Shape of rational-elliptic periodic closed loops depends on two co-prime integers
\item
  Rational cosine loops arise as special case of rational-elliptic loops
\end{itemize}  
In addition to these results, further kinematic properties of the loops will be derived.
Specifically, 
the height and width of soliton loops will be obtained as a function of the loop speed;
the speed of cnoidal loops will be shown to reverse when the loop acquires a pinch; 
the angular speed and size of both types of asymptotically circular loops will be derived;
the condition for a rational-elliptic loop to be closed will be given explicitly. 

The remainder of the paper is organized as follows. 

In section~\ref{sec:travellingwaves},
all of the travelling wave solutions of the mKdV equation are summarized
through a phase plane analysis.
Their conserved quantities are reviewed. 

In section~\ref{sec:curveflows}, 
the connection between the mKdV equation and curve flows in the Euclidean plane
is reviewed, with emphasis on the explanation of how
the integrability structure of the mKdV equation 
emerges purely from the structure equations of a curve flow.
Additionally, the equations for constructing a curve from its curvature function
are also reviewed and applied to arclength-preserving curve flows.

In section~\ref{sec:prelims},
a simple quadrature formula is derived for arclength-preserving curve flows
expressed in complex-coordinate form 
in the case when the curvature has the form of a travelling wave.
This new formula shows that travelling loops and rotating loops are distinguished by
the value of the integration constant in the mKdV travelling-wave equation,
and also provides an explicit expression for the travelling-wave speed
and the rotational frequency.
In addition,
the basic conserved integrals for arclength-preserving curve flows are discussed. 

In section~\ref{sec:loops},
the loop solutions corresponding to each kind of mKdV travelling wave are worked out
by an explicit evaluation of the quadrature formula,
which gives an explicit analytical formula for each type of loop. 
This leads to a considerable simplification in the analysis of rational-elliptic loops.
Features of each type of loop are studied in detail,
and explicit descriptions of the above-highlighted features are provided.

In section~\ref{sec:conclude},
some concluding remarks are made. 

Two appendices contain some additional material.
Appendix~\ref{sec:identities} summarizes the algebraic identities that hold
for Jacobi elliptic functions.
Appendix~\ref{sec:integrability} has a short discussion of deeper aspects of
the link between the structure equations of geometric curve flows in the plane
and the integrability structure of the mKdV equation.

\section{Travelling wave solutions and conservation laws}
\label{sec:travellingwaves}

Travelling waves of the mKdV equation \eqref{mkdv.eqn}
have the form
\begin{equation}\label{travelling.wave}
  \kappa(s,t)  = K(\xi), 
  \quad
  \xi = s + ct,
  \quad
  c=\const
\end{equation}
which are solutions of the nonlinear third-order ordinary differential equation (ODE)
\begin{equation}
-c K' + \tfrac{3}{2} K^2 K' + K''' =0 . 
\end{equation}
Integrating once with respect to the travelling wave variable $\xi$ yields
a nonlinear oscillator equation
\begin{equation}\label{tw.eqn}
  K'' - c K + \tfrac{1}{2} K^3 = C_1 =\const . 
\end{equation}
All solutions can be found explicitly by using the first integral
\begin{equation}\label{1st.integral}
 \tfrac{1}{2} K'{}^2  + \tfrac{1}{8} K^4 - \tfrac{1}{2} c K^2 - C_1 K  =  C_2 = \const, 
\end{equation}
which leads to a separable first-order ODE. 

There are two ways to determine the different types of solutions $K(\xi)$:
an energy analysis, or a phase plane analysis.
Details of the energy analysis are provided in \Ref{Anc.Nay.Rec.2021}.
The phase plane analysis will be summarized here. 

Let $U=K$ and $V=K'$ be the phase plane variables.
The oscillator equation \eqref{tw.eqn} is equivalent to the first-order system
\begin{equation}\label{tw.sys}
  U' = V,
  \quad
  V' = c U - \tfrac{1}{2} U^3 - C_1
\end{equation}
in the $(U,V)$ plane, 
where the first integral \eqref{1st.integral} becomes
\begin{equation} \label{tw.1st.integral}
  V^2  + \tfrac{1}{4} U^4 - c U^2 - 2 C_1 U  =  2C_2 . 
\end{equation}
The equilibria points of the system consist of $V=0$ and $U=U_*$
given by the roots of the cubic $U^3 - 2c U + 2 C_1 =0$.
It has discriminant
\begin{equation}
  \Delta = 32c^3 -108 C_1^2
\end{equation}
whose sign is positive if $C_1^2 < (\tfrac{2}{3}c)^3$,
negative if $C_1^2 > (\tfrac{2}{3}c)^3$,
or zero if $C_1^2 = (\tfrac{2}{3}c)^3$.

\begin{figure}
\includegraphics[width=0.7\textwidth,trim=2cm 12cm 2cm 1cm, clip]{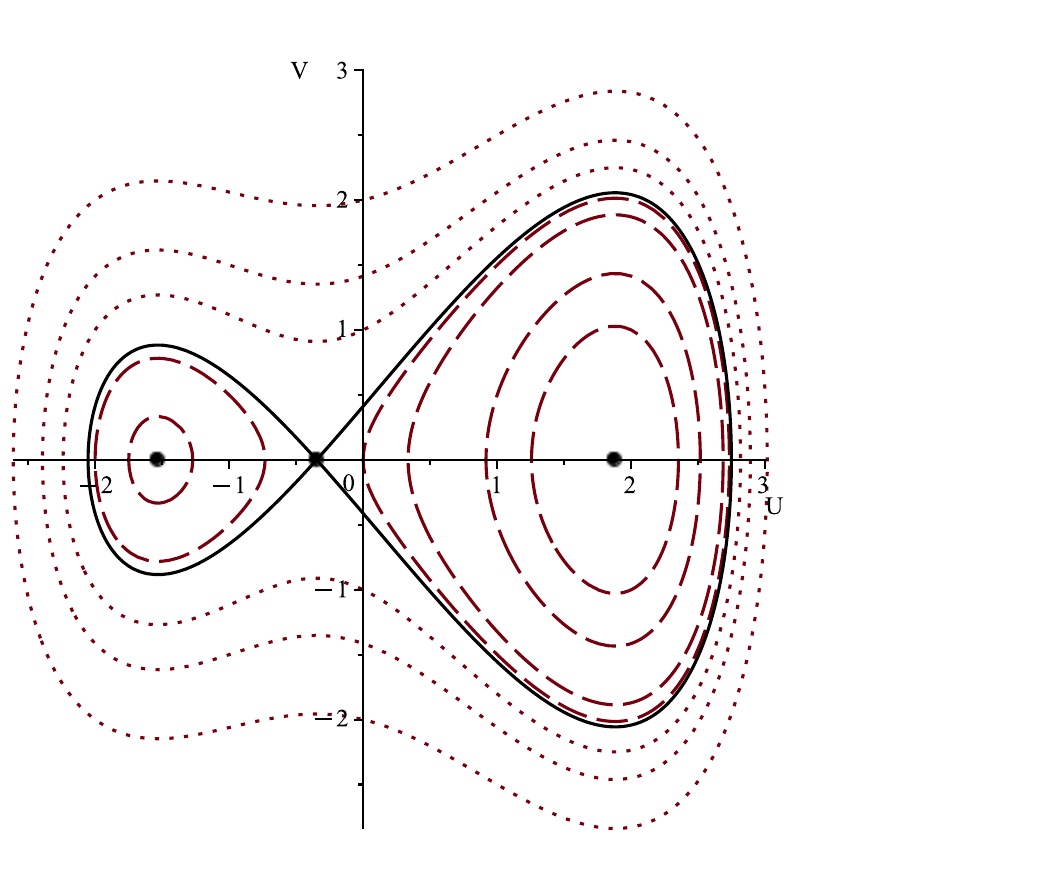}
\caption{Phase plane for $\Delta>0$. $c=3/2$, $C_1=1/2$}
\label{fig:Delta>0}
\end{figure}

\textbf{Case} $\Delta>0$:
In this case, $c>0$ must hold and then the cubic
$U^3 -2c U + 2 C_1 = (U -U_{*1})(U -U_{*2})(U -U_{*3}) =0$
has three distinct real roots, which can be ordered $U_{*1}<U_{*2}<U_{*3}$.
The first integral \eqref{tw.1st.integral} comprises three family of closed curves.
There is one family around each of the points $(U_{*1},0)$ and $(U_{*3},0)$.
A double homoclinic orbit through $(U_{*2},0)$ separates these two families of curves.
Outside of this orbit, a third family surrounds all three points.
See Fig.~\ref{fig:Delta>0}.

The double homoclinic orbit is a pair of bright/dark solitons on a background $U_{2*}$
which can be positive, negative, or zero. 
Each curve in the three families is a periodic wave.

\begin{figure}
\includegraphics[width=0.48\textwidth,trim=2cm 12cm 2cm 1cm, clip]{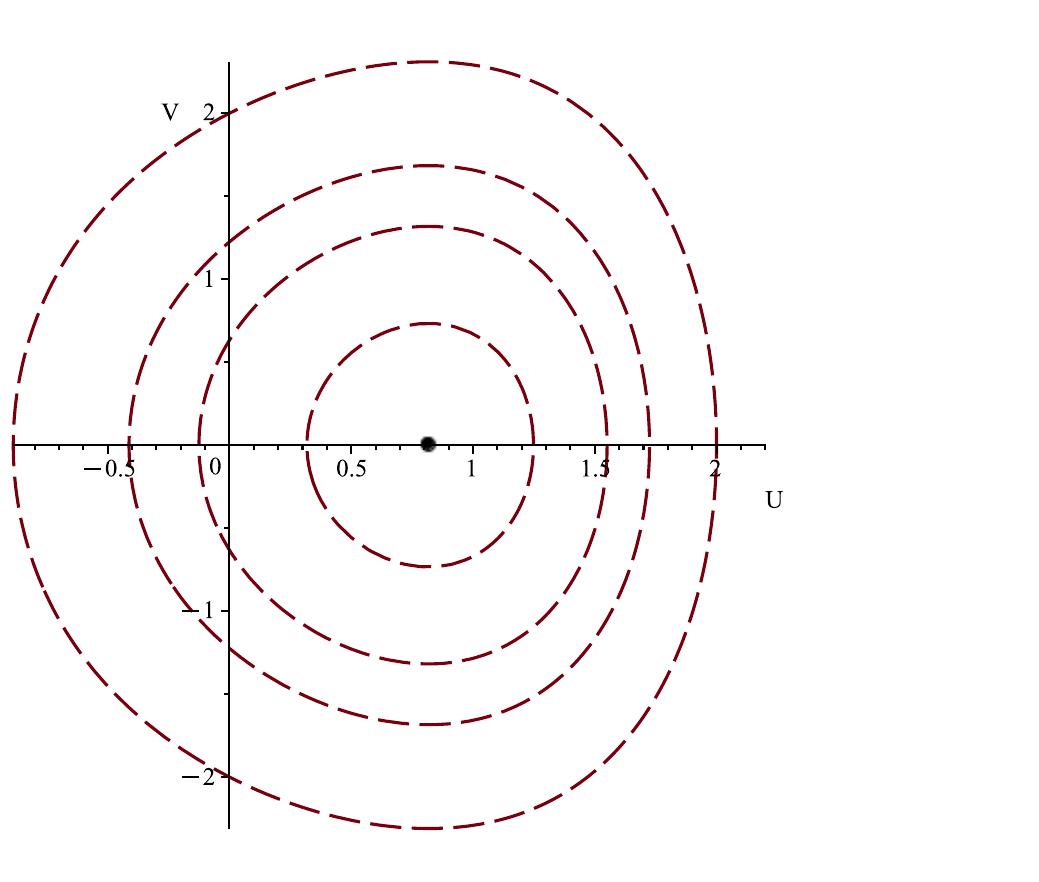}
\includegraphics[width=0.48\textwidth,trim=2cm 12cm 2cm 1cm, clip]{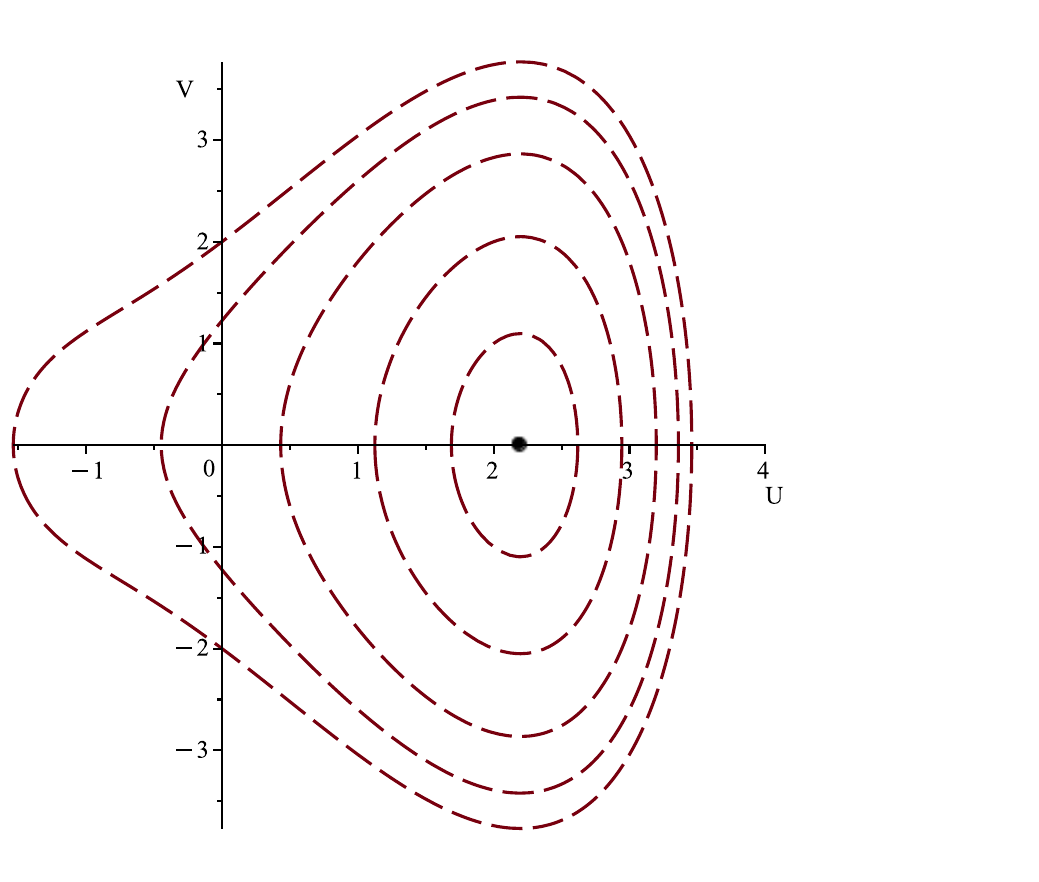}
\caption{Phase plane for $\Delta<0$. 
(Left) $c=-3/2$, $C_1=3/2$
\quad
(Right) $c=3/2$, $C_1=2$}
\label{fig:Delta<0} 
\end{figure}

\textbf{Case} $\Delta<0$:
Here $c$ can have any sign,
while the cubic $U^3 - 2c U + 2 C_1 = (U -U_{*1})(U -U_{*2})(U -\bar{U}_{*2}) =0$
has a single real root $U_{*1}$ and a pair of complex-conjugate roots $U_{*2}$, $\bar{U}_{*2}$.
The first integral \eqref{tw.1st.integral} gives one family of closed curves around
the point $(U_{1*},0)$. 
See Fig.~\ref{fig:Delta<0}.
Each curve is a periodic wave. 

\begin{figure}
\includegraphics[width=0.7\textwidth,trim=1cm 12cm 1cm 1cm]{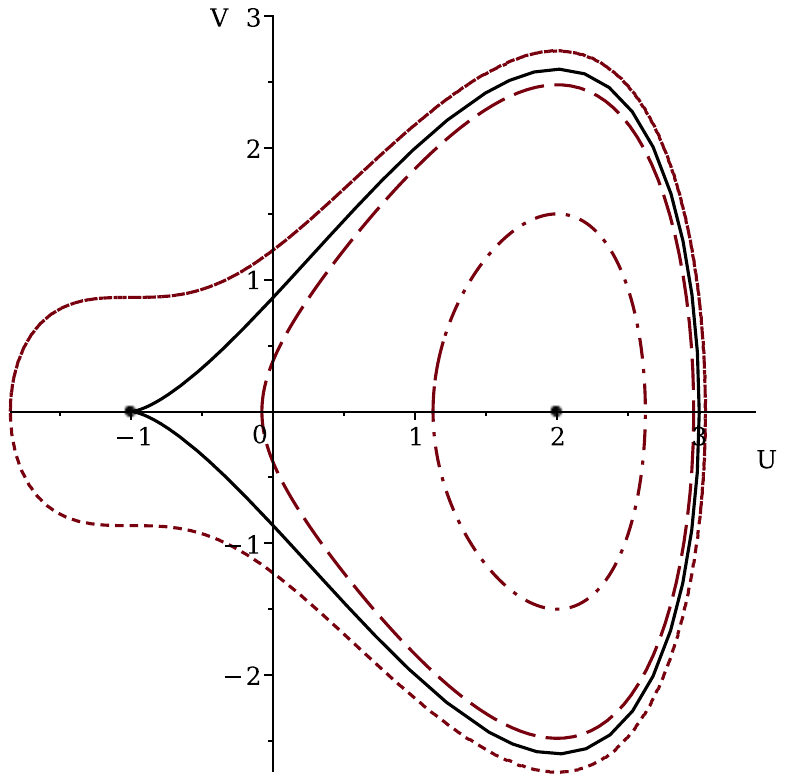}
\caption{Phase plane for $\Delta=0$.
$c=3/2$, $C_1=1$}
\label{fig:Delta=0}
\end{figure}

\textbf{Case} $\Delta=0$:
This case requires $c\geq 0$.
The cubic $U^3 - 2c U + 2 C_1 = (U -U_{*1})^2(U -U_{*2}) =0$
has real roots, one of which is repeated.
The first integral \eqref{tw.1st.integral} comprises two families of closed curves.
There is one family around the point $(U_{*2},0)$
and another family around both points. 
These two families are separated by a homoclinic orbit through the point $(U_{*1},0)$. 
See Fig.~\ref{fig:Delta=0}.

The homoclinic orbit is a rational (heavy-tail) wave on a background $U_{*1}$.
It is bright or dark according to the sign of $U_{*2}-U_{1*}$. 
Each curve in the two families is a periodic wave.

\subsection{Classification of travelling waves}

The preceding phase-plane analysis provides a classification of all travelling waves.
Their explicit form is obtained from the quadrature of the separable ODE \eqref{1st.integral}: 
\begin{equation}\label{quadrature}
  \pm\int_{K_0}^{K(\xi)} \frac{dK}{\sqrt{\tfrac{1}{4} K^4 - c K^2  - 2C_1 K  -2 C_2}} = \xi -\xi_0
\end{equation}
where $K_0$ is the minimum or maximum of the solution, as given by 
$K(\xi_0)=K_0$ and $K'(\xi_0)=0$.
Evaluation of the quadrature leads to the following result.

\begin{thm}\label{thm:travellingwaves}
The travelling waves \eqref{travelling.wave} of the mKdV equation \eqref{mkdv.eqn},
up to a shift in $\xi$, 
consist of:\\
\begin{enumerate}[label=(\roman*)]

\item \emph{pair of bright/dark solitons on a background}
\begin{equation}\label{soliton}
    K(\xi) = b \pm  \frac{2c-3b^2}{\sqrt{c-\tfrac{1}{2}b^2}\,\cosh\Big( \sqrt{c -\tfrac{3}{2}b^2}\,\xi \Big) \pm b},
    \quad
    c> \tfrac{3}{2} b^2
\end{equation}
which have two parameters $b$ and $c$. 
A special case ($b=0$) is the ordinary soliton
\begin{equation}\label{soliton.b=0}
  K(\xi) = \pm 2\sqrt{c}\,\sech(\sqrt{c}\,\xi),
  \quad
  c>0
\end{equation}
with one parameter $c$. 

\item \emph{rational (heavy-tail) wave}
\begin{equation}\label{heavytail}
   K(\xi) = b \Big( 1 - \frac{6}{c\, \xi^2 +\frac{3}{2}} \Big), 
   \quad
   c = \tfrac{3}{2} b^2
\end{equation}
which has one parameter $c$ (or $b$). 

\item \emph{bright/dark periodic waves} 
\begin{subequations}\label{ratcn}
\begin{align}
  K(\xi) & = \pm 2\sqrt{c/F}\, \frac{B\,\CN(\sqrt{cE/F}\,\xi, \sqrt{q}) + A}{C\, \CN(\sqrt{cE/F}\,\xi, \sqrt{q}) + 1},
  \\
\text{(first family)}& \quad
  c\neq 0,
  \quad
  0\leq q< 1,
  \quad
  0\leq C^2< 1
\label{1st.fam}
  \\
\text{(second family)}&\quad
  c>0,
  \quad
  q> 1,
  \quad
  C< -\sqrt{q/(q-1)}\ \text{ or }\ C>-1,
  \label{2nd.fam}
\end{align}
where
\begin{align}
B & = q(1 -C^2) +\tfrac{1}{2}C^2, 
 \quad
A = (q(1 - C^2) +C^2 -\tfrac{1}{2})C
\\
E &= (q(1-C^2) + C^2)(1-C^2) >0 
\label{E.ratcn}
\\
F & =  (2q-1)E +\tfrac{3}{2}C^2 , 
\quad
F/c>0
\label{F.ratcn}
\end{align}
\end{subequations}
which have three parameters $C$, $q$, and $c$. 
A special case ($q=0$ in the first family) is rational cosine waves
\begin{subequations}\label{ratcos}
\begin{align}
  K(\xi) & = \pm \sqrt{c/F_0}\, \frac{C \cos(\sqrt{cE_0/F_0}\,\xi) + 2C^2 -1}{C \cos(\sqrt{cE_0/F_0}\,\xi) + 1},
  \quad
  c >0
 \\
  E_0 &= 1- C^2 >0 ,
  \quad
  F_0 = C^2 +\tfrac{1}{2}
\end{align}
\end{subequations}
with two parameters $C$ and $c$.
Another special case ($C=0$ in both families) is cnoidal and dnoidal waves
\begin{subequations}\label{cn}
\begin{align}
K(\xi) & = \pm 2q\sqrt{c/F_1}\,\CN(\sqrt{cq/F_1}\,\xi, \sqrt{q})
  \\
\text{(cnoidal family)}&\quad
  c\neq 0,
  \quad
  0< q< 1
  \\
\text{(dnoidal family)}&\quad
  c>0,
  \quad
  q> 1
\end{align}
where
\begin{align}
F_1 & =  2q-1, 
\quad
F_1/c>0
\end{align}
\end{subequations}
with two parameters $q$ and $c$.

\end{enumerate}
\end{thm}

\textbf{Remarks}:\\
(1)
In the family of rational $\CN$ waves \eqref{ratcn} for $0<q<1$,
the orientation ``$\pm$'' describes bright/dark waves. 
In the case of cnoidal waves \eqref{cn} for $0<q<1$,
a change of the orientation ``$\pm$'' is equivalent to a shift in $\xi$,
so only the $+$ case needs to be considered.
\\
(2)
The second family of periodic waves can be transformed to a $\DN$ form
\begin{equation}\label{ratdn}
  K(\xi) = \mp 2\sqrt{c/F}\, \frac{B\,\DN(\sqrt{cqE/F}\,\xi, 1/\sqrt{q}) + A}{C\, \DN(\sqrt{cqE/F}\,\xi, 1/\sqrt{q}) + 1}
\end{equation}
via the identity \eqref{scaling},
and likewise in the case $q=0$, 
\begin{equation}\label{dn}
  K(\xi)  = \pm 2q\sqrt{c/F_1}\,\DN(q\sqrt{c/F_1}\,\xi, 1/\sqrt{q}) 
\end{equation}
constitute dnoidal waves. 
The orientations ``$+$'' and ``$-$'' respectively describes bright/dark waves
in the dnoidal family \eqref{dn} and the rational $\DN$ family \eqref{ratdn}. 
\\
(3)
Alternatively,
the second family of periodic waves, including the special case of dnoidal waves, 
can be transformed to a rational $\SN^2$ form
\begin{equation}\label{ratsnsq}
K(\xi)  = \pm 2\sqrt{c/F}\,
\frac{ (A - B)\SN\big( \tfrac{1}{2} p_+\sqrt{cE/F}\, \xi, p_-^2 \big)^2 + (A + B) p_+^2 }
{ (1-C)\SN\big( \tfrac{1}{2} p_+\sqrt{cE/F}\, \xi, p_-^2 \big)^2 + (C + 1) p_+^2 },
\quad
p_\pm = \sqrt{q} \pm \sqrt{q - 1}
\end{equation}
via the identity \eqref{CN.SNsq.CNsq}.
\\
(4)
The first family of periodic waves \eqref{ratcn} 
has been given previously in \Ref{Vas.Djo.Mla.2008}
using a more complicated parameterization,
while the second family of periodic waves was given in rational $\SN^2$ form. 
\\

The period of the rational $\CN$ waves \eqref{ratcn} for $0<q<1$ is given by 
\begin{equation}\label{ratCN.period}
  \Delta\xi = 4\sqrt{F/(cE)}\,\K(\sqrt{q})
\end{equation}
where $\K$ is the complete elliptic integral of the first kind.
In the special case of cnoidal waves \eqref{cn} 
and rational cosine waves \eqref{ratcos}, 
the period becomes, respectively, 
\begin{equation}\label{CN.period}
  \Delta\xi = 4\sqrt{F_1/(cq)}\,\K(\sqrt{q})
\end{equation}
and 
\begin{equation}\label{ratcos.period}
  \Delta\xi = \pi \sqrt{\frac{2(1+2C^2)}{c(1 -C^2)}} . 
\end{equation}
For the respective $\DN$ solutions \eqref{ratdn} and \eqref{dn} with $q>1$, 
the periods are given by
\begin{equation}\label{ratDN.period}
  \Delta\xi = 2\sqrt{F/(cqE)}\,\K(1/\sqrt{q})
\end{equation}
and
\begin{equation}\label{DN.period}
  \Delta\xi = 2\sqrt{F_1/(cq^2)}\,\K(1/\sqrt{q}) . 
\end{equation}

For later use,
the solutions in Theorem~\ref{thm:travellingwaves}
respectively have the following integration constants $C_1$ and $C_2$:
\begin{align}
(i) &\quad 
C_1 = (\tfrac{1}{2}b^2 -c)b , 
\quad
C_2 = (\tfrac{1}{2}c - \tfrac{3}{8} b^2)b^2 
\label{C1.C2.soliton}
\\
(ii) & \quad 
C_1 = - b^3,
\quad
C_2 = \tfrac{3}{8} b^4
\label{C1.C2.heavytail}
\\
(iii) &\quad  
C_1 =\pm \sqrt{c/F}^3\, \big(q(1-C^4) +C^4\big) C,
\label{C1.ratcn}
\\&\quad 
C_2 = 2(c/F)^2\, \big( (C^2 +q(1-C^2))^2 - \tfrac{1}{4}C^2 \big) \big( q(1-q)(1-C^2)^2 +\tfrac{1}{4}C^2 \big)
\label{C2.ratcn}
\end{align}

An explicit identification of the periodic waves that appear in each of the cases
$\Delta>0$, $\Delta<0$, and $\Delta=0$ will now be stated.

\begin{cor}
Let $C_\pm{}^2 = \frac{ 8q(q-1) - 1 \pm \sqrt{1-32q(q-1)}}{4(2q - 1)(q - 1)}$
and $C^\pm{}^2 = \frac{ 16q(q-1)(8q(q-1) -5) - 1  \pm \sqrt{(1-32q(q-1)^3} }{16(q - 1)(2q - 1)^3}$. 
The parameters $C$ and $q$ in the two families of periodic waves \eqref{ratcn}
have the following ranges.
\\\\
Case $\Delta <0$, $c<0$ is comprised by the first family with 
\begin{equation}\label{Del.neg.c.neg.conds}
  0 < q < \tfrac{1}{2},
  \quad
  0\leq |C| < C_+ . 
\end{equation}
Case $\Delta <0$, $c>0$ is comprised by the first family with 
\begin{subequations}\label{Del.neg.c.pos.conds}
\begin{align}
  & 0 < q < \tfrac{1}{2} -\tfrac{\sqrt{3}}{4},
  \quad
  C^+ < |C| < C_+;
  \\
  & \tfrac{1}{2}-\tfrac{\sqrt{3}}{4} \leq q < \tfrac{1}{2},
  \quad
  C_+ < |C| < 1;
  \\
  & \tfrac{1}{2} <q < \tfrac{1}{2}+\tfrac{\sqrt{3}}{4},
  \quad
  C^- < |C| < 1 . 
\end{align}
\end{subequations}
Case $\Delta =0$:
Outside the homoclinic orbit comprises the first family with 
\begin{equation}\label{Del.zero.outside.conds}
  c>0,
  \quad
  \tfrac{1}{2} < q < \tfrac{1}{2} +\tfrac{\sqrt{3}}{4},
  \quad
  |C|= C^- . 
\end{equation}
Inside the homoclinic orbit comprises the first family with
\begin{equation}\label{Del.zero.inside.conds}
  c>0, 
  \quad
  0 < q < \tfrac{1}{2}  - \tfrac{\sqrt{3}}{4},
  \quad
  |C|= C^+ . 
\end{equation}
Case  $\Delta >0$:
Outside the double homoclinic orbit comprises the first family with 
\begin{subequations}\label{Del.pos.outside.conds}
\begin{align}
  & c>0,
  \quad
  \tfrac{1}{2} < q < \tfrac{1}{2} +\tfrac{\sqrt{3}}{4},
  \quad
  |C|< C^-;
  \\
  & c>0,
  \quad
  \tfrac{1}{2} +\tfrac{\sqrt{3}}{4} \leq q < 1,
  \quad
  0\leq |C|< 1 . 
\end{align}
\end{subequations}
Inside the double homoclinic orbit comprises the second family with 
\begin{subequations}\label{Del.pos.inside.conds}
\begin{align}
   & 0 \leq q < \tfrac{1}{2} - \tfrac{\sqrt{3}}{4},
   \quad 
   C^+ < |C| < 1 ;
  \\
  & q> 1, 
  \quad
  |C|<1 ;
  \\
  & 1< q < C^2/(C^2-1),
  \quad
  |C|> 1 . 
  \end{align}
\end{subequations}
\end{cor}

The proof of this result uses
a characterization of the phase plane regions for each case of $\Delta$
in terms of the constants $C_1$ and $C_2$.
To begin, observe that
\begin{equation}\label{sgn.Del}
  \sgn(\Delta) = \sgn((\tfrac{2}{3}c)^3 - C_1^2) . 
\end{equation}
The periodic waves \eqref{ratcn} have 
$\sgn(\Delta) = \sgn\big( (q-1)(2q-1)(C - C^+)(C-C^-) \big)$,
which can be positive, negative, or zero, depending on $q$ and $C$.
Note that these waves must obey the inequalities $E>0$ and $F/c>0$. 

The case $\Delta<0$ corresponds to $|C_1|\geq 0$ when $c<0$
and to $|C_1|> (\tfrac{2}{3}c)^3$ when $c>0$.
For $c<0$, 
the condition $\sgn(\Delta) =-1$ holds identically,
and then the inequalities $E>0$ and $F<0$ 
have the solution \eqref{Del.neg.c.neg.conds}. 
For $c>0$,
the condition $\sgn(\Delta) =-1$ combined with the inequalities $E>0$ and $F>0$
have the solutions \eqref{Del.neg.c.pos.conds}. 

The case $\Delta=0$ corresponds to $|C_1| = (\tfrac{2}{3}c)^3$,
which requires $c>0$. 
Combining the condition $\Delta =0$ and the inequalities $E>0$ and $F>0$
yields the solutions \eqref{Del.zero.outside.conds}--\eqref{Del.zero.inside.conds}. 
Since the homoclinic orbit has $C_2 = \tfrac{1}{6}c^2$,
the orbits outside and inside can be distinguished by whether 
$C_2$ is greater or less than the value $\tfrac{1}{6}c^2$ given by the rational wave
(cf \eqref{C1.C2.heavytail}). 
Expression \eqref{C2.ratcn} gives the ratio
\begin{equation}
C_2/(\tfrac{1}{6}c^2) = 
\frac{ 12( ((q - 1)C^\pm{}^2 - q)^2 -  \frac{1}{4}C^\pm{}^2 ) ( q(1-q)(C^\pm{}^2 - 1)^2 + \frac{1}{4}C^\pm{}^2 )}{( (2q-1)((q-1)C^\pm{}^2 - q) (C^\pm{}^2 - 1) + \frac{3}{2} C^\pm{}^2 )^2}
\end{equation}
after use of $|C|= C^\pm$ as given by $\Delta =0$. 
It is straightforward to show that this ratio is
greater than $1$ when inequalities \eqref{Del.zero.outside.conds} hold, 
and less than $1$ when inequalities \eqref{Del.zero.inside.conds} hold. 

The case $\Delta>0$ corresponds to $0\leq |C_1| < (\tfrac{2}{3}c)^3$,
which again requires $c>0$.
Inequalities $E>0$ and $F>0$ combined with the condition $\sgn(\Delta) =1$ 
give solutions \eqref{Del.pos.outside.conds} and \eqref{Del.pos.inside.conds}.
Orbits outside and inside the double homoclinic orbit can be distinguished by
whether $C_2$ is greater or less than the value $(\tfrac{1}{2}c - \tfrac{3}{8} b^2)b^2$
given by the soliton with background $b$
(cf \eqref{C1.C2.soliton}). 
This value can be expressed as
$-\tfrac{1}{8} b^4 (1+ 4\tilde C)$, 
where $\tilde C = C_1/b^3=  \tfrac{1}{2}-c/b^2$ replaces $c$. 
Similarly, expression \eqref{C2.ratcn} can be written in terms of $\tilde C$. 
This yields the ratio
\begin{equation}\label{C2.ratio}
C_2/((\tfrac{1}{2}c - \tfrac{3}{8} b^2)b^2)
= - \frac{16 (\tilde C -\frac{1}{2})^2 \left( (C^2 +q(1-C^2))^2 - \frac{1}{4}C^2 \right) \left( q(1-q)(1-C^2)^2 +\frac{1}{4}C^2 \right)}{F^2(4\tilde C + 1)} . 
\end{equation}
The double homoclinic orbit is fixed by the relation 
$\tilde C^2 = ((q - 1)C^4 - q)^2 C^2 (1 - 2\tilde C)^3/(8F^3)$
which comes directly from writing expression \eqref{C1.ratcn}
for $C_1$ in terms of $\tilde C$, after $c$ again has been replaced. 
This gives a cubic equation that can be solved for $\tilde C$,
which is then substituted into the ratio \eqref{C2.ratio}.
A graphical analysis reveals that the ratio is
greater than $1$ when inequalities \eqref{Del.pos.outside.conds} hold, 
and less than $1$ when inequalities \eqref{Del.pos.inside.conds} hold.

This completes the proof.

\subsection{Conserved quantities}
\label{sec:conserved}

The mKdV equation \eqref{mkdv.eqn} is well known to possess
an infinite hierarchy of local conservation laws $D_t \Psi^t + D_s \Psi^s =0$
for solutions $\kappa(s,t)$,
where $\Psi^t$ is the conserved density, and $\Psi^s$ is the flux.
(Here $D_t$ and $D_s$ denote total derivatives with respect to $t$ and $s$
acting via the chain rule.)
On a domain $\Omega$ for $s$,
every local conservation law gives rise to a conserved integral
\begin{equation}\label{conserved.integral}
  \C = \int_\Omega \Psi^t\,ds
  \end{equation}
satisfying
\begin{equation}\label{balance}
  \frac{d\C}{dt} = -\Psi^s\Big|_{\partial\Omega} . 
\end{equation}
This is a balance equation which states that
the rate of change of the quantity $\mathcal{C}$ on the domain $\Omega$
is equal to the net flux escaping through the boundary $\partial\Omega$. 

The hierarchy starts with the basic physical conservation laws \cite{Miu.Gar.Kru}
for mass, momentum, and energy:
\begin{gather}
  \Psi^t = \kappa,
  \quad
  \Psi^s = -\big(\tfrac{1}{2}\kappa^3 + \kappa_{ss}\big) ; 
  \label{mass}
  \\
  \Psi^t = \tfrac{1}{2}\kappa^2, 
  \quad
  \Psi^s = -\big(\tfrac{3}{8}\kappa^4 -\tfrac{1}{2}\kappa_s^2 +  \kappa\kappa_{ss}\big) ; 
  \label{momentum}
  \\
  \Psi^t = -\tfrac{1}{2}\kappa_s^2 + \tfrac{1}{8} \kappa^4, 
  \quad
  \Psi^s = -\tfrac{1}{2}\big(\kappa_{ss} + \tfrac{1}{2}\kappa^3\big)^2
  + (\kappa_{sss} + \tfrac{3}{2}\kappa^2 \kappa_s) \kappa_s . 
   \label{energy}
\end{gather}
A recursion formula exists to obtain the entire hierarchy which consists of higher-order versions of the energy (see e.g. \cite{Anc.2003}). 
In addition, there is an extra conservation law that is like a Galilean energy
\cite{Miu.Gar.Kru}. 

It is straightforward to evaluate the resulting conserved integrals \eqref{conserved.integral}
for all travelling wave solutions,
using an appropriate domain $\Omega$.
See Table~\ref{table:conserved}.

Ordinary solitons \eqref{soliton.b=0} have exponential decay for large $s$,
and thus the conserved densities also decay exponentially.
Hence, the conserved integrals are finite over the domain $\Omega = (-\infty,\infty)$.
For solitons \eqref{soliton} on a non-zero background, $b\neq0$,
each conserved density decays to a non-zero constant.
Subtraction of this constant gives a renormalized density which decays exponentially to zero,
without changing the flux.
The corresponding (renormalized) conserved integrals are then finite over the domain $\Omega = (-\infty,\infty)$.
A similar renormalization works for the rational (heavy-tail) waves \eqref{heavytail}.
For all of the preceding solutions,
the conserved integrals have zero flux at infinity,
$\Psi^s\big|_{\pm\infty}=0$,
and hence each one constitutes a constant of motion, $\frac{d\C}{dt} =0$. 

For periodic waves \eqref{ratcn} and their specialization to cnoidal waves \eqref{cn}, 
the domain for the conserved integrals is most usefully taken to be a finite interval
$\Omega = (-\ell,\ell)$ where $2\ell$ denotes the period of the wave \eqref{ratCN.period}.
The resulting flux turns out to vanish, $\Psi^s\big|_{-\ell}^{\ell}=0$. 
Hence, each conserved integral is a constant of motion.

\begin{table}[H]
\hbox{\hspace{-0.58in}
\begin{tabular}{l|c|c|c}
\hline
& $\C_0$ (mass)
& $\C_1$ (momentum)
& $\C_2$ (energy)
\\
\hline
\hline
soliton \eqref{soliton.b=0}
& $\scriptstyle \pm 2\pi$
& $\scriptstyle 4\sqrt{c}$
& $\scriptstyle \tfrac{4}{3}\sqrt{c}^3$
\\
\hline
$\begin{aligned}
&\text{soliton on non-zero}\\&\text{background \eqref{soliton}}
\end{aligned}$
& 
$\scriptstyle \pm 8\arctan\left( \frac{\sqrt{c -\frac{1}{2}b^2} \mp b}{\sqrt{c -\frac{3}{2}b^2}} \right)$
& $\scriptstyle 4\sqrt{c -\frac{3}{2}b^2}$
& $\scriptstyle \frac{4}{3}\sqrt{c -\frac{3}{2}b^2}\, c$
\\
\hline
rational wave \eqref{heavytail}
& $\scriptstyle -\sgn(b)\, 4 \pi$
& $0$
& $0$
\\
\hline
cnoidal wave \eqref{cn}
& $0$
& $\scriptstyle 8(q - 1)\sqrt{\frac{c}{2q - 1}}\,\K(\sqrt{q})$
& $\scriptstyle \frac{8}{3}(3q - 1)(q - 1) \sqrt{\frac{c}{2q - 1}}^3 \K(\sqrt{q})$
\\
\hline
dnoidal wave \eqref{dn}
& $\scriptstyle \pm 4 \AM(\K(\frac{1}{\sqrt{q}}), \frac{1}{\sqrt{q}})$
& $\scriptstyle 4\sqrt{\frac{cq}{2q - 1}}\, \E(\frac{1}{\sqrt{q}})$
& $\scriptstyle \frac{4}{3} \sqrt{q} \frac{\sqrt{c}^3}{\sqrt{2q - 1}} \big(
\E(\frac{1}{\sqrt{q}})  + \frac{q-1}{2q-1}\K(\frac{1}{\sqrt{q}}) \big)$
\\
\hline
rational cn wave \eqref{ratcn}
& $\scriptstyle \mp \frac{8B}{C\sqrt{E}} \K(\sqrt{q})$
& $\scriptstyle \frac{2(4(q - 1) E + C^2)}{\sqrt{E}} \sqrt{\frac{c}{F}} \K(\sqrt{q})$
& $\begin{aligned}&\scriptstyle
\big( ( \frac{8}{3}(1-q)(1-3C^2)(3E +C^2-1) - 4C^2 ) \\&\scriptstyle 
  \frac{8L^3 C^2}{\sqrt{E}}(C^2 L  - 4 E) \frac{\sqrt{E}^3}{1-C^2}
  \big) \sqrt{\frac{c}{F}}^3 \K(\sqrt{q})
  \end{aligned}$
\\
\hline
rational dn wave \eqref{ratdn}
& $\begin{aligned}&\scriptstyle
  \frac{\pm4}{C\sqrt{q}} \big( -\frac{B}{\sqrt{E}}\K(\frac{1}{\sqrt{q}}) \\&\scriptstyle
  + \frac{\sqrt{E}}{1-C^2} \Pi(\frac{C^2}{q(C^2 - 1)}, \frac{1}{\sqrt{q}}) \big)
\end{aligned}$
& $\begin{aligned}&\scriptstyle
\sqrt{\frac{c}{qF}}\big( 4\sqrt{E} (q\E(\frac{1}{\sqrt{q}}) +1-q) \\&\scriptstyle
+ (4(q - 1)E + C^2)\K(\frac{1}{\sqrt{q}}) \big)
\end{aligned}$
& $\begin{aligned}&\scriptstyle
\big( 
((\frac{4}{3}(1-3q)(1-q) E + 2(q - 1)C^2)\sqrt{E} + \frac{1}{4\sqrt{E}}C^4)
\K(\frac{1}{\sqrt{q}})
\\&\scriptstyle
+  (\frac{4}{3}(2q-1) E + 2C^2)\sqrt{E} (q\E(\frac{1}{\sqrt{q}}) - q + 1)
  \big) \sqrt{\frac{c}{F}}^3 \frac{1}{\sqrt{q}}
  \end{aligned}$
\\
\hline
$\begin{aligned}
  &\text{rational cosine}\\&\text{wave \eqref{ratcos}}
\end{aligned}$
& $\scriptstyle \mp \big(2 - \frac{1}{\sqrt{1-C^2}}\big)2\pi$
& $\scriptstyle   \frac{\sqrt{2c}}{\sqrt{(1 +2C^2)(1-C^2)}}\pi$
& $\scriptstyle \frac{\sqrt{c}^3}{\sqrt{1+2C^2}^3\sqrt{2(1-C^2)}}\pi$
\\
\hline
\end{tabular}
}
\caption{Conserved integrals for mKdV travelling waves.}
\label{table:conserved}
\end{table}

\section{Curve flows in the Euclidean plane}
\label{sec:curveflows}

In the Euclidean plane $\Rnum^2$,
consider a smooth curve (open or closed) $\vec{r}(s) = \big(x(s),y(s)\big)$ 
in Cartesian coordinates $(x,y)$.
Choose the parameter $s$ to be the arclength, namely
\begin{equation}\label{arclength} 
  \Big|\frac{d\vec{r}(s)}{ds}\Big| =1 . 
\end{equation}
The Frenet frame \cite{Gug-book} along the curve is given by
the tangent vector 
\begin{equation}\label{tangent.vec}
\T = \frac{d\vec{r}(s)}{ds}=\big(x'(s),y'(s)\big)
\end{equation}
and the normal vector 
\begin{equation}\label{normal.vec}
\N = \star\frac{d\vec{r}(s)}{ds}=\big(-y'(s),x'(s)\big)
\end{equation}
where $\star$ denotes the Hodge dual
which is a counter-clockwise rotation through an angle of $90^\circ$. 
This pair of vectors satisfies
\begin{equation}
\T\cdot\T=1,
\quad
\N\cdot\N=1, 
\quad
\T\cdot\N=0 . 
\end{equation}
The Frenet equations describe how the frame vectors change with position $s$ along the curve:  
\begin{equation}\label{Frenet.eqns}
\frac{d\T}{ds}=\kappa(s)\N,
\quad
\frac{d\N}{ds}=-\kappa(s)\T
\end{equation}
where $\kappa(s)$ defines the curvature.
This function describes the amount of bending in the curve. 
It can be directly expressed in terms of the frame vectors by the relation 
\begin{equation}\label{curvature}
\kappa(s)=\N\cdot\frac{d\T}{ds} . 
\end{equation}

\subsection{Non-stretching curve flows}

Now suppose the arclength-parameterized curve undergoes a flow 
$\vec{r}(s,t)$, with $t$ denoting time, 
such that the arclength is locally preserved at each point on the curve. 
This means the curve does not stretch as it moves 
(such motion is also called \emph{inelastic}). 
The non-stretching (inelastic) condition is given by 
\begin{equation}
\frac{\d}{\d t}\Big|\frac{\d\vec{r}(s,t)}{\d s}\Big|=0 , 
\end{equation}
which is equivalent to 
\begin{equation}
\T\cdot\frac{\d\T}{\d t}=0
\end{equation}
as expressed in terms of the Frenet frame. 
Note that the Frenet equations \eqref{Frenet.eqns}--\eqref{curvature} hold at each time $t$. 

Under the flow, the change in the Frenet frame is described by the equations
\begin{equation}\label{t.der.frame}
\frac{\d\T}{\d t}=\omega(s,t)\N,
\quad
\frac{\d\N}{\d t}=-\omega(s,t)\T
\end{equation} 
where 
\begin{equation}\label{omega}
\omega(s,t) = \N\cdot\frac{\d\T}{\d t} . 
\end{equation}
This function describes the amount of bending in the curve caused by its motion. 
It is related to the curvature $\kappa$ through the compatibility relation
between $s$-derivatives and $t$-derivatives.
In particular, from equations \eqref{curvature} and \eqref{omega}, 
$\dfrac{\d\kappa}{\d t} - \dfrac{\d\omega}{\d s}
= -\omega \T\cdot \dfrac{\d\T}{\d s} +\kappa \T\cdot \dfrac{\d\T}{\d t}
=0$
holds after use of the Frenet equations \eqref{Frenet.eqns} and their temporal counterpart \eqref{t.der.frame}.
Thus, 
\begin{equation}\label{eqn1}
\dfrac{\d\kappa}{\d t} = \dfrac{\d\omega}{\d s}
\end{equation}
holds in an arclength-preserving flow. 

The motion of the curve itself can be formulated as a differential equation
\begin{equation}\label{r.flow.eqn}
\vec{r}_t = h_\pa \T + h_\pe \N 
\end{equation}
where the scalar functions $h_\pa$ and $h_\pe$
represent the tangential and normal components of the flow.
Because the motion preserves arclength locally,
these components satisfy a compatibility relation.
Specifically, from the $s$-derivative of equation \eqref{r.flow.eqn}
combined with the $t$-derivative of the tangent vector \eqref{tangent.vec}, 
$\dfrac{\d}{\d s}\Big( h_\pa \T + h_\pe \N \Big) = \dfrac{\d \T}{\d t}$
implies that
\begin{equation}
\Big( \dfrac{\d h_\pa}{\d s} -\kappa h_\pe \Big)\T
+\Big( h_\pa \kappa + \dfrac{\d h_\pe}{\d s} -\omega \Big)\N
=0
\end{equation}
after use of the Frenet equations \eqref{Frenet.eqns} and their temporal counterpart \eqref{t.der.frame}.
This yields the equations
\begin{equation}\label{eqn2}
\omega =\frac{\d h_\pe}{\d s} + \kappa h_\pa 
\end{equation}
and
\begin{equation}\label{eqn3}
\frac{\d h_\pa}{\d s} = \kappa h_\pe . 
\end{equation}

The three equations \eqref{eqn1}, \eqref{eqn2}, \eqref{eqn3}
together constitute the structure equations
for an arclength-preserving flow of a curve \eqref{r.flow.eqn} 
in the Euclidean plane \cite{Gol.Pet.1993,Anc.2008}. 
It is important to note that no other conditions have been imposed on the flow so far.

These structure equations can be written in matrix form
\begin{equation}
\dfrac{\d}{\d s}\begin{pmatrix}  \T \\ \N \end{pmatrix}
=  \begin{pmatrix} 0 & \kappa \\ -\kappa & 0 \end{pmatrix}
\begin{pmatrix}  \T \\ \N \end{pmatrix},
\quad
\dfrac{\d}{\d t}\begin{pmatrix}  \T \\ \N \end{pmatrix}
=  \begin{pmatrix} 0 & \omega \\ -\omega & 0 \end{pmatrix}
\begin{pmatrix}  \T \\ \N \end{pmatrix} , 
\end{equation}
which is analogous to the linear isospectral equations \cite{Abl.Cla-book}
in the Lie algebra $\mk{so}(2)\subset \mk{sl}(2,\Rnum)$
with the spectral parameter set equal to zero.
In essence,
as is well known from AKNS soliton theory \cite{Abl.Cla-book}, 
this structure encodes the integrability features of
the mKdV hierarchy of integrable PDEs.

Combining the structure equations leads to an evolution equation for the curvature 
\begin{equation}\label{curvature.evolution}
\kappa_t = h_\pe{}_{ss} + \kappa^2 h_\pe + \kappa_s\int \kappa h_\pe\,ds 
\end{equation}
in terms of the normal component $h_\pe$ of the flow.
The right-hand side of this evolution equation defines an operator
\begin{equation}\label{R.op}
\Rop =\partial_s^2 +\kappa^2 +\kappa_s \partial^{-1}_s \kappa  , 
\end{equation}
which turns out to be precisely the recursion operator for symmetries 
of the mKdV hierarchy \cite{Olv-book}. 
(Further integrability features are discussed in the appendix.)

\subsection{Geometric realization of mKdV equation}

There is a simple choice of $h_\pe$ for which the curvature evolution equation \eqref{curvature.evolution} 
becomes the mKdV equation \eqref{mkdv.eqn}.
Comparison with the number of $s$-derivatives that appear in the operator $\Rop$
indicates that $h_\pe$ should accordingly be given by 
\begin{equation}\label{hperp.flow}
h_\perp=\kappa_s . 
\end{equation}
(A deeper motivation based on integrability features is explained in the appendix.)
This yields 
\begin{equation}
\kappa_t = \Rop(\kappa_s) = \tfrac{3}{2}\kappa^2 \kappa_s +\kappa_{sss} , 
\end{equation}
and hence the mKdV equation arises as the curvature evolution equation \eqref{curvature.evolution}
for the curve flow generated by the normal flow component \eqref{hperp.flow}.
The tangential flow component is given by
\begin{equation}\label{hpar.flow}
h_\pa = \partial^{-1}_s(\kappa \kappa_s) = \tfrac{1}{2} \kappa^2 . 
\end{equation}

Substitution of the components \eqref{hpar.flow} and \eqref{hperp.flow} 
into equation \eqref{r.flow.eqn} gives the geometric curve flow \eqref{mkdv.curveflow.eqn}
for $\vec{r}(s,t)$. 
Both expressions $\kappa^2$ and $\kappa_s$
can be written more directly in terms of $\vec{r}$ as follows. 
For $\kappa^2$,
the dot product of the first Frenet equation \eqref{Frenet.eqns} with itself 
yields
\begin{equation}\label{kappasq}
\kappa^2 = \Big|\frac{\d\T}{\d s}\Big|^2 = \big|\vec{r}_{ss}\big|^2
\end{equation}
after substitution of expression \eqref{tangent.vec} for the tangent vector. 
For $\kappa_s$, 
integration by parts on $\kappa_s\N$ followed by
substitution of both Frenet equations \eqref{Frenet.eqns}
gives 
\begin{equation}\label{kappa_sN}
\kappa_s\N 
=\frac{\d(\kappa \N)}{\d s} -\kappa \frac{\d\N}{\d s}
=\frac{\d^2\T}{\d s^2}+ \kappa^2\T 
= \vec{r}_{sss} + u^2\vec{r}_s . 
\end{equation}

Thus, the geometric curve flow \eqref{mkdv.curveflow.eqn} becomes 
\begin{equation}\label{mkdv.r.curveflow}
\vec{r}_t =\tfrac{3}{2}\big|\vec{r}_{ss}\big|^2\vec{r}_s +\vec{r}_{sss}
\end{equation}
expressed entirely in terms of $s$-derivatives of $\vec{r}$.
This flow is called the \emph{mKdV geometric map equation in the Euclidean plane}.
(A generalization exists for many other geometries \cite{Anc.2008}.)

\subsection{Non-stretching curve motions from mKdV solutions}

Every solution of the mKdV equation \eqref{mkdv.eqn} for $\kappa(s,t)$ 
yields a solution of the curve flow equation \eqref{mkdv.r.curveflow} for $\vec{r}(s,t)$. 
To reconstruct the curve $\vec{r}(s,t)$ from its curvature $\kappa(s,t)$, 
a parametric representation in terms of Cartesian coordinates $x(s,t)$ and $y(s,t)$
will be stated. 

\begin{lem}\label{lem:x.y.from.kappa}
A general arclength-parameterized curve flow \eqref{r.flow.eqn}
with curvature $\kappa(s,t)$ in $\Rnum^2$
has Cartesian coordinates $\vec{r}(s,t) =\big(x(s,t),y(s,t)\big)$ given by
\begin{subequations}\label{curveflow.x.y}
\begin{align}
x(s,t) & 
= \int_0^s \cos(\theta)\,ds 
+\int_0^t \big(h_\pa \cos(\theta) -h_\pe \sin(\theta)\big)\big|_{s=0}\, dt 
+ x_{0}
\label{x.eqn}
\\
y(s,t) &
=\int_0^s \sin(\theta)\,ds 
+\int_0^t\big( h_\pa\sin(\theta)+h_\pe \cos(\theta)\big)\big|_{s=0}\, dt 
+ y_{0}
\label{y.eqn}
\end{align}
\end{subequations}
where 
\begin{equation}\label{theta}
\theta(s,t) 
= \int_0^s \kappa\, ds + \int_0^t \big( h_\pe{}_s +h_\pa\kappa \big)\big|_{s=0}\, dt . 
\end{equation}
This is the angle of the tangent vector of the curve with respect to the $x$-axis. 
\end{lem}

The derivation of this lemma will be given in the next subsection.
A useful development is that the coordinate equations \eqref{x.eqn}--\eqref{y.eqn}
can be combined into a single equation using the complex coordinate $z= x+iy$.
In this way, a curve flow for $\vec{r}(s,t)$ in the Euclidean plane $\Rnum^2$
equivalently becomes
a curve flow $z(s,t)$ in the complex plane $\Cnum= (\Rnum^2, i)$.

\begin{thm}\label{thm:complex.curve.flow}
A general arclength-parameterized curve flow \eqref{r.flow.eqn} in $\Rnum^2$ 
is equivalent to a curve flow in the complex plane $\Cnum$
under the correspondence 
\begin{equation}\label{r.z.relation}
  \vec{r}(s,t) = \big(x(s,t),y(s,t)\big)
\quad \longleftrightarrow \quad
  z(s,t) = x(s,t) + i y(s,t)
\end{equation}
where 
\begin{equation}\label{z_t}
  z_t = (h_\pa + ih_\pe) z_s . 
\end{equation}
The condition of local arclength preservation \eqref{arclength} is given by
\begin{equation}
  |z_s| = 1 , 
\end{equation}
whereby
\begin{equation}\label{z_s.expitheta}
  z_s = e^{i\theta}
\end{equation}
with the angle $\theta$ being related to the curvature $\kappa$ of the curve by
\begin{equation}
  \theta_s = \kappa . 
\end{equation}
For a given $\theta(s,t)$,
the curve flow $z(s,t)$ is determined from the integral 
\begin{equation}\label{curveflow.z}
z(s,t) = \int_0^s e^{i\theta} \,ds 
+\int_0^t \big( (h_\pa + i h_\pe ) e^{i\theta} \big)\big|_{s=0}\, dt 
+ z_{0} . 
\end{equation}
\end{thm}

The proof of this theorem will be given in the next subsection.

A main consequence of Theorem~\ref{thm:complex.curve.flow} and Lemma~\ref{lem:x.y.from.kappa}
is to provide a direct way for constructing
an arclength-parameterized curve flow from its curvature.
This will now be applied to the mKdV equation \eqref{mkdv.eqn}. 

For any given mKdV solution $\kappa(s,t)$,
the corresponding mKdV curve flow can be constructed 
in an equivalent complex variable form 
\begin{equation}\label{r.z}
  \vec{r}(s,t) = \big(\mathrm{Re}\, z(s,t),\mathrm{Im}\,z(s,t)\big)
\end{equation}
by evaluating the integral \eqref{curveflow.z}
after using expressions \eqref{hperp.flow} and \eqref{hpar.flow}
for $h_\pe$ and $h_\pa$. 
Specifically, first, $\theta(s,t)$ is obtained from 
\begin{equation}\label{theta.mkdv}
\theta(s,t) 
= \int_0^s \kappa\, ds + \int_0^t \big( \kappa_{ss} +\tfrac{1}{2}\kappa^3 \big)\big|_{s=0}\, dt
\end{equation}
and then, $z(s,t)$ is obtained from
\begin{equation}\label{z.mkdv}
  z(s,t) =
  \int_0^s e^{i\theta}\,ds 
+\int_0^t \big( (\tfrac{1}{2}\kappa^2 + i \kappa_s) e^{i\theta} \big)\big|_{s=0}\, dt 
+ z_{0} . 
\end{equation}
The resulting mKdV curve flow \eqref{r.z} 
will satisfy the geometric equation \eqref{mkdv.curveflow.eqn}
which has the equivalent form
\begin{equation}\label{z_t.mkdv}
  z_t = (\tfrac{1}{2}\theta_s^2 + i\theta_{ss}) z_s
\end{equation}
in terms of the angle $\theta$ of $z_s$ on the unit circle $|z_s| =1$.
Through the latter relation $z_s = e^{i\theta}$, 
and its derivatives $z_{ss} = i\theta_s e^{i\theta}$
and $z_{sss} = (i\theta_{ss} - \theta_s^2) e^{i\theta}$, 
the geometric equation \eqref{z_t.mkdv} can be expressed explicitly in terms of $z$ as
\begin{equation}\label{mkdv.z.eqn}
  z_t = z_{sss} - \tfrac{1}{2}|z_{ss}|^2 z_s . 
\end{equation}
This will be called the \emph{mKdV geometric map equation in $\Cnum$}.
The angle $\theta$ correspondingly satisfies the equation
\begin{equation}\label{mkdv.theta.eqn}
\theta_t = \theta_{sss} +\tfrac{1}{2}\theta_s^3
\end{equation}
which is recognized to be the mKdV equation in potential form.

\subsection{Derivation of the curve construction lemma and its complex variable formulation }

As the first step in the derivation of Lemma~\ref{lem:x.y.from.kappa}, 
observe that the arclength relation \eqref{arclength} 
expressed in Cartesian coordinates is 
\begin{equation}\label{x.y.arclength}
x_s^2+ y_s^2 = 1 . 
\end{equation}
This allows expressing 
\begin{equation}\label{x.y.cos.sin}
x_s = \cos(\theta), 
\quad
y_s = \sin(\theta)
\end{equation}
where $\theta(s,t)$ is the angle of the tangent vector \eqref{tangent.vec}, namely 
\begin{equation}\label{x.y.tangent.vec}
\T = \big(\cos(\theta), \sin(\theta)\big) .
\end{equation}
For later, it is useful to note that the normal vector \eqref{normal.vec} is given by
\begin{equation}\label{x.y.normal.vec}
\N = \big({-}\sin(\theta),\cos(\theta)\big) .
\end{equation}
Furthermore, substituting these expressions into the curvature relation \eqref{curvature}
shows that 
\begin{equation}\label{theta_s.curvature}
\kappa = \theta_s .
\end{equation}
Thus, $\theta(s,t)$ is also a potential for the curvature. 

For the second step,
start by integrating equation \eqref{x.y.cos.sin} with respect to $s$ at fixed $t$,
which yields
\begin{equation}\label{x.y.theta}
x(s,t) = \int_0^s \cos(\theta)\, ds + C_1(t),
\quad
y(s,t) = \int_0^s \sin(\theta)\, ds + C_2(t). 
\end{equation}
To determine $C_1(t)$ and $C_2(t)$,
differentiate the preceding expressions with respect to $t$
and substitute the Cartesian components of the curve flow \eqref{r.flow.eqn}, 
\begin{equation}\label{x_t.y_t}
  x_t = h_\pa \cos(\theta) - h_\pe \sin(\theta) , 
  \quad
  y_t = h_\pa \sin(\theta)  +h_\pe \cos(\theta) .
\end{equation}
Then, evaluate at $s=0$, yielding
\begin{equation}
C_1'(t) = \big( h_\pa \cos(\theta) - h_\pe \sin(\theta) \big)\big|_{s=0}, 
  \quad
C_2'(t) =   \big( h_\pa \sin(\theta)  +h_\pe \cos(\theta) \big)\big|_{s=0} .
\end{equation}
Integration of this pair of equations leads directly to the integrals \eqref{curveflow.x.y}.

The last step consists of deriving the integral for $\theta$.
Differentiating the flow components \eqref{x_t.y_t} with respect to $s$, 
followed by substituting expressions \eqref{x.y.cos.sin}, 
yields 
\begin{align}
& \big( h_\pa{}_s - h_\pe\theta_s \big)  \cos(\theta)
- \big( h_\pe{}_s + h_\pa\theta_s \big) \sin(\theta) 
=   -\sin(\theta) \theta_t,
\\
& \big( h_\pa{}_s - h_\pe\theta_s \big)\sin(\theta)
+\big( h_\pe{}_s +h_\pa\theta_s \big) \cos(\theta)
= \cos(\theta) \theta_t .
\end{align}
The first term in each of these equations vanishes through
the arclength-preservation condition \eqref{eqn3}
combined with relation \eqref{theta_s.curvature}.
Consequently, both equations reduce to 
\begin{equation}\label{theta_t}
\theta_t = h_\pe{}_s +h_\pa\theta_s 
.\end{equation}
This is simply the potential form of the curvature evolution equation \eqref{curvature.evolution}.
Integration of the relation \eqref{theta_s.curvature} with respect to $s$ gives
\begin{equation}
\theta =  \int_0^s \kappa \,ds + C_0(t) .
\end{equation}
Substituting this integral into equation \eqref{theta_t} and evaluating at $s=0$,
yields
\begin{equation}
C_0'(t) = \big( h_\pe{}_s +h_\pa\kappa \big)\big|_{s=0}
\end{equation}
after use of relation \eqref{theta_s.curvature}.
Finally, integration with respect to $t$ leads to the integral \eqref{theta}.

This completes the derivation. 

The proof of Theorem~\ref{thm:complex.curve.flow} amounts to transcribing
the above steps into complex variable notation. 
Details will be omitted.

\section{Curve flows arising from travelling waves} 
\label{sec:prelims}

For a given mKdV travelling wave \eqref{travelling.wave},
the parametric integrals \eqref{theta.mkdv}--\eqref{z.mkdv}
yield the corresponding loop solutions in the complex-coordinate form
$z(s,t) = x(s,t) + iy(s,t)$,
with the curve flow in $\Rnum^2$ being given by the algebraic relation \eqref{r.z}.
These parametric integrals turn out to simplify considerably when evaluated for travelling waves,
leading to the following result. 

\begin{thm}\label{thm:mkdv.loops}
Up to a shift in parameterization,
the loop solution for a mKdV travelling wave \eqref{travelling.wave}
is given by
\begin{subequations}\label{z.loop.eqn}
\begin{numcases}
{z(s,t) =}
J(s) e^{i(C_1t + \vartheta(s))} +z_0, 
& $C_1\neq 0$
\label{z.loop.eqn.C1not0}
\\&\nonumber\\
\int_0^{s} e^{i\vartheta(\xi)}\,d\xi +\nu t +z_0,
& $C_1=0$
\label{z.loop.eqn.C1is0}
\end{numcases}
\end{subequations}
with
\begin{subequations}
\begin{align}
J(s) & =  \tfrac{1}{C_1}\big( K'(s)  + i(c -\tfrac{1}{2} K(s)^2 ) \big),
\label{J}
\\
\nu & = -c + \tfrac{1}{2} K(0)^2 +i K'(0),
\label{nu}
\end{align}
\end{subequations}
where 
\begin{equation}\label{vartheta}
\vartheta(s) = \int_0^{s} K(\xi)\,d\xi , 
\end{equation}
and where 
$C_1$ is the integration constant in the travelling wave equation \eqref{tw.eqn}
and $z_0$ is an arbitrary constant.
For $C_1\neq0$,
$z(s,t)$ has the form of a rotating loop with angular speed $C_1$, 
while for $C_1=0$, 
$z(s,t)$ has the form of a travelling loop
with speed $|\nu|$ and direction angle
$\arctan(\mathrm{Im}\,\nu/\mathrm{Re}\,\nu)$.
\end{thm}

\begin{rem}
In the case $C_1=0$,
if the mKdV travelling wave \eqref{travelling.wave} is shifted to satisfy $K'(0) =0$,
then the loop solution travels in the $x$-direction with speed $\tfrac{1}{2} K(0)^2 -c$.
\end{rem}

The subsequent subsections will evaluate the loop expression \eqref{z.loop.eqn}
for each type of mKdV travelling wave,
using the classification given in Theorem~\ref{thm:travellingwaves}. 
For each kind of resulting loop solution,
its dynamics and other notable features
--- such as whether the loop is open or closed, and the winding number of the loop --- 
will be discussed.

Theorem~\ref{thm:mkdv.loops} will now be proved.

Let $\kappa(s,t) = K(s+ct)$ be a mKdV travelling wave \eqref{travelling.wave}.
First, consider the parametric integral \eqref{theta.mkdv},
which becomes
\begin{equation}
\begin{aligned}
\theta(s,t) 
& = \int_0^s K(s+ct)\, ds + \int_0^t \big( K''(ct) +\tfrac{1}{2} K'(ct)^3 \big)\, dt
\\
& = \int_{ct}^{s+ct} K(\xi)\, d\xi + \frac{1}{c}\int_0^{ct} \big( K''(\xi) +\tfrac{1}{2} K'(\xi)^3 \big)\, d\xi
\\
& = \int_{ct}^{s+ct} K(\xi)\, d\xi + \frac{1}{c}\int_0^{ct} \big( C_1 + c K(\xi) \big)\, d\xi
\\
& = \int_{0}^{s+ct} K(\xi)\, d\xi + C_1 t
\end{aligned}
\end{equation}
after use of the travelling wave equation \eqref{tw.eqn}. 
Hence, 
\begin{equation}
\theta(s,t) = \vartheta(s+ct) + C_1 t , 
\end{equation}
from which it follows that
\begin{equation}\label{vartheta_s.curvature}
  \kappa(s,t) = \vartheta'(s+ct) . 
\end{equation}
Next, the other parametric integral \eqref{z.mkdv} similarly becomes
\begin{equation}
\begin{aligned}
z(s,t) 
& = \int_0^s e^{i(\vartheta(s+ct) + C_1t)}\, ds
+ \int_0^t \big( \tfrac{1}{2} K(ct)^2 + i K'(ct) \big)e^{i(\vartheta(ct) + C_1t)}\, ds\, dt
\\
& = e^{iC_1t} \int_{ct}^{s+ct} e^{i\vartheta(\xi)}\, d\xi
+ \frac{1}{c}\int_0^{ct} \big( \tfrac{1}{2} K(\xi)^2 + i K'(\xi) \big)e^{i(\vartheta(\xi) + \frac{C_1}{c}\xi)}\, d\xi , 
\end{aligned}
\end{equation}
and thus
\begin{equation}\label{z.f}
z(s,t) = e^{iC_1t} \int_{0}^{s+ct} e^{i\vartheta(\xi)}\, d\xi +f(t) 
\end{equation}
where
\begin{equation}
  f(t) =   \int_0^{ct} \Big(
  c^{-1}\big( \tfrac{1}{2} K(\xi)^2 + i K'(\xi) \big) e^{i \frac{C_1}{c}\xi} - e^{iC_1t}
  \Big) e^{i\vartheta(\xi)}\, d\xi . 
\end{equation}
A differential equation for this function $f(t)$ can be derived from the curve flow equation \eqref{z_t.mkdv}.
Differentiation of expression \eqref{z.f} yields
\begin{equation}
  z_s = e^{iC_1t} e^{i\vartheta(s+ct)},
  \quad
  z_t =   e^{iC_1t}\Big( iC_1 \int_{0}^{s+ct} e^{i\vartheta(\xi)}\, d\xi + c\,e^{i\vartheta(s+ct)} \Big)   +f'(t) . 
\end{equation}
Substitution of these expressions into equation \eqref{z_t.mkdv}, 
followed by use of relation \eqref{vartheta_s.curvature},
gives
\begin{equation}\label{f.F}
  -e^{-iC_1t} f'(t)  = F(s+ct)
\end{equation}
where 
\begin{equation}\label{F.rel}
  F(s+ct) =  iC_1 \int_{0}^{s+ct} e^{i\vartheta(\xi)}\, d\xi + \big( c - \tfrac{1}{2} K(s+ct)^2 - i K'(s+ct) \big) e^{i\vartheta(s+ct)}
\end{equation}
after use of relation \eqref{vartheta_s.curvature} again. 
Equation \eqref{f.F} is separable with respect to $\xi=s+ct$ and $t$,
and therefore
\begin{equation}\label{F.eqn}
F(s+ct) = F_0
\end{equation}
and
\begin{equation}\label{f.eqn}
  f'(t) =   -e^{iC_1t} F_0
\end{equation}
where the separation constant is given by 
\begin{equation}
  F_0 = F(0)  = c - \tfrac{1}{2} K(0)^2 - i K'(0) . 
\end{equation}
Next, integration of equation \eqref{f.eqn} yields
\begin{equation}
  f(t) =
  \begin{cases}
    i (F_0/C_1) e^{iC_1t} + f_0, & C_1\neq 0
    \\
    -F_0 t, & C_1 =0 . 
  \end{cases}
\end{equation}
Substitution of this expression into equation \eqref{z.f} gives
\begin{equation}
z(s,t) =
\begin{cases}
\displaystyle
e^{iC_1t}\Big( \int_0^{s+ct} e^{i\vartheta(\xi)}\,d\xi +i F_0/C_1 \Big) + z_0, 
& C_1\neq 0
\\
\displaystyle
\int_0^{s+ct} e^{i\vartheta(\xi)}\,d\xi -F_0 t +z_0,
& C_1=0 . 
\end{cases}
\end{equation}
This directly yields the result \eqref{z.loop.eqn.C1is0} 
after the re-parameterization $s+ct\to s$. 
For the case $C_1\neq0$,
the result \eqref{z.loop.eqn.C1not0} is obtained 
by combining equations \eqref{F.rel} and \eqref{F.eqn} to get the relation
\begin{equation}
  \int_0^{s+ct} e^{i\vartheta(\xi)}\,d\xi +i F_0/C_1
  = \tfrac{1}{C_1} \big( K'(s+ct) + i(c - \tfrac{1}{2} K(s+ct)^2 )  \big) e^{i\vartheta(s+ct)}, 
\end{equation}
followed by re-parameterizing $s+ct\to s$. 

This completes the proof.

\subsection{Local and global conserved geometric quantities} 

The non-stretching curve flows \eqref{r.z} that arise from mKdV solutions $\kappa(s,t)$
possess an infinite number of conserved geometric integrals. 
When these integrals are expressed as local conservation laws
\begin{equation}\label{local.conslaw}
  D_t \Psi^t + D_s \Psi^s =0,
  \end{equation}
they are applicable to any finite portion of a given curve.

Arclength $\int ds$, which is inherently conserved for any non-stretching curve, 
corresponds to
\begin{equation}
  \Psi^t = 1,
  \quad
  \Psi^s=0 . 
\end{equation}
Area enclosed by a curve,
$\oint \tfrac{1}{2} {*\vec{r}} \cdot \vec{r}_s\,ds
= \mathrm{Re}\oint \tfrac{1}{2} i z\bar z_s\,ds$, 
is well known to be conserved (see e.g. \cite{Gol.Pet.1993}).
It corresponds to 
\begin{equation}
\Psi^t = \tfrac{1}{2}\mathrm{Re}( i z\bar z_s ) = -\tfrac{1}{2}\mathrm{Im}( z\bar z_s ),
  \quad
\Psi^s= \mathrm{Im} ( \tfrac{1}{2}  z\bar z_t  + (\ln z_s)_s ) . 
\end{equation}
These expressions follow from the proof of area conservation in local form:
\begin{equation}
  \begin{aligned}
\partial_t \big( \tfrac{1}{2} \mathrm{Re}(i z\bar z_s) \big) 
&   = \tfrac{1}{2}\mathrm{Re}\big( i\big( z_t\bar z_s - z_s\bar z_t + (z\bar z_t)_s\big) \big)
\\
& = \tfrac{1}{2}\mathrm{Re} \big( {-2}\theta_{ss} + (z\bar z_t)_s \big)
\\
& = \partial_s \big( {-\theta_s}  -\tfrac{1}{2}\mathrm{Im}(z\bar z_t) \big)
  \end{aligned}
\end{equation}
using the mKdV curve flow equation \eqref{z_t.mkdv}
together with the angle relation \eqref{z_s.expitheta}. 

Thus, local arclength and local ``area'' are conserved quantities.
For a closed curve,
they can evaluated over the entire curve to obtain finite conserved integrals
--- namely, constants of motion ---
yielding total arclength and area.
For an open curve,
they provide either
finite conserved integrals if suitable asymptotic conditions hold as $s \to\pm\infty$,
or balance equations \eqref{balance} on any finite portion of the curve. 

In addition, every local conservation law holding for the mKdV equation
gives rise to a corresponding local conservation law
for the mKdV curve flow equation \eqref{z_t.mkdv}.
The conservation laws discussed in section~\ref{sec:conserved} yield
a hierarchy of curvature quantities,
all of which can be expressed in terms of the angle $\theta$ of the tangent vector. 

The integral for mass conservation \eqref{mass} is given by
\begin{equation} 
\C_0(s_1,s_2) =  \int_{s_1}^{s_2} \theta_s \,ds = \theta\Big|_{s_1}^{s_2}
\end{equation}
between any two points $z(s_1)$ and $z(s_2)$ on a curve (at fixed time $t$). 
This is the net change in the turning angle in the tangent vector
from $z(s_1)$ to $z(s_2)$,
which after division by $2\pi$ is called the \emph{winding number},
\begin{equation}\label{winding.number}
N:= \tfrac{1}{2\pi} \theta\Big|_{s_1}^{s_2} = \tfrac{1}{2\pi} \C_0 . 
\end{equation}
It is a constant of motion for closed curves
or for open curves that have suitable asymptotic conditions as $s\to \pm\infty$.
It is locally conserved (in the sense of a balance equation)
on any finite portion of an open or closed curve.

Momentum conservation \eqref{momentum} similarly yields the integral
\begin{equation} 
\C_1(s_1,s_2) =  \int_{s_1}^{s_2} \tfrac{1}{2} \theta_s^2 \,ds
 = \tfrac{1}{2} \int_{\theta_1}^{\theta_2} \theta_s \,d\theta  . 
\end{equation}
This is proportional to the average rate of turning from $z(s_1)$ to $z(s_2)$,
which can be thought of as defining a \emph{mean-winding rate} of the tangent vector:
\begin{equation} 
W:= \tfrac{1}{2\pi} \int_{\theta_1}^{\theta_2} \theta_s \,d\theta = \tfrac{1}{\pi}\C_1 . 
\end{equation}\label{winding.rate}
Likewise, the integral for energy conservation \eqref{energy} can be written 
\begin{equation}
\C_2(s_1,s_2)=  \int_{s_1}^{s_2} ( -\tfrac{1}{2}\theta_{ss}^2 + \tfrac{1}{8} \theta_s^4 )\,ds
  = -\tfrac{1}{2} \Big( \theta_s \Big|_{s_1}^{s_2} -   \tfrac{1}{4} \int_{\theta_1}^{\theta_2} \theta_s^3 \,d\theta \Big) , 
\end{equation}
which is proportional to the net change in turning rate minus a power of the winding rate. 

When a curve asymptotically winds around a circle of radius $R$, 
the winding number and mean-winding rate will be infinite.
However, the contribution due to the asymptotic circle can be removed,
yielding finite conserved quantities: 
\begin{equation}\label{renormalized.winding.number}
  \mathcal{N}:  = \tfrac{1}{2\pi}  \int_{-\infty}^{\infty} (\theta_s -1/R)\,ds
\end{equation}
which is a renormalized winding number;
and
\begin{equation}
  \mathcal{W}:  = \tfrac{1}{2\pi} \int_{-\infty}^{\infty} (\theta_s -1/R)\,d\theta 
\end{equation}\label{renormalized.winding.rate}
which is a renormalized mean-winding rate.

\section{mKdV loop solutions}
\label{sec:loops}

The loop solutions given by the non-stretching curve flows
that arise from each kind of mKdV travelling wave
will now be worked out by means of Theorem~\ref{thm:mkdv.loops}. 
Cnoidal mKdV waves \eqref{cn} and rational $\CN$ waves \eqref{ratcn}
for $0<q<1$ will be considered separately
from dnoidal waves \eqref{dn} and rational $\DN$ waves \eqref{ratdn}
for which $q>1$.
The main features of each kind of loop solution,
highlighted in section~\ref{sec:intro},
will be discussed.

\subsection{Soliton loops}
\label{sec:soliton.loops}

Ordinary mKdV solitons \eqref{soliton.b=0}
are bright/dark, corresponding to the $\pm$ sign. 
Since the constant $C_1$ given by expression \eqref{C1.C2.soliton} is zero when $b=0$, 
the resulting loop solution is obtained by evaluation of
the parametric integrals \eqref{vartheta} and \eqref{z.loop.eqn.C1is0}
plus expression \eqref{nu}.

The first parametric integral yields
\begin{equation}
  \vartheta(s) = \pm\big( 4\arctan(\exp(\sqrt{c}\,s)) - \pi \big) . 
\end{equation}
This gives
\begin{equation}
e^{i\vartheta(s)} = (1 \pm i \sinh(\sqrt{c}\,s))^2 \sech(\sqrt{c}\,s)^2 . 
\end{equation}
Then the second parametric integral gives 
\begin{equation}
\int_0^{s} e^{i\vartheta(s)}\, ds
= \tfrac{2}{\sqrt{c}}\big( \tanh(\sqrt{c}\,s) \mp i\, \sech(\sqrt{c}\,s) \big) -s \pm\tfrac{2}{\sqrt{c}}i . 
\end{equation}
Evaluation of expression \eqref{nu} yields
$\nu = -c + \tfrac{1}{2}(2\sqrt{c})^2 = c$. 
Substitution into the loop equation \eqref{z.loop.eqn.C1is0} produces
\begin{equation}\label{soliton.b=0.loop}
z(s,t) = \tfrac{2}{\sqrt{c}}\big( \tanh(\sqrt{c}\,s) \mp i\, \sech(\sqrt{c}\,s) \big)  - s +ct
  + z_0
\end{equation}
after absorbing constant terms into $z_0$.
This is a travelling loop that has the same speed as the mKdV soliton.

\begin{figure}[h]
\centering
\includegraphics[width=0.45\textwidth,trim=2cm 16cm 6cm 6cm,clip]{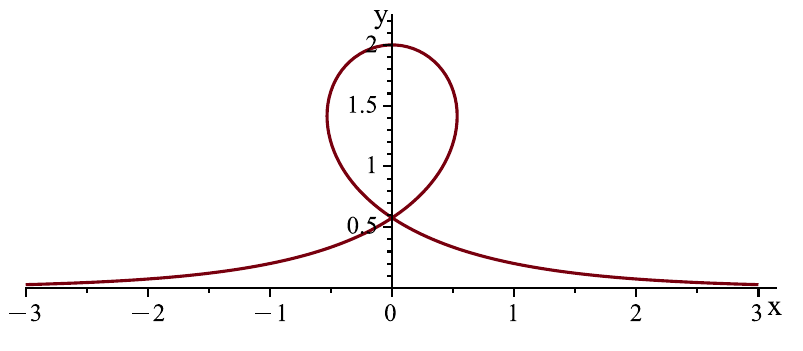}
\hfill
\includegraphics[width=0.45\textwidth,trim=2cm 16cm 6cm 6cm,clip]{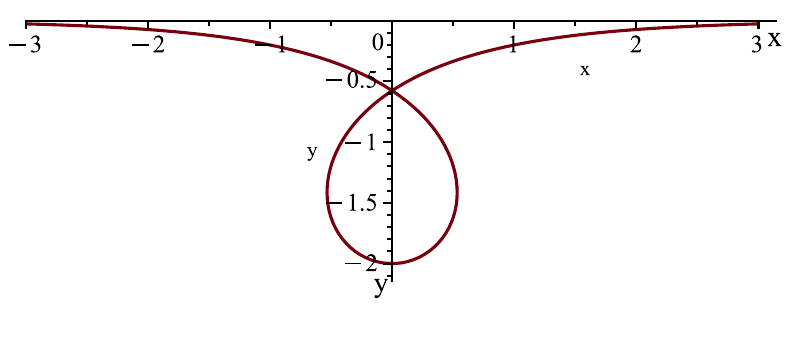}
\caption{mKdV soliton loop, $c=1$.}
\label{fig:soliton_loop}
\end{figure}

The corresponding geometric curve flow \eqref{r.z} is given by
\begin{equation}\label{soliton.b=0.curveflow}
\vec{r}_\mp(s,t)  =
  \big( ct - s + \tfrac{2}{\sqrt{c}}\,\tanh(\sqrt{c}\,s) , \mp\tfrac{2}{\sqrt{c}}\,\sech(\sqrt{c}\,s) \big) . 
\end{equation}
This flow is a superposition of
a static loop
\begin{equation}\label{soliton.b=0.staticloop}
\vec{r}_\mp(s,0)  =
\big({-s} + \tfrac{2}{\sqrt{c}}\,\tanh(\sqrt{c}\,s) , \mp\tfrac{2}{\sqrt{c}}\,\sech(\sqrt{c}\,s) \big) 
\end{equation}
plus a translational motion $\big( ct , 0 \big)$ 
with speed $c$ in the $x$-direction.
A plot of the static loop is shown in Fig.~\ref{fig:soliton_loop}

The loop \eqref{soliton.b=0.staticloop} is open, has a single peak at $s=0$, 
and asymptotically approaches a straight line as $|s| \to \infty$.
Note that $s=\infty$ corresponds to $(-\infty,0)$ 
and $s=-\infty$ corresponds to $(\infty,0)$. 
These features are an immediate consequence of the basic features of its curvature
given by the mKdV soliton \eqref{soliton.b=0},
namely it has a single peak at $\xi=0$ and decays exponentially to zero as $|\xi|\to \infty$.
However, the left tail of the soliton ($\xi\to -\infty$)
corresponds to the right tail of the loop ($s\to +\infty$), 
and vice versa.
Thus, the loop is oriented down/up according to the soliton being bright/dark.

The height $h$ of the loop is the vertical distance
between its peak and the asymptotic line,
which is given by $y(0,0)$. 
This yields
\begin{equation}
  h = \mp 2/\sqrt{c}
\end{equation}
where the sign is the orientation of the loop. 

The width $w$ of the loop is the distance $|x(s,0) - x(-s,0)|$
between the two points closest to the peak
such that the tangent vector is parallel to the $y$-axis,
as determined by $x_s(|s|,0)=0$.
This gives
\begin{equation}
  w = \ln(3+2\sqrt{2})/\sqrt{c} . 
\end{equation}

From the results in Table~\ref{table:conserved}
combined with equations \eqref{winding.number} and \eqref{winding.rate},
the loop has winding number $N=1$
and mean-winding rate $W = 4\sqrt{c}/\pi$.

\subsection{Rational loops}
\label{sec:rational.loops}

For heavy-tail (rational) mKdV waves\eqref{heavytail}, 
it will be convenient to use the parameterization
\begin{equation}
\kappa = K(\xi) = b\Big( 1 - \frac{4}{b^2 \xi^2 + 1}\Big)
\end{equation}
which involves only the background constant,
where the wave speed is $c = \tfrac{3}{2}b^2>0$
while the background can be positive or negative, $b\gtrless 0$. 

The loop solution is obtained by evaluation of
the parametric integral \eqref{vartheta}
and the algebraic formula \eqref{z.loop.eqn.C1not0}, 
plus expression \eqref{J}. 
First, the integral is given by 
\begin{equation}
\vartheta(s) = b s - 4\arctan(bs) , 
\end{equation}
which yields 
\begin{equation}
 e^{i\vartheta(s)} = e^{i bs} \dfrac{(bs + i)^2}{(bs -i)^2}
=  e^{i bs}\dfrac{b^4 s^4 - 6b^2 s^2 + 1 + i  4b s (b^2 s^2 - 1)}{(b^2 s^2 + 1)^2} . 
\end{equation}
Next, expression \eqref{J} yields 
\begin{equation}
  J(s) = \frac{(8b s + i(b^4 s^4 + 6 b^2 s^2  -3))b^2}{C_1(b^2 s^2 + 1)^2} 
\end{equation}
with $C_1$ given by \eqref{C1.C2.heavytail}.
Thus, the loop equation \eqref{z.loop.eqn.C1not0} produces
\begin{equation}\label{heavytail.loop}
  z(s,t) = \dfrac{4b s + i(3 -b^2 s^2)}{b(b^2 s^2 + 1)} e^{ibs} e^{-ib^3 t}
  + z_0 . 
\end{equation}
This is a rotating loop whose angular speed is $-b^3$. 

\begin{figure}[h]
\centering
 \begin{subfigure}[t]{0.9\textwidth}
\centering
\includegraphics[width=0.3\textwidth,trim=4cm 12cm 8cm 2cm,clip]{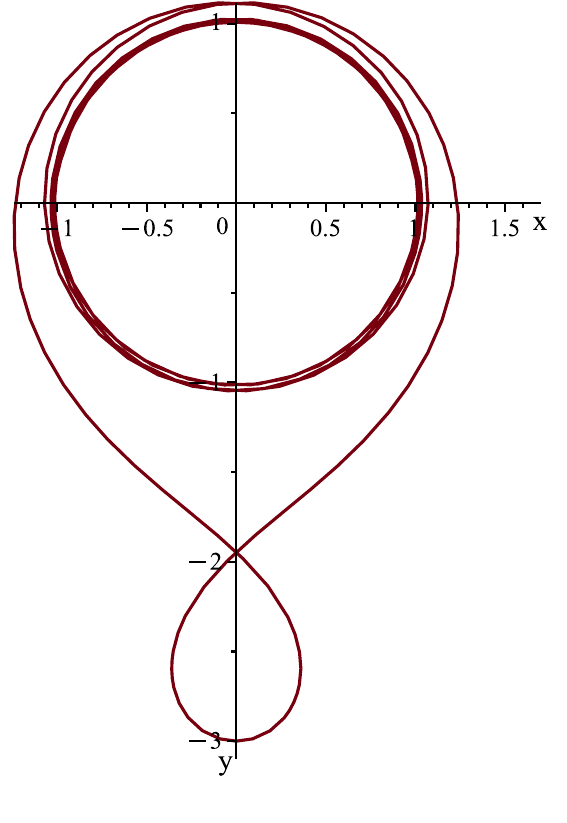}
\hfill
\includegraphics[width=0.3\textwidth,trim=4cm 12cm 8cm 2cm,clip]{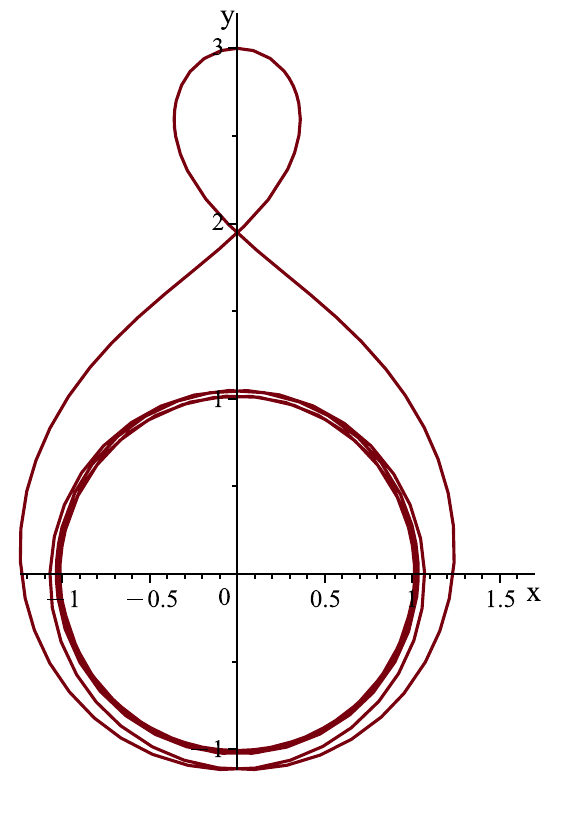}
\caption{Left: $b=-1$. Right: $b=1$.}
\end{subfigure}

\begin{subfigure}[t]{0.9\textwidth}
\centering
\includegraphics[width=0.3\textwidth,trim=4cm 12cm 8cm 2cm,clip]{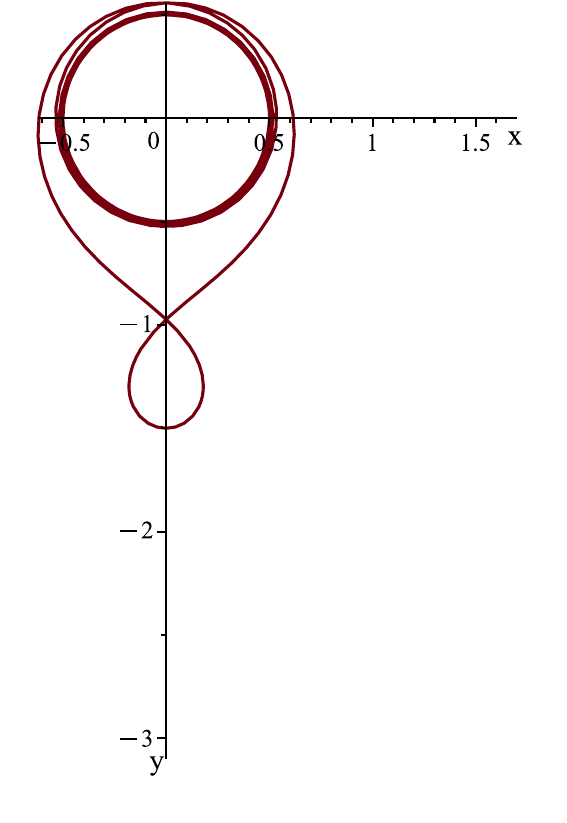}
\hfill
\includegraphics[width=0.3\textwidth,trim=4cm 12cm 8cm 2cm,clip]{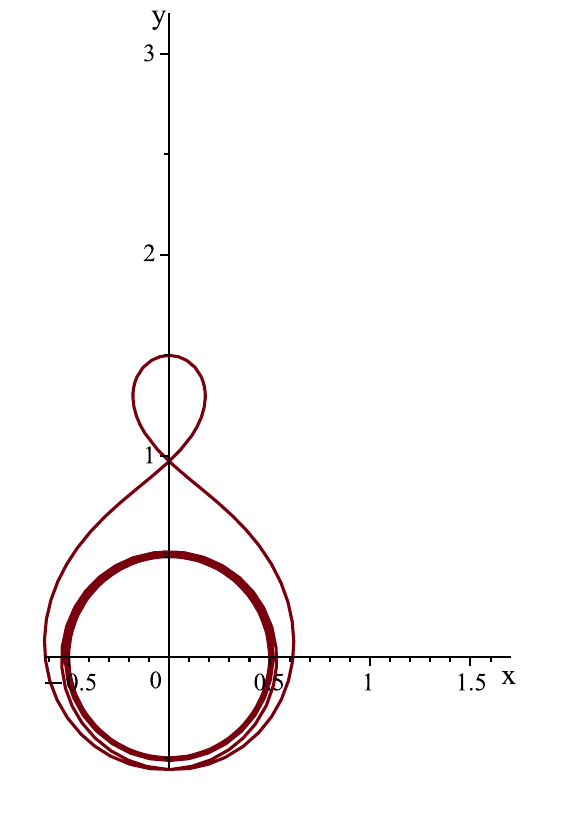}
\caption{Left: $b=-2$. Right: $b=2$.}
\end{subfigure}
\caption{mKdV rational loop}
\label{fig:rational-loop}
\end{figure}

The resulting geometric curve flow \eqref{r.z} is given by
\begin{equation}\label{heavytail.curveflow}
\begin{aligned}
\vec{r}(s,t)  =
\frac{1}{b^2 s^2 + 1} & \Big( 
4 s\cos(b(s -b^2 t)) +(b s^2 - 3/b) \sin(b(s-b^2 t)) ,
\\&\quad
4s\sin(b(s-b^2 t)) -(b s^2 - 3/b)\cos(b(s-b^2 t)) 
\Big) . 
\end{aligned}
\end{equation}
This flow is the composition of a static loop
\begin{equation}\label{heavytail.staticloop}
\vec{r}(s,0)  =
\frac{1}{b^2 s^2 + 1} \Big( 
4 s\cos(bs) +(b s^2 - 3/b) \sin(bs) ,
4s\sin(bs) -(b s^2 - 3/b)\cos(bs) 
\Big)
\end{equation}
plus a rotational motion with angular speed $-b^3$. 
A plot of the static loop is shown in Fig.~\ref{fig:rational-loop}

The static loop has the following features.
It asymptotically winds around a circle of radius $1/|b|$ as $|s|\to \infty$,
as shown by 
\begin{equation}
|\vec{r}(s,0)| = \frac{1}{|b|} \sqrt{\frac{b^2 s^2 + 9}{b^2 s^2 + 1}}
= \frac{1}{|b|} + O(1/s^2),
\quad
\vartheta_s = K(s) = b + O(1/s^2) . 
\end{equation}
This winding is counter-clockwise for $b>0$, and clockwise for $b<0$.
The loop has a single twist that forms a smaller self-intersecting loop 
whose peak is at $s=0$. 
Its height $h$ is given by
\begin{equation}
h=  \sgn(b)\, /|b|
\end{equation}
where the sign is the up/down orientation of the loop. 
Its width $w$ is the distance $|x(s,0) - x(-s,0)|$
between the closest pair of points around $s=0$ 
at which the tangent vector is parallel to the $y$-axis,
as given by $x_s(|s|,0)=\cos(\vartheta(|s|))=0$.
This determines $s= \zeta/b$ where $\zeta$ is the smallest positive root of
$4\tan(\zeta) = (\zeta^4 - 6\zeta^2 + 1)/(\zeta(\zeta^2 - 1))$.
Numerically, $\zeta \approx 0.605$.
Evaluating the $x$-component of expression \eqref{heavytail.staticloop}
with $s\approx 0.605/b$ and substituting into $|x(s,0) - x(-s,0)|$
then yields
\begin{equation}
w\approx 0.721/|b| . 
\end{equation}

The winding number \eqref{winding.number} of the whole loop is infinite,
due to the winding around the asymptotic circle.
When the contribution from the asymptotic circle is removed,
the renormalized winding number \eqref{renormalized.winding.number} is 
\begin{equation}\label{heavytail.winding.number}
  \mathcal{N}  = -\sgn(b) 2 . 
\end{equation}
Similarly, the renormalized mean-winding rate \eqref{renormalized.winding.rate} is 
\begin{equation}
  \mathcal{W}  = 2 |b| . 
\end{equation}
The result \eqref{heavytail.winding.number} can be visualized in the case $b>0$
by considering a point on the loop just above the asymptotic circle,
with the twisted (self-intersecting) portion of the loop lying directly above the point.
As the point moves counterclockwise around the bottom of the loop
and up to the top of the twist, the tangent vector has turned through an angle $-\pi$;
by symmetry, the tangent vector turns an additional angle of $-\pi$ 
as the point moves counterclockwise back to its original position. 
Hence, the net change in angle is $-2\pi$.

\subsection{Asymptotically circular solitary loops}
\label{sec:nonzerobc.soliton.loops}

For bright/dark mKdV solitons on a non-zero background \eqref{soliton}, 
the loop solutions are obtained similarly to the rational loop solution.
Note that $c$ and $b$ obey the inequality $c>\tfrac{3}{2}b^2 >0$.

First, the parametric integral \eqref{vartheta} is given by 
\begin{equation}
  \vartheta(s) = b s \pm 4\arctan\Bigg(
  \frac{\sqrt{c -\tfrac{1}{2}b^2} \mp b}{\sqrt{c -\tfrac{3}{2}b^2}} \tanh\big(\tfrac{1}{2}\sqrt{c -\tfrac{3}{2}b^2}\, s\big)
  \Bigg) , 
\end{equation}
and hence
\begin{equation}
  e^{i\vartheta(s)} =
  \frac{\Big( \sqrt{B \mp b}\,\tanh\big(\tfrac{1}{2} F s\big) \mp i \sqrt{B\pm b} \Big)^2  e^{ibs}}
  {\Big( \sqrt{B \mp b}\,\tanh\big(\tfrac{1}{2} F s\big) \pm i \sqrt{B\pm b} \Big)^2}
\end{equation}
where
\begin{equation}
  B=\sqrt{c -\tfrac{1}{2}b^2},
  \quad
  F = \sqrt{c -\tfrac{3}{2}b^2} . 
\end{equation}
Note that $B>|b|$ holds as a consequence of the inequality on $c$ and $b$. 
Next, expression \eqref{J} is given by 
\begin{equation}
  J(s) = \tfrac{1}{C_1} \bigg(
  \mp \frac{2 B F^3 \sinh(F s)}{\big( B\cosh(F s) \pm b\big)^2}
  + i\Big( c - \tfrac{1}{2}\Big( b \pm  \frac{2F^2}{B\cosh( F s) \pm b} \Big)^2 \Big)
  \bigg)
\end{equation}
where $C_1 = -bB^2$ from expression \eqref{C1.C2.soliton}. 
Thus, the loop equation \eqref{z.loop.eqn.C1not0} produces
\begin{equation}\label{soliton.loop}
\begin{aligned}
z(s,t) = & 
\frac{1}{bB^2} \Bigg(
\pm \frac{2 B F^3 \sinh(F s)}{\big( B\cosh(F s) \pm b\big)^2}
- i\Big( c - \tfrac{1}{2}\Big( b \pm  \frac{2F^2}{B\cosh( F s) \pm b} \Big)^2 \Big)
\Bigg)
\\&\qquad\times
\frac{\Big( \sqrt{B \mp b}\,\tanh\big(\tfrac{1}{2} F s\big) \mp i \sqrt{B\pm b} \Big)^2}{\Big( \sqrt{B \mp b}\,\tanh\big(\tfrac{1}{2} F s\big) \pm i \sqrt{B\pm b} \Big)^2}\,
   e^{ib(s - B^2 t)}
  + z_0 . 
 \end{aligned} 
\end{equation}
This is a rotating loop whose angular speed is $-bB^2$.

The corresponding geometric curve flow \eqref{r.z} is given by
the real and imaginary parts of the loop solution \eqref{soliton.loop},
$\vec{r}(s,t)  =\big(\mathrm{Re}\,z(s,t),\mathrm{Im}\,z(s,t)\big)$. 
This can be written in a succinct form in terms of $\vartheta(s)$ and $K(s)$: 
\begin{equation}\label{soliton.curveflow}
\begin{aligned}
  \vec{r}_\pm(s,t)  =\frac{1}{bB^2} & \Big(
  K'(s) \cos(\vartheta(s) -bB^2 t) + \big(c -\tfrac{1}{2}K(s)^2\big) \sin(\vartheta(s) -bB^2t) ,
  \\&\qquad
  K'(s) \sin(\vartheta(s) -bB^2 t) - \big(c -\tfrac{1}{2}K(s)^2\big) \cos(\vartheta(s) -bB^2t)
  \Big)
\end{aligned}
\end{equation}
where
\begin{equation}
  K(s) = b \pm  \frac{2}{B\cosh( F s) \pm b} ,
  \quad
  K'(s) = \pm \frac{2 B F^3 \sinh(F s)}{\big( B\cosh(F s) \pm b\big)^2} . 
\end{equation}

\begin{figure}[h]
\centering
 \begin{subfigure}[t]{0.9\textwidth}
\centering
\includegraphics[width=0.35\textwidth,trim=2cm 12cm 6cm 2cm,clip]{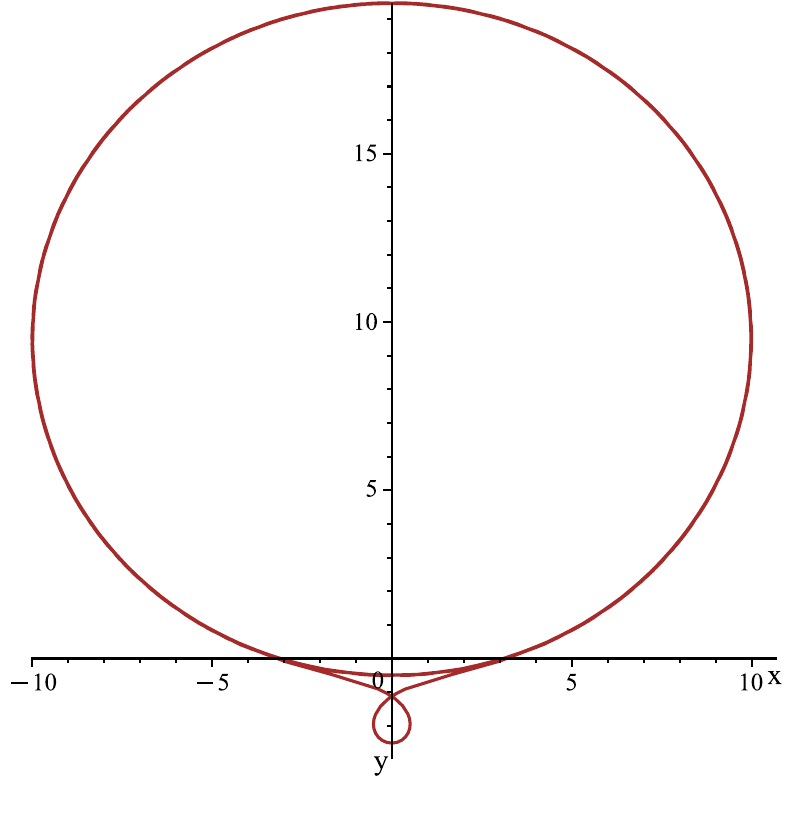}
\hfill
\includegraphics[width=0.35\textwidth,trim=2cm 12cm 6cm 2cm,clip]{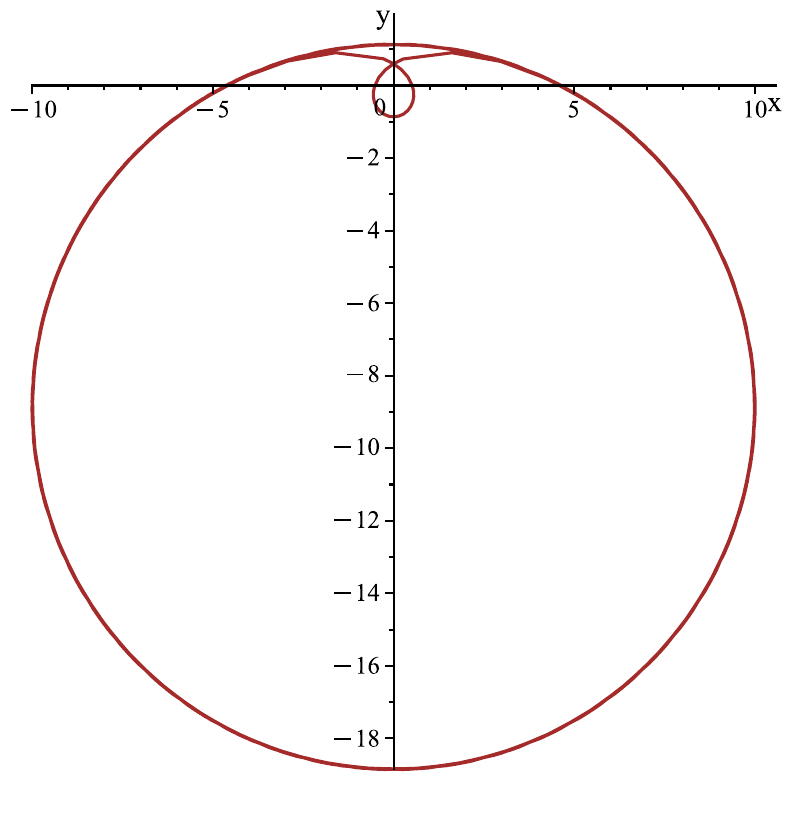}
\caption{Left: $b=-0.1$. Right: $b=0.1$.}
\end{subfigure}

\begin{subfigure}[t]{0.9\textwidth}
\centering
\includegraphics[width=0.35\textwidth,trim=2cm 12cm 6cm 2cm,clip]{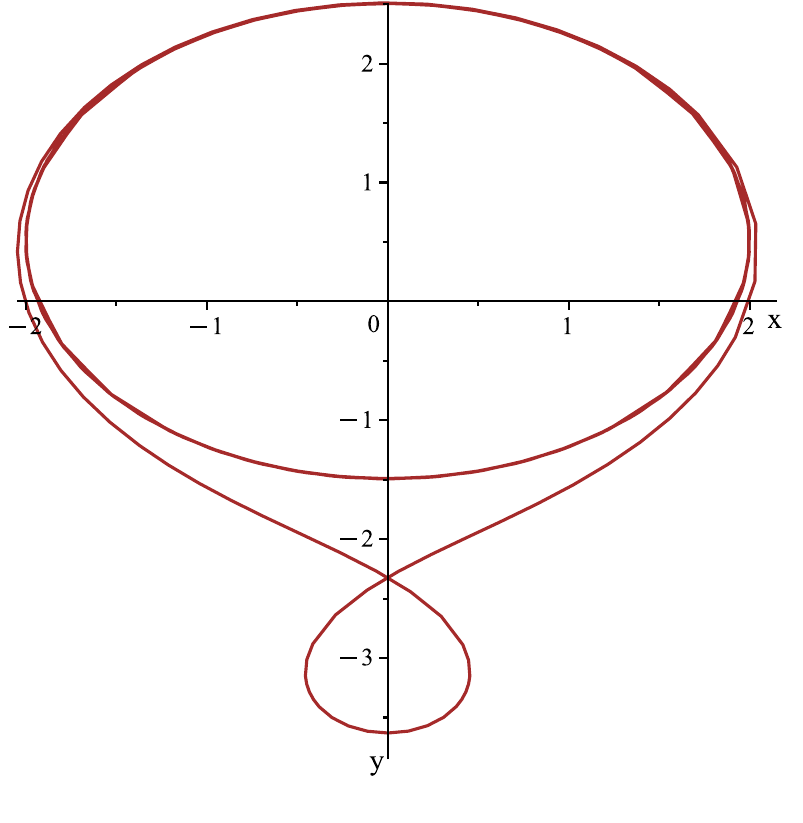}
\hfill
\includegraphics[width=0.35\textwidth,trim=2cm 12cm 6cm 2cm,clip]{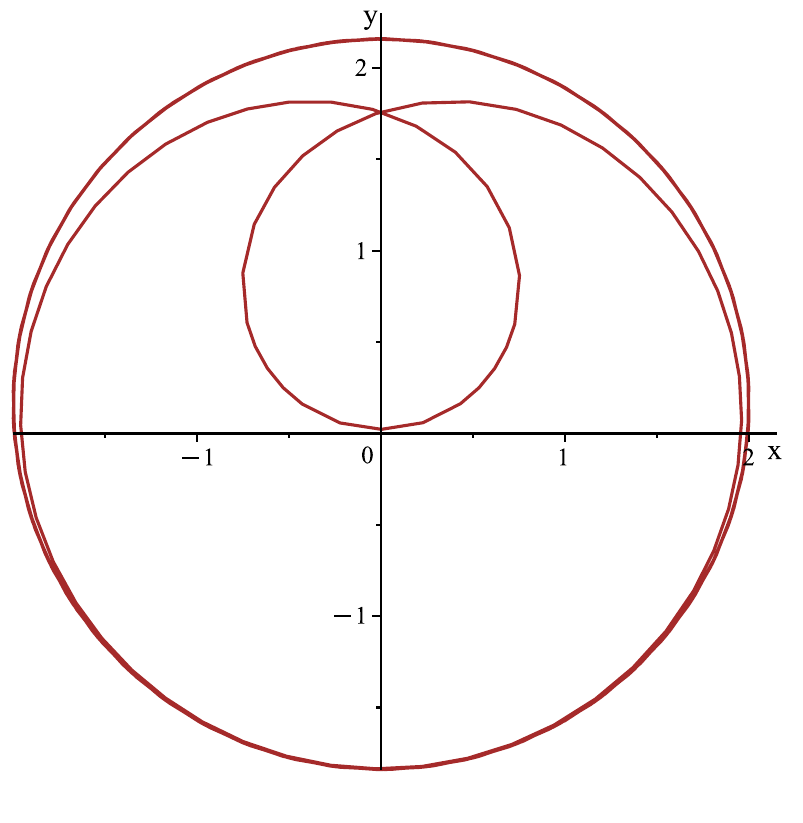}
\caption{Left: $b=-0.5$. Right: $b=0.5$.}
\end{subfigure}
    
\caption{mKdV soliton on a non-zero background, $c=1$. }
\label{fig:soliton_nzbc_loop}
\end{figure}

This flow is the composition of a static loop
\begin{equation}\label{soliton.staticloop}
\vec{r}_\pm(s,0)  =\frac{1}{bB^2} \Big(
  K'(s) \cos\vartheta(s) + \big(c -\tfrac{1}{2}K(s)^2\big) \sin\vartheta(s),
  K'(s) \sin\vartheta(s) - \big(c -\tfrac{1}{2}K(s)^2\big) \cos\vartheta(s)
\Big)
\end{equation}
plus a rotational motion with angular speed $-bB^2$. 
A plot of the static loop is shown in Fig. ~\ref{fig:soliton_nzbc_loop}.

The static loop has the following features, which are similar to those of
the rational loop \eqref{soliton.staticloop}.
It asymptotically winds around a circle of radius $1/|b|$ as $|s|\to \infty$,
as shown by 
\begin{equation}
  |\vec{r}_\pm(s,0)| =
  \frac{1}{|b|B^2}\sqrt{K'(s)^2 + (\tfrac{1}{2}K(s)^2 - c)^2}
  = \frac{1}{|b|} + O(1/s^2), 
\quad
\vartheta_s = K(s) = b + O(1/s^2)
\end{equation}
where the winding is counter-clockwise for $b>0$, and clockwise for $b<0$.
There is a choice of $\pm$ sign in the loop expression \eqref{soliton.staticloop}, 
and independently a choice of $b>0$ or $b<0$. 
In all cases,
the static loop contains a single twist that forms a smaller self-intersecting loop.
For $b<0$, this smaller loop sits inside the asymptotic circle,
whereas for $b>0$, it sits outside.
The peak of this inner/outer loop is at $s=0$
and it is oriented down in the $+$ sign case and up in the $-$ sign case,
corresponding to the soliton being bright/dark. 

The height $h$ of the peak relative to the asymptotic circle is given by
$y(0,0) - 1/|b|$ when $b>0$ and $y(0,0) + 1/|b|$ when $b<0$,
which yields 
\begin{equation}
h=  \mp \frac{2}{\sqrt{c-\tfrac{1}{2}b^2}}
\end{equation}
where the sign indicates the up/down orientation of the loop. 
The inner/outer loop has width $w = |x(s,0) - x(-s,0)|$
which is the distance between the closest pair of points around $s=0$
such that $x_s(|s|,0)=\cos(\vartheta(|s|))=0$.
This determines $s=\zeta/b$
where $\zeta$ is the smallest positive root of the equation
\begin{equation}
 \frac{\pm\sgn(b)\sqrt{a^2 + 1}  - 1}{a} \tanh(\tfrac{1}{2}a\zeta)
  = \tan(\tfrac{1}{8}(\pi - 2\zeta)),
  \quad
  a= F/|b| =\sqrt{(c/b^2) - \tfrac{3}{2}} > 0 . 
\end{equation}
Substituting $s=\zeta/b$ into $|x(s,0) - x(-s,0)|=2|x(s)|$
given by the $x$-component of expression \eqref{soliton.staticloop}
then yields
\begin{equation}
w=  \frac{2c -K(\zeta/b)^2}{|b|(c- \tfrac{1}{2}b^2)} . 
\end{equation}

Similarly to situation for the rational loop solution,
the asymptotically circular loop has an infinite winding number.
The contribution from the inner/outer loop as given by
the renormalized winding number \eqref{renormalized.winding.number} is
\begin{equation}
  \mathcal{N} = \pm \frac{4}{\pi}\arctan\Big( \big(\sqrt{c -\tfrac{1}{2}b^2} \mp b\big)\big/\sqrt{c -\tfrac{3}{2}b^2} \Big) , 
\end{equation}
while the renormalized mean-winding rate \eqref{renormalized.winding.rate} is
\begin{equation}
  \mathcal{W} = \frac{4}{\pi}\Big( \sqrt{c -\tfrac{3}{2}b^2} 
  \pm b\, \arctan\Big( \big(\sqrt{c -\tfrac{1}{2}b^2} \mp b\big)\big/\sqrt{c -\tfrac{3}{2}b^2} \Big)
  \Big) . 
\end{equation}

\subsection{cnoidal loops}
\label{sec:cn.loops}

Cnoidal mKdV waves \eqref{cn} for $0<q<1$
obey $u\to -u$ when they are shifted by a half period \eqref{CN.period},
so only the $+$ case in expression \eqref{cn} needs to be considered. 
These waves have $C=0$,
so the constant $C_1$ given by expression \eqref{C1.ratcn} is zero.
Consequently, the loop solutions are obtained by evaluation of
the parametric integrals \eqref{vartheta} and \eqref{z.loop.eqn.C1is0}
together with expression \eqref{nu}.
Note that $q$ and $c$ obey the inequality
\begin{equation}\label{ineqn.cn}
    q< \tfrac{1}{2},
    \quad
    c<0;
    \qquad
    \tfrac{1}{2} < q< 1,
    \quad
    c>0 . 
\end{equation}

The first parametric integral yields
\begin{equation}
  \vartheta =
  2\arctan\bigg( \sqrt{q}\, \frac{\SN(\sqrt{cq/F_1}\,s, \sqrt{q})}{\DN(\sqrt{cq/F_1}\,s, \sqrt{q})} \bigg) , 
\end{equation}
which gives
\begin{equation}
  e^{i\vartheta(s)} = 
  \frac{ -\sqrt{q}\, \SN(\sqrt{cq/F_1}\, s, \sqrt{q}) +i \DN(\sqrt{cq/F_1}\, s, \sqrt{q})  }
       { \sqrt{q}\, \SN(\sqrt{cq/F_1}\, s, \sqrt{q}) +i \DN(\sqrt{cq/F_1}\, s, \sqrt{q}) } . 
\end{equation}
Then the second parametric integral gives 
\begin{equation}
 \int_0^s e^{i\vartheta(s)}\,ds = 
 -s + 2 \sqrt{F_1/(cq)}\, \E\big(\SN(\sqrt{cq/F_1}\,s, \sqrt{q}), \sqrt{q}\big)
   -2 i \sqrt{F_1/c} \big( \CN(\sqrt{cq/F_1}\,s, \sqrt{q}) - 1 \big)
\end{equation}
where $\E$ denotes the incomplete elliptic integral of the second kind. 
Evaluation of expression \eqref{nu} yields
$\nu = \frac{2q}{2q - 1}c -c = c/(2q-1)$
which is positive due to the inequality \eqref{ineqn.cn}. 

Substitution into the loop equation \eqref{z.loop.eqn.C1is0} produces
\begin{equation}
  z(s,t) =
  \tfrac{1}{2q-1} ct   -s 
  + 2 \sqrt{F_1/(cq)}\Big( \E\big(\SN(\sqrt{cq/F_1}\,s, \sqrt{q}), \sqrt{q}\big)
    - i \sqrt{q} \big( \CN(\sqrt{cq/F_1}\,s, \sqrt{q}) - 1 \big) \Big) . 
\end{equation}
This describes a travelling loop whose speed is positive,
which differs from mKdV cnoidal wave speed $c$ by the factor $1/(2q-1)$.
In particular, when $q<\tfrac{1}{2}$,
the travelling loop moves in the opposite direction.

The corresponding geometric curve flow \eqref{r.z} is given by
\begin{equation}\label{cn.curveflow}
  \vec{r}(s,t)  = \big(
  \tfrac{1}{2q-1} ct   -s 
  + 2 \sqrt{F_1/(cq)}\, \E\big(\SN(\sqrt{cq/F_1}\,s, \sqrt{q}), \sqrt{q}\big), 
 -2 \sqrt{F_1/c} \big( \CN(\sqrt{cq/F_1}\,s, \sqrt{q}) - 1 \big)
\big) . 
\end{equation}
It is a superposition of 
a static loop
\begin{equation}\label{cn.staticloop}
\vec{r}(s,0)  =\big(
  {-s} + 2 \sqrt{F_1/(cq)}\, \E\big(\SN(\sqrt{cq/F_1}\,s, \sqrt{q}), \sqrt{q}\big), 
 - 2 \sqrt{F_1/c} \big( \CN(\sqrt{cq/F_1}\,s, \sqrt{q}) - 1 \big)
\big)
\end{equation}
plus a translation motion $\big( \tfrac{1}{2q-1}ct , 0 \big)$ in the $x$-direction.

For a general value of $q$,
the static loop is open and periodic under translation in $x$.
The first property follows from $x(s,0) = \mathrm{Re}\, z(s,0)$
being an unbounded function of $s$,
while the periodicity is a consequence of both $x_s(s,0)$ and $y(s,0)=\mathrm{Im}\, z(s,0)$ being Jacobi elliptic functions having the same period
\begin{equation}\label{cn.staticloop.per}
  s_\text{per.} = 4\sqrt{F_1/c}\, \K(\sqrt{q}) 
\end{equation}
where $\K$ is the complete elliptic integral of the first kind.
Thus, the $x$-period of the loop is given by 
\begin{equation}
  \Delta x =  4 \sqrt{F_1/c}\, \big( 2\E(1, \sqrt{q}) - \K(\sqrt{q}) \big) . 
\end{equation}

\begin{figure}[h]
\centering
\includegraphics[width=0.3\textwidth,trim=2cm 12cm 6cm 2cm,clip]{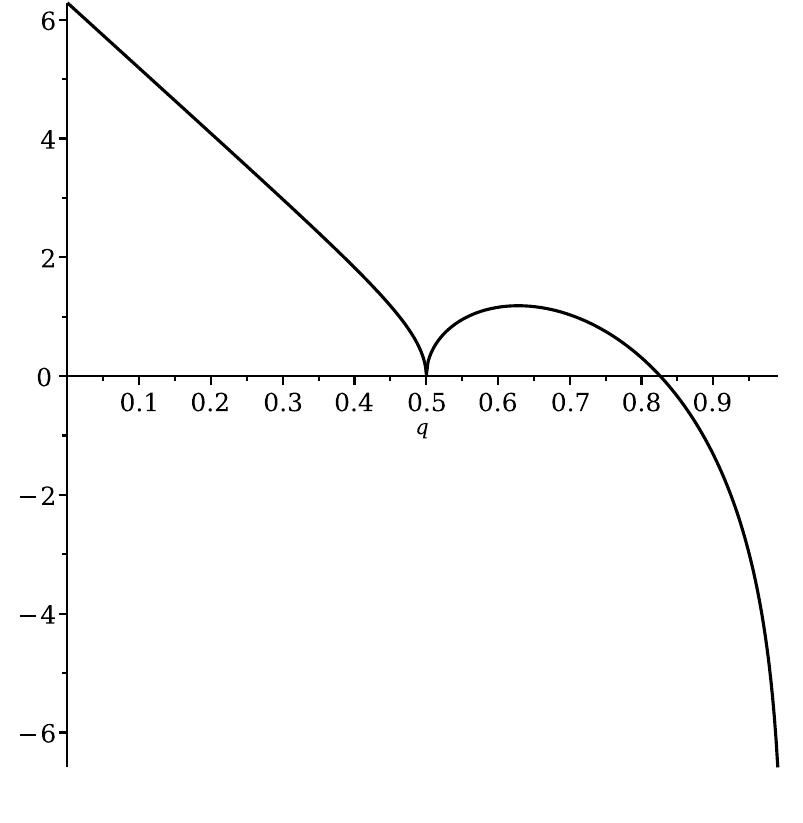}
\caption{$x$-period of cnoidal loop for $|c|=1$.}
\label{fig:cnoidal_x_per}
\end{figure}

The shape of the static loop over one period, referred to as a cycle,
is primarily dependent on $q$,
while $|c|$ acts as a scaling factor.
Hereafter, $|c|=1$ will be fixed.
The $x$-period as a function of $q$ can be positive or negative, as
seen in Fig.~\ref{fig:cnoidal_x_per}.

\begin{figure}[h]
\centering
 \begin{subfigure}[t]{0.9\textwidth}
\centering
\includegraphics[width=0.35\textwidth,trim=2cm 12cm 6cm 2cm,clip]{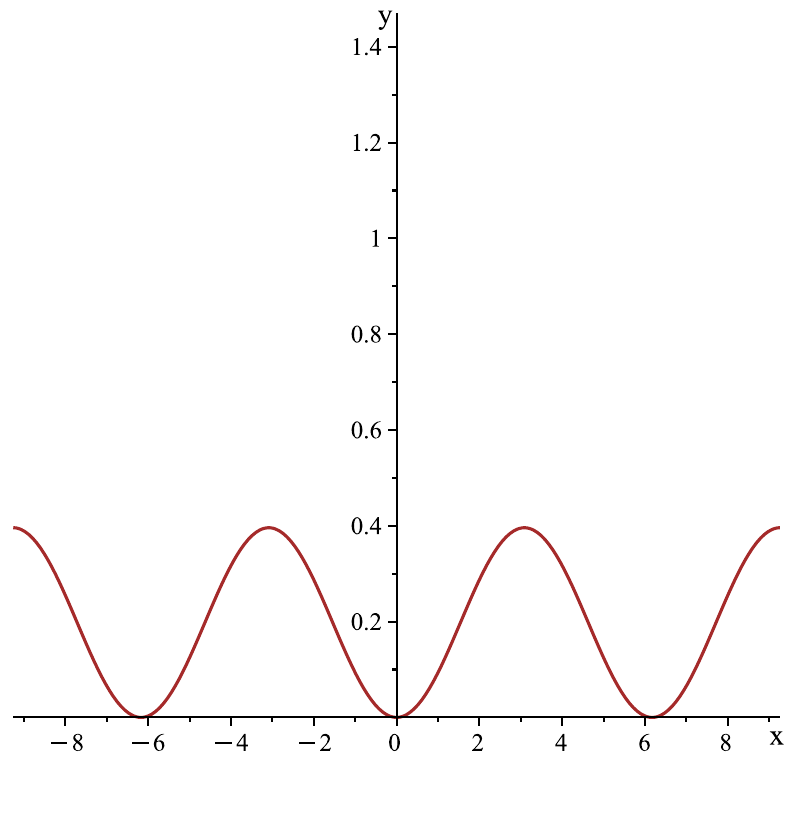}
\hfill
\includegraphics[width=0.35\textwidth,trim=2cm 12cm 6cm 2cm,clip]{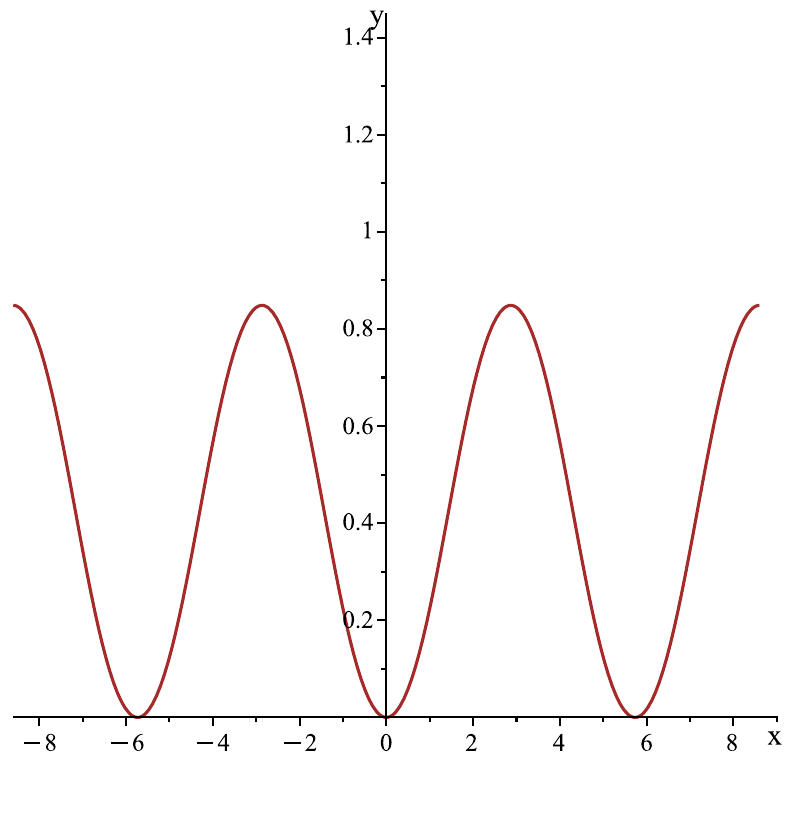}
\caption{Left: $q=0.01$. Right: $q=0.05$.}
\end{subfigure}

\begin{subfigure}[t]{0.9\textwidth}
\centering
\includegraphics[width=0.35\textwidth,trim=2cm 12cm 6cm 2cm,clip]{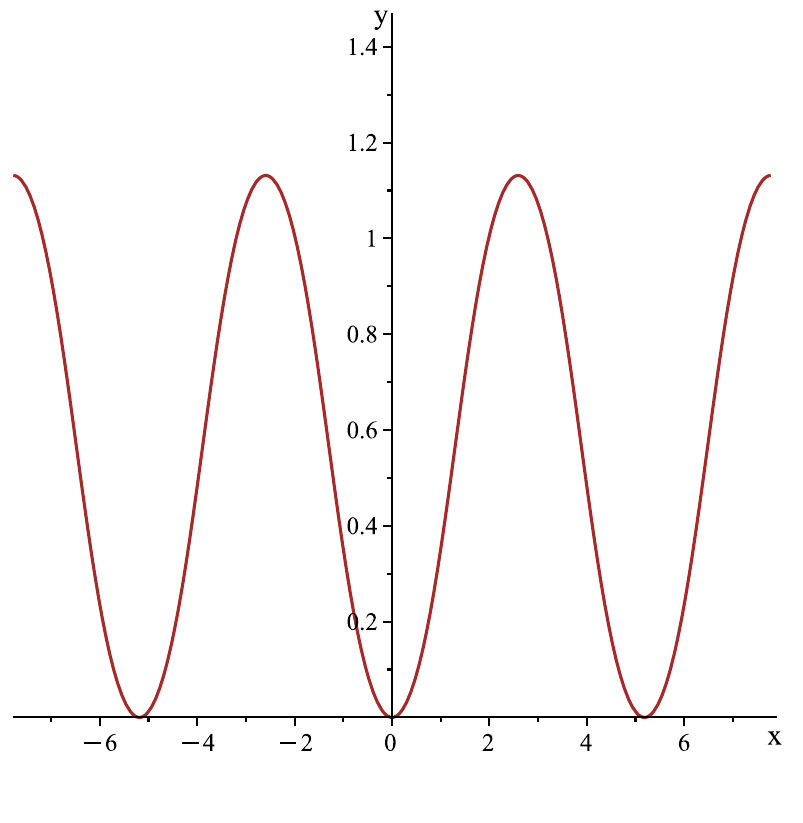}
\hfill
\includegraphics[width=0.35\textwidth,trim=2cm 12cm 6cm 2cm,clip]{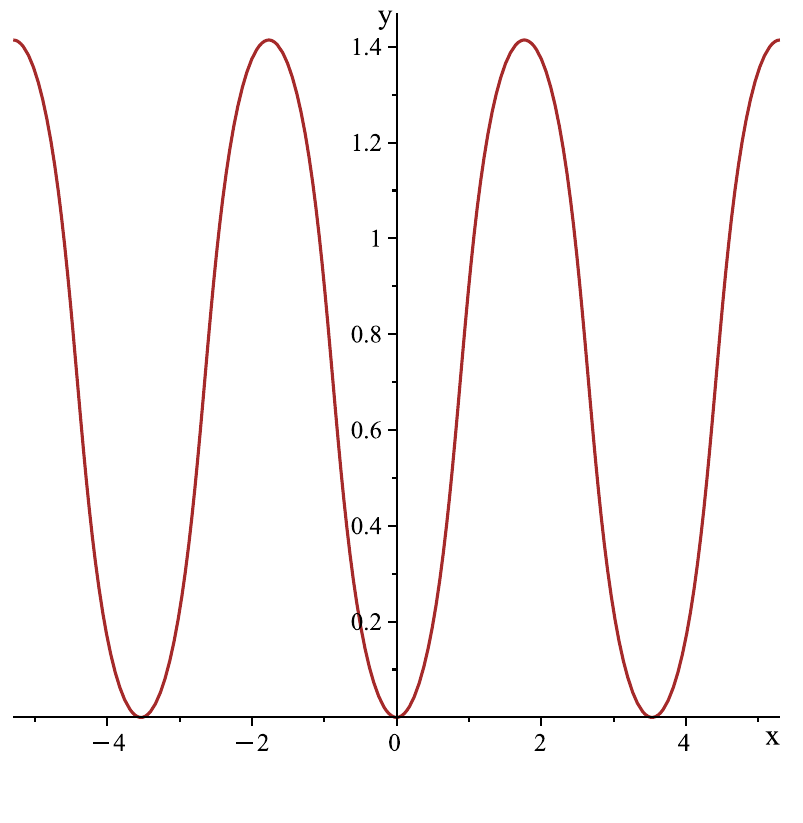}
\caption{Left: $q=0.20$. Right: $q=0.25$.}
\end{subfigure}
\caption{cnoidal loop, $c=-1$.}
\label{fig:cnoidal_loop_c<0}
\end{figure}

For $q<\tfrac{1}{2}$,
the cycle is an untwisted loop that is symmetric under $s\to -s$. 
It has maximum amplitude $4 \sqrt{F_1/c}$,
while its minimum amplitude is $0$.
In addition, its speed is negative. 
See Fig.~\ref{fig:cnoidal_loop_c<0}.

\begin{figure}[h]
\centering
\includegraphics[width=0.35\textwidth,trim=2cm 13cm 6cm 2cm,clip]{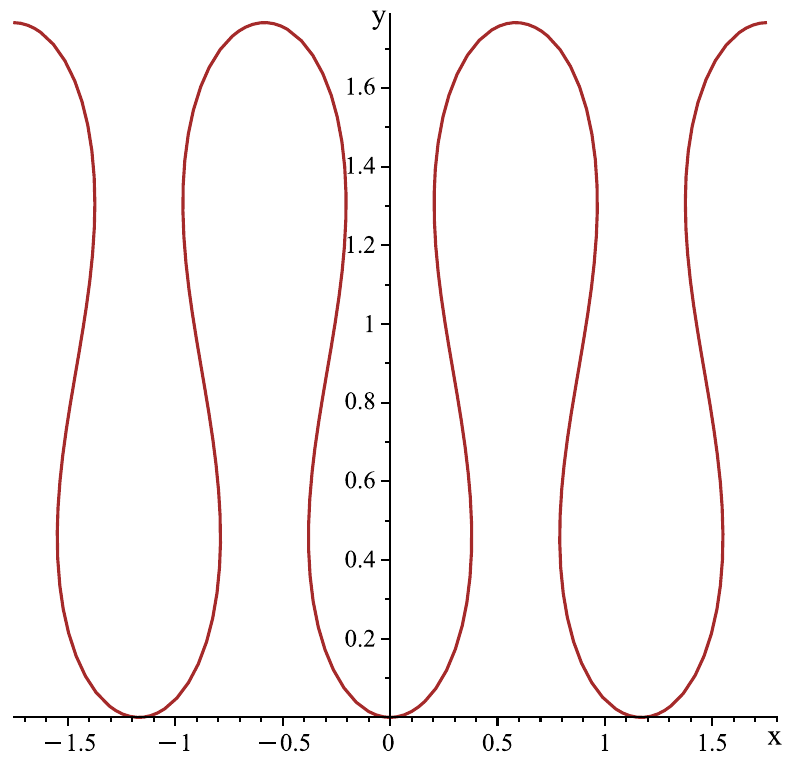}
\caption{cnoidal loop, $c=1$, $q= 0.65$.}
\label{fig:cnoidal_loop_c=1_q_0.65}
\end{figure}

\begin{figure}[h]
\centering
\includegraphics[width=0.35\textwidth,trim=2cm 12cm 6cm 2cm,clip]{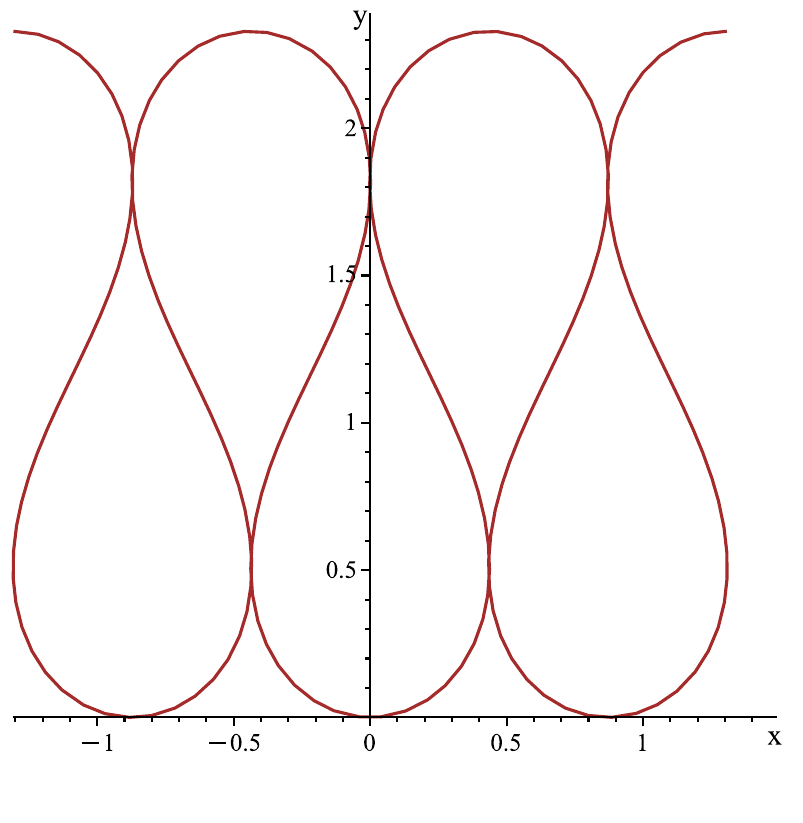}
\caption{cnoidal loop, $c=1$, $q= q_{*1}\simeq 0.7317$.}
\label{fig:cnoidal_loop_qtouch}
\end{figure}

For $q>\tfrac{1}{2}$,
the speed becomes positive and the shape of the loop
is more complicated but remains symmetric.
See Fig.~\ref{fig:cnoidal_loop_c=1_q_0.65}. 
As $q$ increases above $\tfrac{1}{2}$,
the loop begins to pinch together
until the two halves touch at $x=0$ when $q=q_{*1}$. 
The touching condition is $x(s_*,0)=0$ and $x_s(s_*,0)=0$ with $s_*\neq0$,
which gives two equations that determine $s_*$ and $q_{*1}$. 
This yields
$s_*\simeq 2.195$ and $q_{*1} \simeq 0.7317$.
See Fig.~\ref{fig:cnoidal_loop_qtouch}. 

\begin{figure}[h]
\centering
\includegraphics[width=0.35\textwidth,trim=2cm 12cm 6cm 2cm,clip]{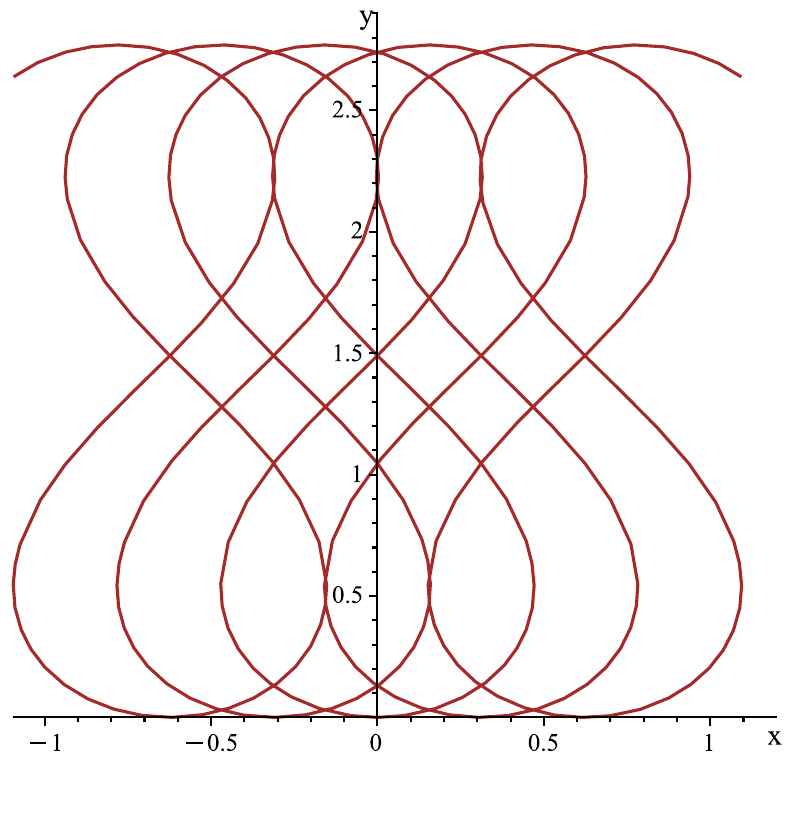}
\hfill
\includegraphics[width=0.35\textwidth,trim=2cm 12cm 6cm 2cm,clip]{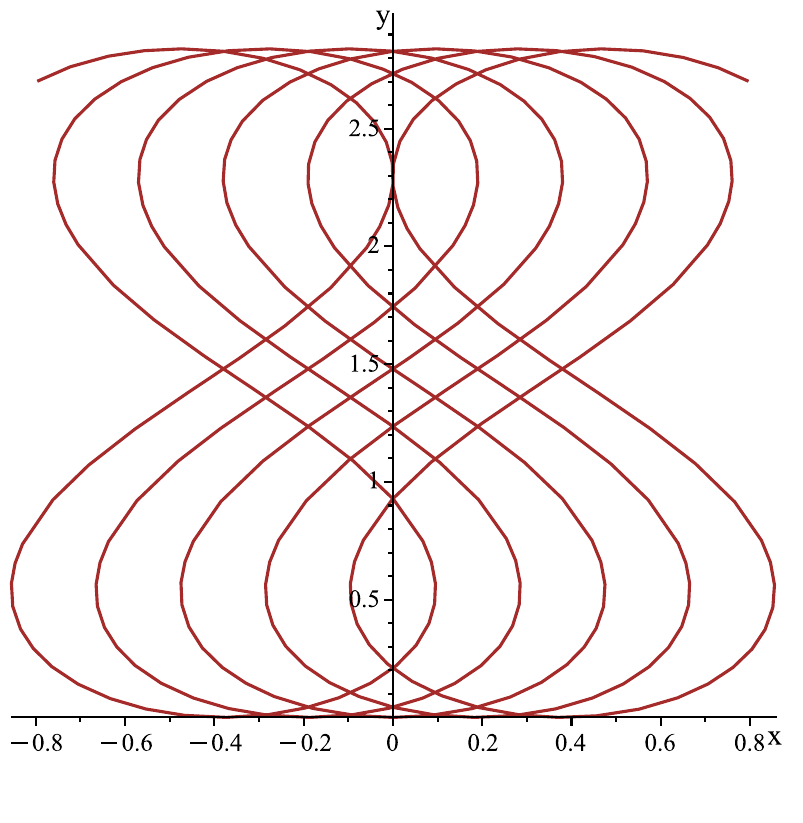}
\caption{cnoidal loop, $c=1$. Left: $q\simeq 0.799$. Right: $q \simeq 0.811$.}
\label{fig:cnoidal_loop_secondtouch}
\end{figure}

For $q>q_{*1}$,
the loop acquires a twist,
such that the two halves are overlapping, 
with the winding number remaining zero.
If multiple cycles are considered,
then the two halves of the loop in each successive cycle
touch at $x=0$ for certain values $q=q_{*n}$, $n=2,3,\ldots$. 
See Fig.~\ref{fig:cnoidal_loop_secondtouch}. 

\begin{figure}[h]
\centering
\includegraphics[width=0.3\textwidth,trim=2cm 12cm 6cm 2cm,clip]{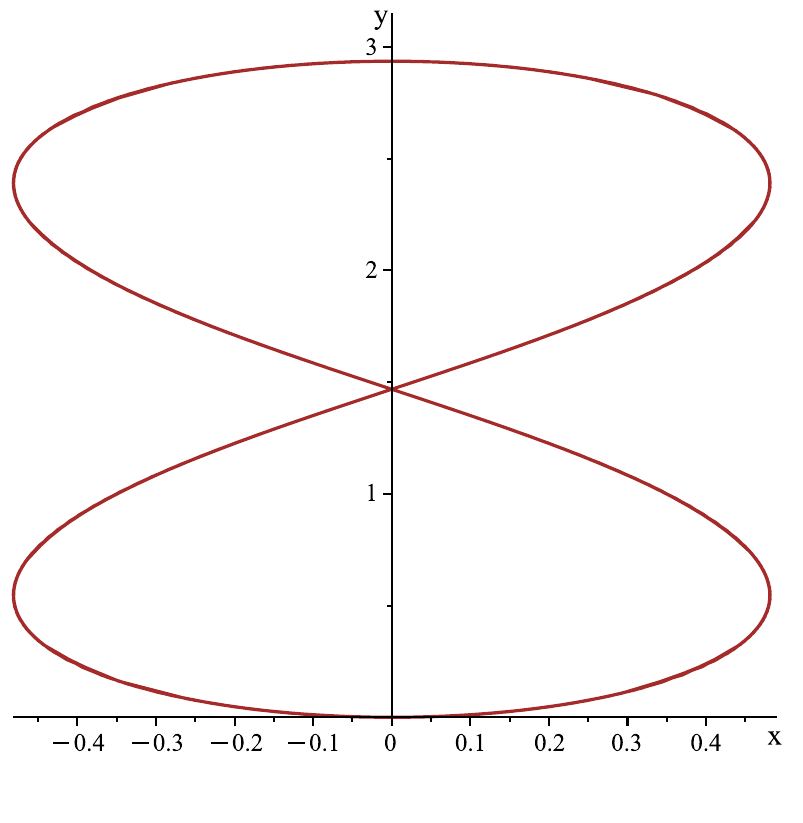}
\caption{closed cnoidal loop: $q= q_{closed} \simeq 0.8261$, $c=1$.}
\label{fig:cnoidal_loop_closed}
\end{figure}

The two halves of the loop will exactly overlap such that a closed loop is formed
when $x(s,0)$ returns to the same value after one period \eqref{cn.staticloop.per},
namely $x(0,0)=x(4\K(\sqrt{cq/(2q - 1)}),0)$ so that the $x$-periodic is zero. 
This condition yields $q=q_\text{closed} \simeq 0.8261$.
The resulting closed loop has the shape of a figure 8
with height $\simeq 2.936$ and width $\simeq 0.546$, respectively. 
See Fig.~\ref{fig:cnoidal_loop_closed}. 

\begin{figure}[h]
\centering
\includegraphics[width=0.35\textwidth,trim=2cm 12cm 6cm 2cm,clip]{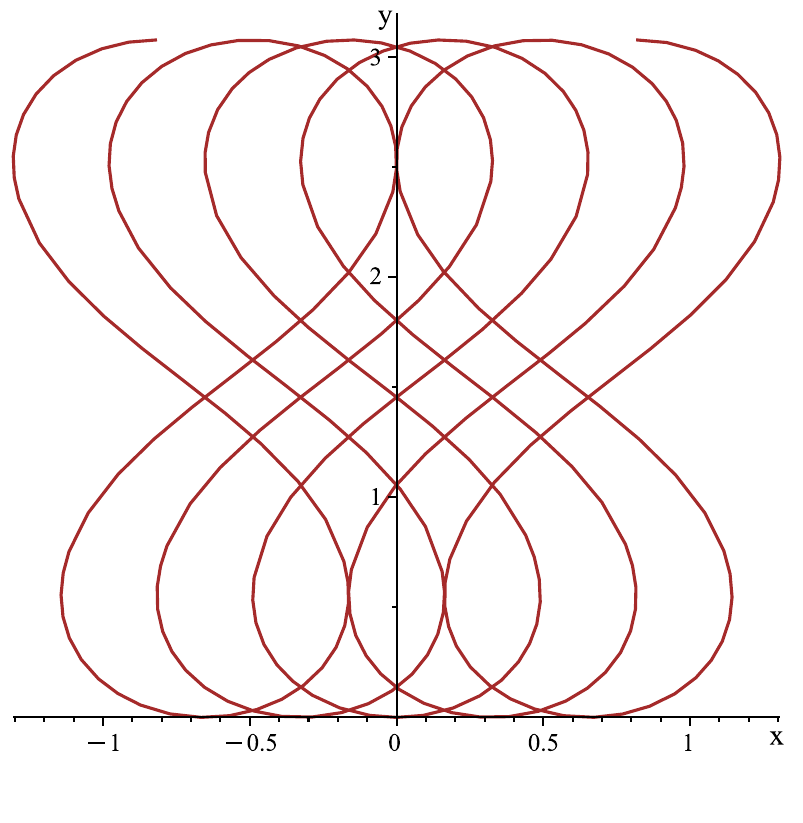}
\hfill
\includegraphics[width=0.35\textwidth,trim=2cm 12cm 6cm 2cm,clip]{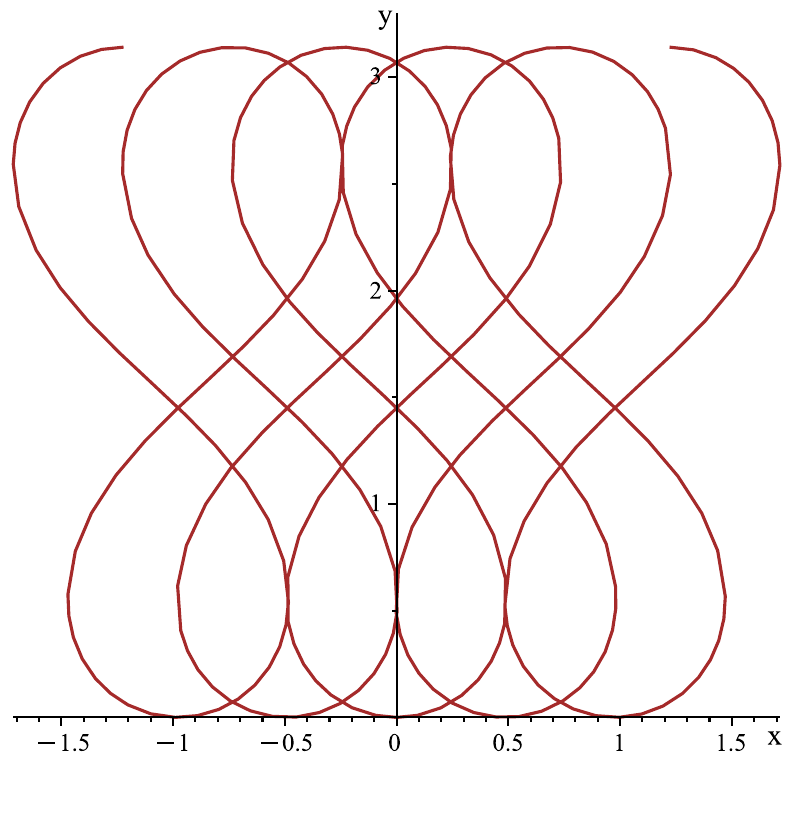}
\caption{cnoidal loop, $c=1$. Left: $ q\simeq 0.849$. Right: $q \simeq 0.859$.}
\label{fig:cnoidal_loop_2ndlast_touch}
\end{figure}

For $q>q_\text{closed}$, the $x$-period becomes negative.
As $q$ increases, 
the two halves of the loop again touch at $x=0$ in successive cycles
for certain values $q=q^*_{n}$, $n=1,2,\ldots$,
as shown in Fig.~\ref{fig:cnoidal_loop_2ndlast_touch}.

\begin{figure}[h]
\centering
\includegraphics[width=0.35\textwidth,trim=2cm 12cm 6cm 2cm,clip]{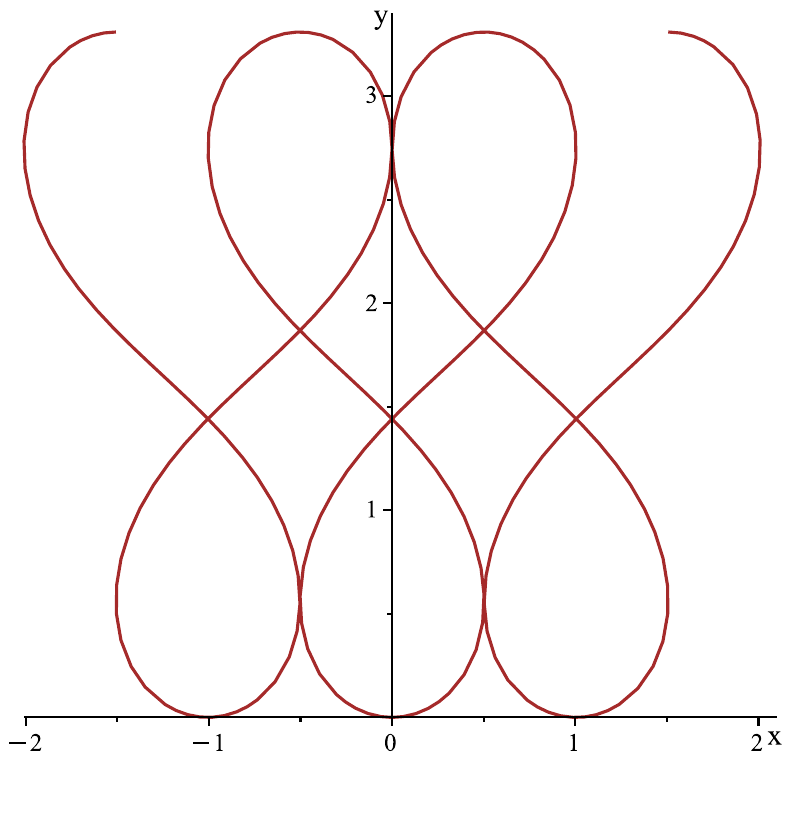}
\caption{cnoidal loop $q=q^*_{1}\simeq 0.886$, $c=1$.}
\label{fig:cnoidal_loop_last_touch}
\end{figure}

The last touch occurs for $q= q^*_{1} \simeq 0.886$.
See Fig.~\ref{fig:cnoidal_loop_last_touch}. 

\begin{figure}[h]
\centering
\includegraphics[width=0.35\textwidth,trim=2cm 12cm 6cm 2cm,clip]{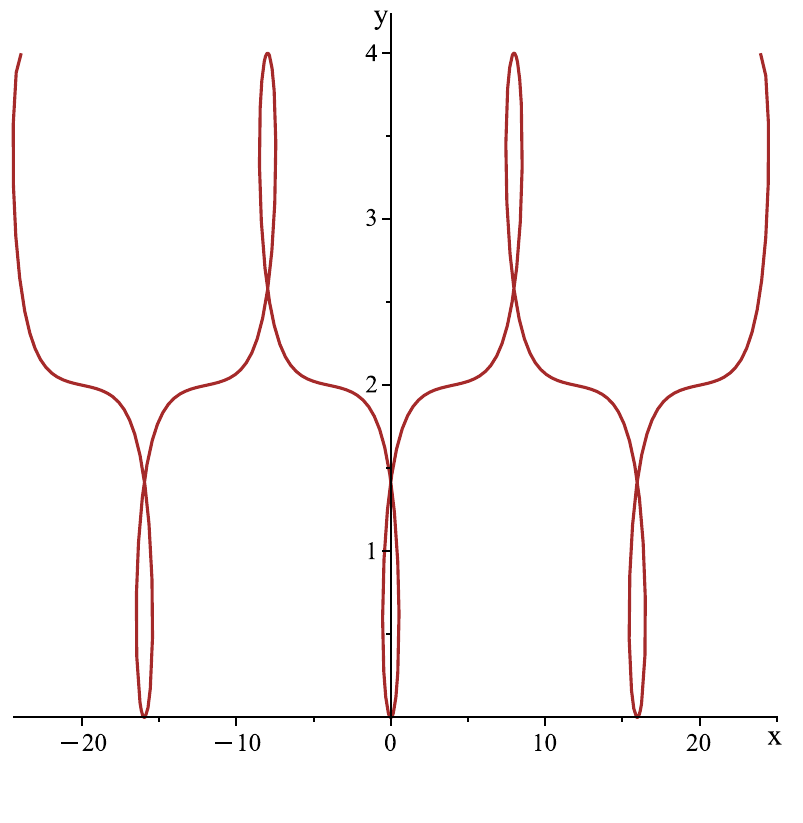}
\caption{cnoidal loop, $c=1$, $q\simeq 0.999$.}
\label{fig:cnoidal_loop_after_last_touch}
\end{figure}

As $q$ increases past $q^*_{1}$ to $1$, the loops separate.
See Fig.~\ref{fig:cnoidal_loop_after_last_touch}. 

In all cases,
the loop has winding number $N=0$
while the mean-winding rate is
\begin{equation}
  W=\frac{8}{\pi}(q - 1)\sqrt{\frac{c}{2q - 1}}\,\K(\sqrt{q})
\end{equation}
which follows by using the results in Table~\ref{table:conserved}
combined with equations \eqref{winding.number} and \eqref{winding.rate}.

\subsection{dnoidal loops}
\label{sec:dn.loops}

Dnoidal mKdV waves \eqref{dn} are bright/dark corresponding to the $\pm$ sign. 
Similarly to the cnoidal solutions,
these waves \eqref{dn} yield loop solutions obtained by evaluation of
the parametric integrals \eqref{vartheta} and \eqref{z.loop.eqn.C1is0}
together with expression \eqref{nu},
where $C=0$.
Note that $q$ and $c$ obey the inequality
\begin{equation}\label{ineqn.dn}
    q> 1, 
    \quad
    c>0 . 
\end{equation}

The first parametric integral yields
\begin{equation}
  \vartheta =
  \pm 2\arctan\bigg( \frac{\SN(q\sqrt{c/F_1}\,s, 1/\sqrt{q})}{\CN(q\sqrt{c/F_1}\,s, 1/\sqrt{q})} \bigg)
  = \pm 2\, \AM(q\sqrt{c/F_1}\,s, 1/\sqrt{q})
\end{equation}
where $\AM$ denotes the Jacobi amplitude function.
This gives
\begin{equation}
  e^{i\vartheta(s)} = 
  \frac{ \mp \SN(q\sqrt{c/F_1}\, s, 1/\sqrt{q}) +i \CN(q\sqrt{c/F_1}\, s, 1/\sqrt{q})  
  }{ \pm \SN(q\sqrt{c/F_1}\, s, 1/\sqrt{q}) +i \CN(q\sqrt{c/F_1}\, s, 1/\sqrt{q})  } . 
\end{equation}
Then the second parametric integral gives 
\begin{equation}
  \begin{aligned}
 \int_0^s e^{i\vartheta(s)}\,ds = 
 (1-2q)s + & 2 \sqrt{F_1/c}\Big( \E\big(\SN(q\sqrt{c/F_1}\,s, 1/\sqrt{q}), 1/\sqrt{q}\big)
 \\&\qquad
 \mp i (1/q^2) \big( \DN(q\sqrt{c/F_1}\,s, 1/\sqrt{q}) - 1 \big) \Big)
 \end{aligned}
\end{equation}
where $\E$ denotes the incomplete elliptic integral of the second kind. 
Evaluation of expression \eqref{nu} yields
$\nu = c/(2q-1)$
which is positive due to the inequality \eqref{ineqn.dn}. 

Substitution into the loop equation \eqref{z.loop.eqn.C1is0} produces
\begin{equation}
  \begin{aligned}
  z(s,t) =
  \tfrac{1}{2q-1} ct  -(2q-1)s & 
  + 2 \sqrt{F_1/c}\Big( \E\big(\SN(q\sqrt{cF_1}\,s, 1/\sqrt{q}), 1/\sqrt{q}\big)
  \\&\qquad
  \mp i \tfrac{1}{q^2} \big( \DN(q\sqrt{c/F_1}\,s, 1/\sqrt{q}) - 1 \big) \Big) . 
  \end{aligned}
\end{equation} 
This describes a travelling loop with positive speed.
The corresponding geometric curve flow \eqref{r.z} is given by
\begin{equation}\label{dn.curveflow}
  \begin{aligned}
  \vec{r}_\pm(s,t)  = & \big(
  \tfrac{1}{2q-1} ct  -(2q-1)s 
  + 2 \sqrt{F_1/c}\, \E\big(\SN(q\sqrt{c/F_1}\,s, 1/\sqrt{q}), 1/\sqrt{q}\big),
  \\&\qquad
\pm 2 \sqrt{F_1/(cq^2)} \big( 1 - \DN(q\sqrt{c/F_1}\,s, 1/\sqrt{q}) \big)
\big) . 
   \end{aligned}
\end{equation}
It is a superposition of 
a static loop
\begin{equation}\label{dn.staticloop}
  \begin{aligned} 
\vec{r}_\pm(s,0)  =& \big(
    {-(2q-1)s} + 2 \sqrt{F_1/c}\, \E\big(\SN(q\sqrt{c/F_1}\,s, 1/\sqrt{q}), 1/\sqrt{q}\big),
    \\&\qquad
\pm 2 \sqrt{F_1/(cq^2)} \big( 1- \DN(q\sqrt{c/F_1}\,s, 1/\sqrt{q}) \big)
\big) 
   \end{aligned}
\end{equation}
plus a translation motion $\big( \tfrac{1}{2q-1}ct , 0 \big)$ in the $x$-direction.

\begin{figure}[h]
\centering
\includegraphics[width=0.3\textwidth,trim=2cm 12cm 6cm 2cm,clip]{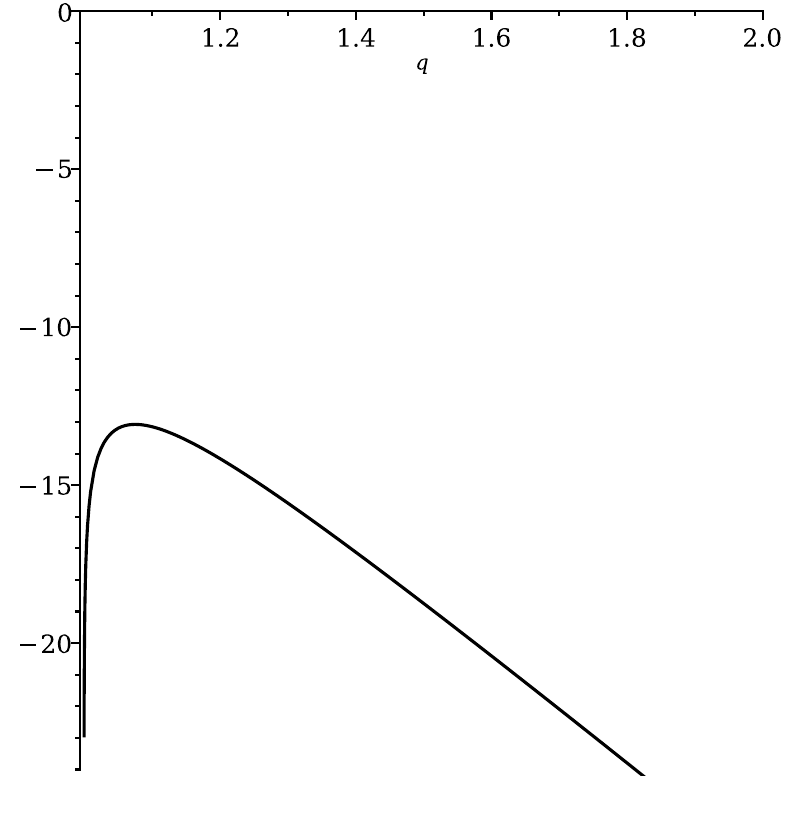}
\caption{$x$-period of dnoidal loop for $c=1$.}
\label{fig:dnoidal_x_per}
\end{figure}

The static loop is open and periodic under translation in $x$, 
with the period being 
\begin{equation}
  \Delta x  =    -4 \K(1/\sqrt{q}) \sqrt{2q - 1}^3/\sqrt{cq}
\end{equation}
which is negative, as seen in Fig.~\ref{fig:dnoidal_x_per},
where $\K$ is the complete elliptic integral of the first kind.

\begin{figure}[h]
\centering
 \begin{subfigure}[t]{0.9\textwidth}
\centering
\includegraphics[width=0.35\textwidth,trim=2cm 12cm 6cm 2cm,clip]{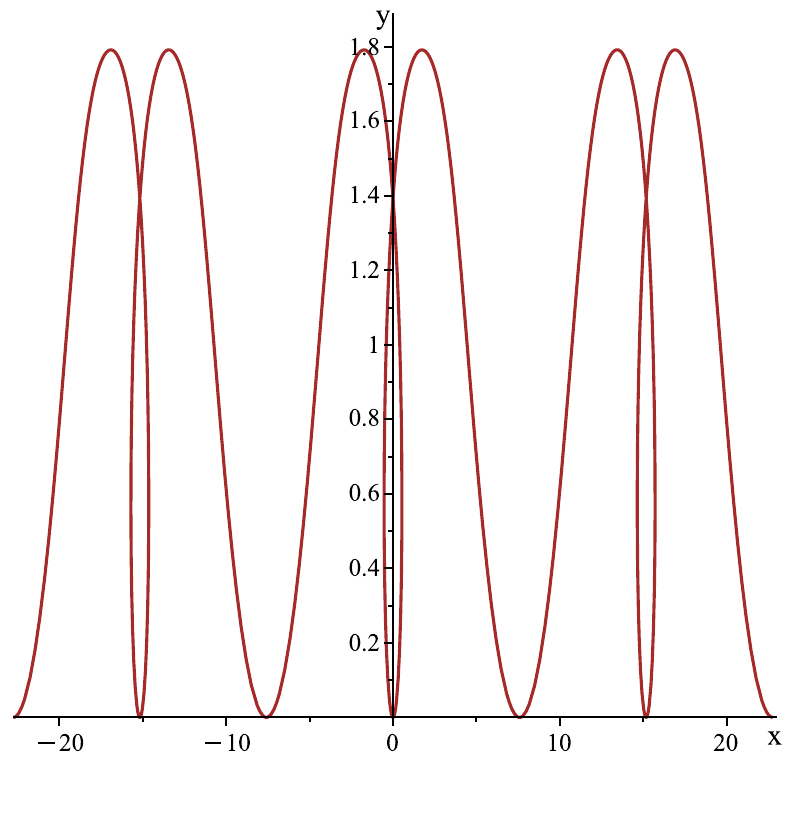}
\hfill
\includegraphics[width=0.35\textwidth,trim=2cm 12cm 6cm 2cm,clip]{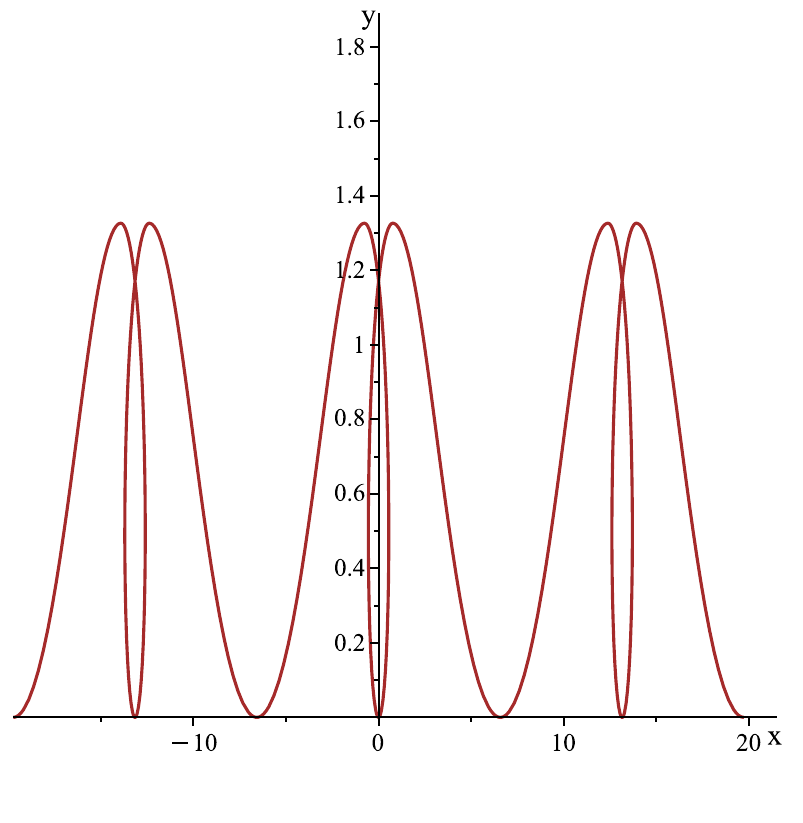}
\caption{Left: $q=1.01$. Right: $q=1.1$.}
\end{subfigure}

\begin{subfigure}[t]{0.9\textwidth}
\centering
\includegraphics[width=0.35\textwidth,trim=2cm 12cm 6cm 2cm,clip]{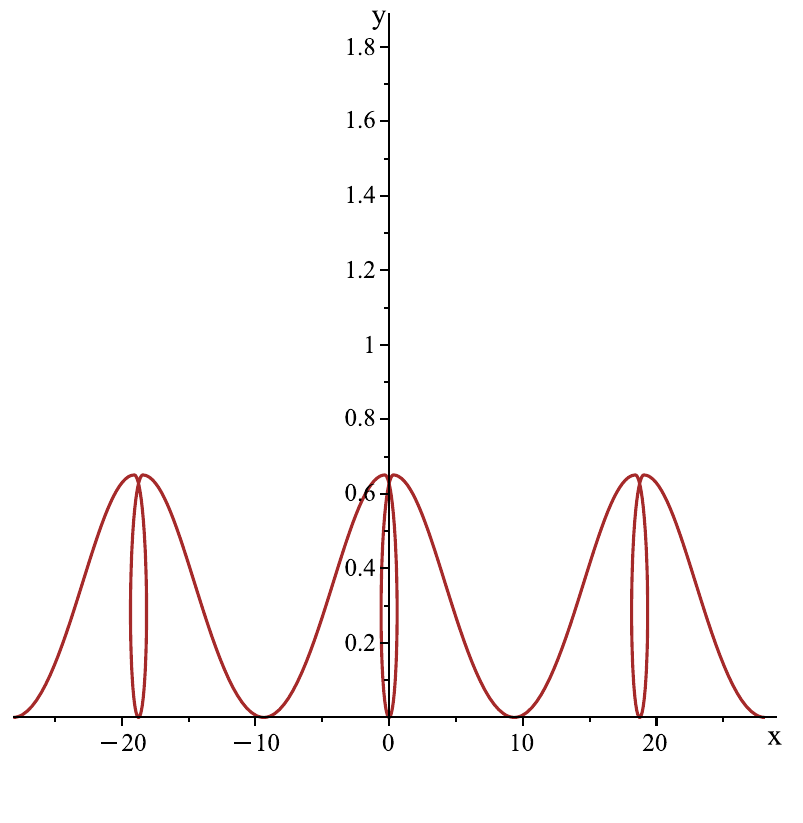}
\hfill
\includegraphics[width=0.35\textwidth,trim=2cm 12cm 6cm 2cm,clip]{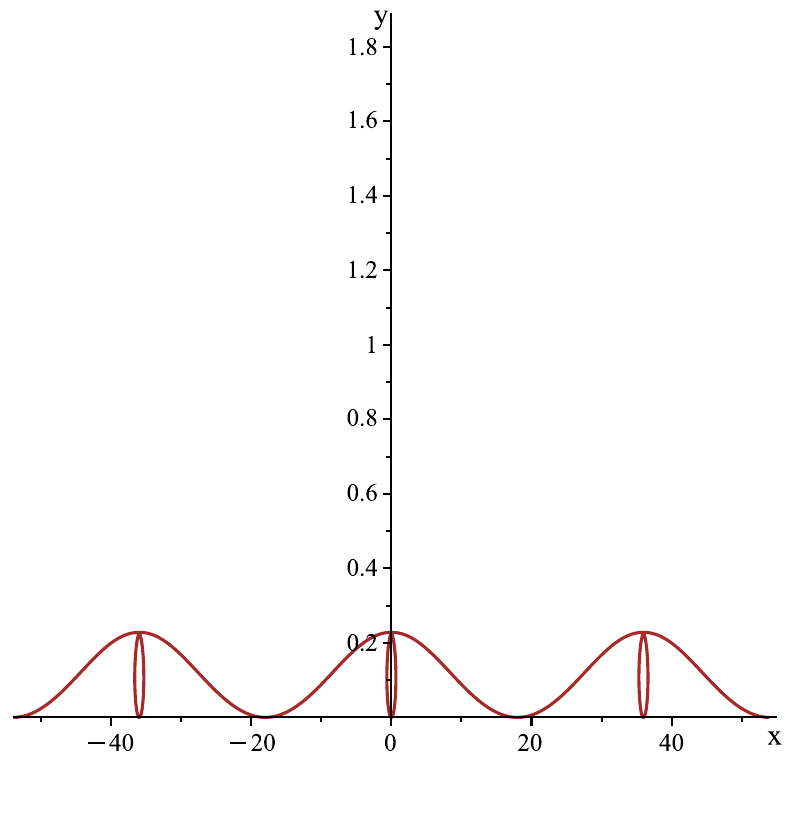}
\caption{Left: $q=1.5$. Right: $q=2.5$.}
\end{subfigure}
\caption{dnoidal loop, $c=1$.}
\label{fig:dnoidal_loop}
\end{figure}

The shape of the static loop over one period, referred to as a cycle,
depends primarily on $q$,
while $c$ acts as a scaling factor.
Hereafter, $c=1$ will be fixed.
Each cycle consists of a twisted loop with an untwisted loop on both sides.
The peaks in the cycle are oriented up/down corresponding to the $\pm$ sign. 
See Fig.~\ref{fig:dnoidal_loop} for the up case. 

From the results in Table~\ref{table:conserved}, 
combined with equations \eqref{winding.number} and \eqref{winding.rate},
the loop has winding number
\begin{equation}
  N = \pm \frac{2}{\pi} \AM(\K(1/\sqrt{q}), 1/\sqrt{q})
\end{equation}
and winding rate
\begin{equation} 
  W = \frac{4}{\pi} \sqrt{\frac{cq}{2q - 1}}\, \E(1/\sqrt{q}) . 
\end{equation}

\subsection{Rational cn loops}
\label{sec:ratcn.loops}

Periodic rational-cn mKdV waves \eqref{ratcn} with $0<q<1$
have bright/dark peaks, which is determined by the $\pm$ sign in expression \eqref{ratcn}.
The resulting loop solutions are obtained similarly to the asymptotically circular soliton loops,
by evaluation of the parametric integral \eqref{vartheta}
and the algebraic formula \eqref{z.loop.eqn.C1not0}, 
in addition to expression \eqref{J}.
Note that $q$, $C$, and $c$ obey the inequalities
\begin{equation}
  \begin{aligned}
    \frac{\sqrt{1+32q(1-q)} - 8q(1-q) -1}{4(1-2q)(1-q)} < C^2 < 1,
    \quad
    0 < q < \tfrac{1}{2},
    \quad
    c>0 ; 
    \\
    0 < C^2 < \frac{\sqrt{1+32q(1-q)} - 8q(1-q) -1}{4(1-2q)(1-q)}, 
    \quad
    0 < q < \tfrac{1}{2},
    \quad
    c<0 ; 
     \\
     0 < C^2 < 1,
     \quad
     \tfrac{1}{2} < q < 1, 
     \quad
     c>0 . 
  \end{aligned}
\end{equation}

It will be convenient to let
\begin{equation}\label{ratcn.G}
  G = \sqrt{E}/(1-C^2) = \sqrt{(q(1-C^2)+C^2)/(1-C^2)}
\end{equation}
which is positive. 
The parametric integral \eqref{vartheta} is then given by 
\begin{equation}\label{vartheta.ratcn}
\begin{aligned}
\vartheta(s) = 
\pm 2\Bigg( &
-\frac{B}{C} \sqrt{c/F}\, s
- \arctan\bigg(G\,\frac{\SN(\sqrt{cE/F}\, s, \sqrt{q})}{\DN(\sqrt{cE/F}\, s, \sqrt{q})}\bigg)
\\&\qquad
+ \frac{G}{C}\, \Pi\big( \SN(\sqrt{cE/F}\, s, \sqrt{q}), C^2/(C^2 - 1), \sqrt{q} \big)
    \Bigg)
\end{aligned}
\end{equation}
where $\Pi$ denotes the elliptic integral of the third kind. 
This yields 
\begin{equation}
  \begin{aligned}
    e^{i\vartheta(s)} = & 
    \exp\bigg( \pm i   \frac{2}{C}\Big(
    -B\sqrt{c/F}\, s + G \Pi\big( \SN(\sqrt{cE/F}\, s, \sqrt{q}), C^2/(C^2 - 1), \sqrt{q} \big) \Big) 
\bigg)
    \\&\qquad \times
    \bigg( \frac{\mp G\, \SN(\sqrt{cE/F}\,s, \sqrt{q}) + i\, \DN(\sqrt{cE/F}\, s, \sqrt{q})}{\pm G\, \SN(\sqrt{cE/F}\,s, \sqrt{q}) + i\, \DN(\sqrt{cE/F}\, s, \sqrt{q})} \bigg) . 
  \end{aligned}
\end{equation}
Expression \eqref{J} is given by 
\begin{equation}\label{ratcn.J}
  \begin{aligned}
J(s) =
\frac{2\sqrt{F/c}}{(q-(q - 1)C^4)C} \bigg( &
\sqrt{E}^3\, \frac{\DN(\sqrt{cE/F}\, s, \sqrt{q})\, \SN(\sqrt{cE/F}\, s, \sqrt{q})}{\big( C\, \CN(\sqrt{cE/F}\, s, \sqrt{q}) + 1\big)^2}
\\&\quad
\pm i \bigg(
\tfrac{1}{2} F - \bigg(\frac{B\, \CN(\sqrt{cE/F}\, s, \sqrt{q}) +LC}{C\, \CN(\sqrt{cE/F}\, s, \sqrt{q}) + 1}\bigg)^2 \bigg)
\bigg)
  \end{aligned}
\end{equation}
with
\begin{equation}\label{ratcn.L}
  L =(1-q)(1-C^2) - \tfrac{1}{2}, 
\end{equation}
using $C_1 = \pm (q + (1-q)C^4)C \sqrt{c/F}^3$
from expression \eqref{C1.ratcn}.

The loop equation \eqref{z.loop.eqn.C1not0} is thus 
\begin{equation}\label{ratcn.loop}
z(s,t) = J(s)  e^{i\vartheta(s)} e^{\pm i (q+ (1-q)C^4)C \sqrt{c/F}^3 t}  + z_0 . 
\end{equation}
This describes a rotating loop whose angular speed is 
$\pm (q + (1-q)C^4)C \sqrt{c/F}^3$.
The resulting geometric curve flow \eqref{r.z} is given by 
the real and imaginary parts of the loop solution \eqref{ratcn.loop},
$\vec{r}_\pm(s,t)  =\big(\mathrm{Re}\,z(s,t),\mathrm{Im}\,z(s,t)\big)$, 
which is the composition of a static loop $\vec{r}_\pm(s,0)$ plus a rotational motion. 

\begin{figure}[H]
\centering
\begin{subfigure}[t]{0.9\textwidth}
\centering
\includegraphics[width=0.4\textwidth,trim=2cm 12cm 6cm 2cm,clip]{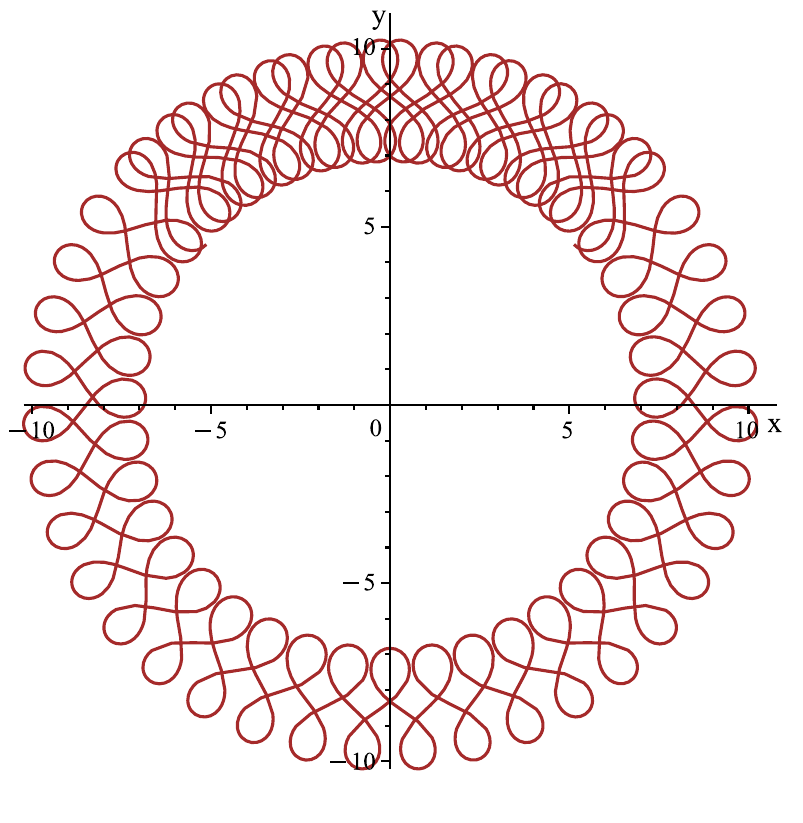}
\hfill
\includegraphics[width=0.4\textwidth,trim=2cm 12cm 6cm 2cm,clip]{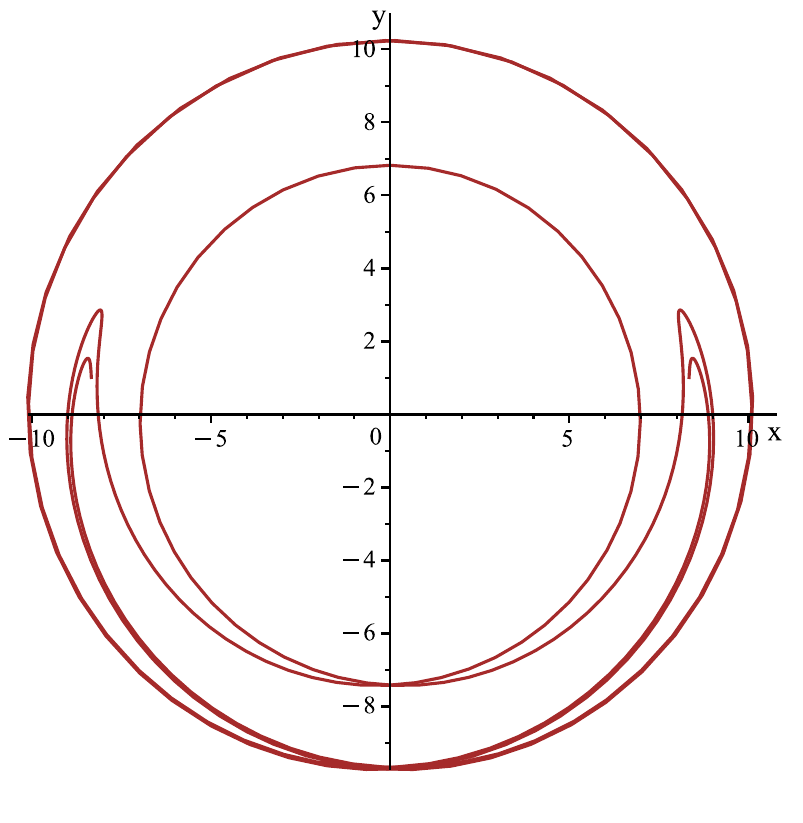}
\caption{$C=0.1$. Left: $\sigma=1$.  Right: $\sigma=-1$.}
\end{subfigure}

\vspace{0.5em}

\begin{subfigure}[t]{0.9\textwidth}
\centering
\includegraphics[width=0.4\textwidth,trim=2cm 12cm 6cm 2cm,clip]{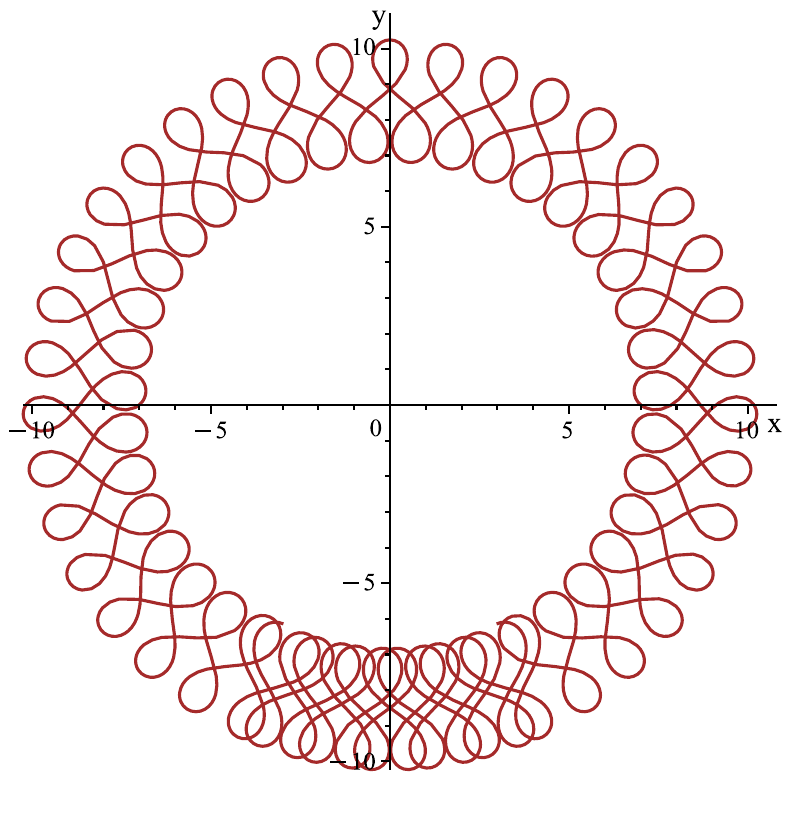}
\hfill
\includegraphics[width=0.4\textwidth,trim=2cm 12cm 6cm 2cm,clip]{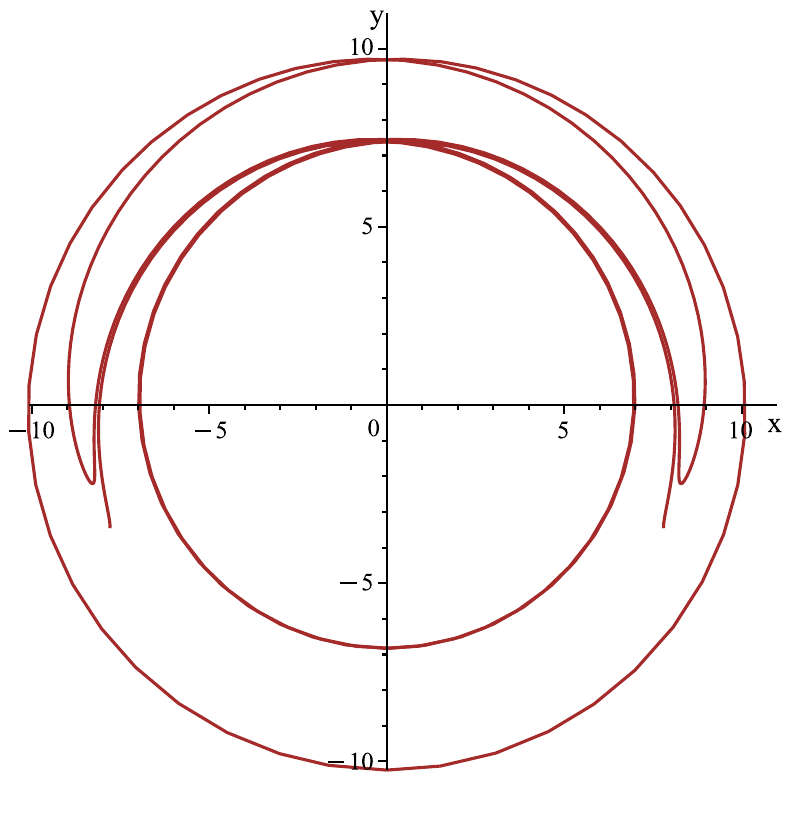}
\caption{$C=-0.1$. Left: $\sigma=1$.  Right: $\sigma=-1$.}
\end{subfigure}
\caption{rational $\mathrm{CN}$ open loops with $c=1, q=0.9$.}
\label{fig:ratcn_loop_open_curve1}
\end{figure}

\begin{figure}[H]
\centering
\begin{subfigure}[t]{0.9\textwidth}
\centering
\includegraphics[width=0.4\textwidth,trim=2cm 12cm 6cm 2cm,clip]{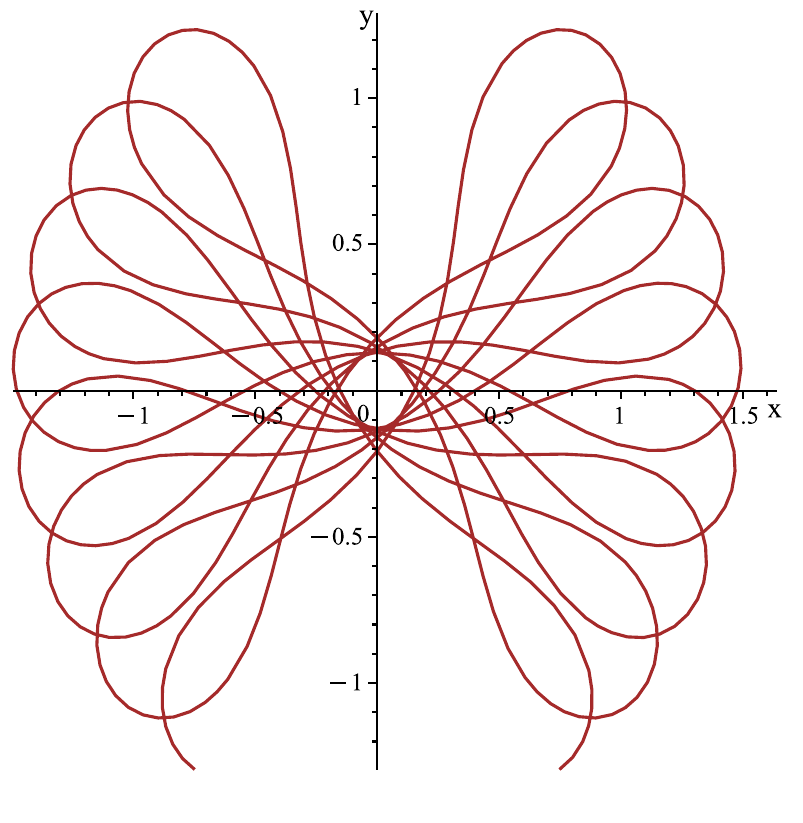}
\hfill
\includegraphics[width=0.4\textwidth,trim=2cm 12cm 6cm 2cm,clip]{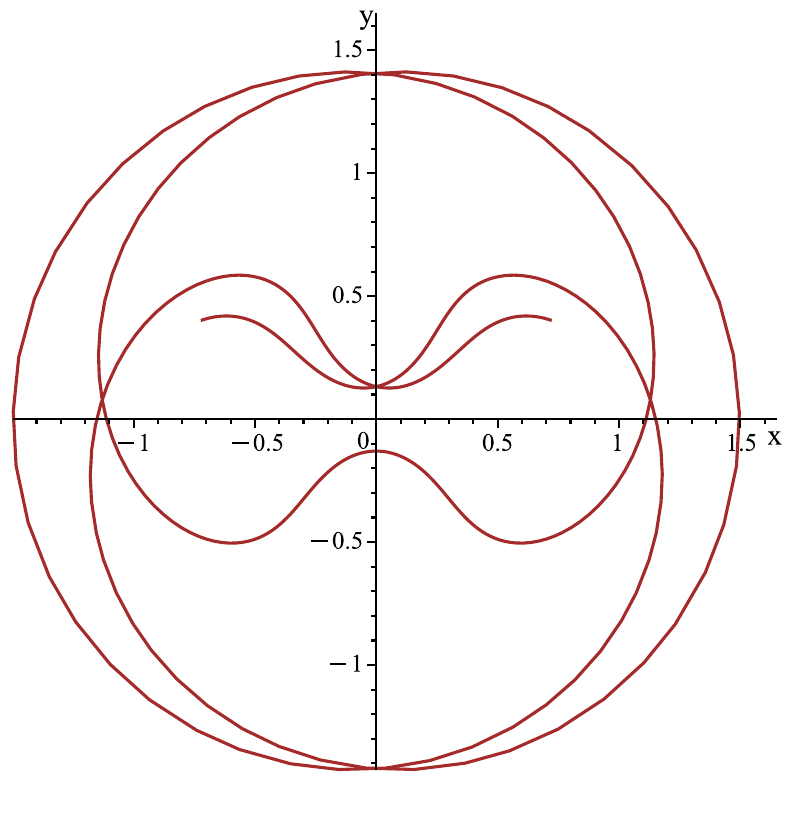}
\caption{$C = 0.4$. Left: $\sigma = 1$.  Right: $\sigma=-1$.}
\end{subfigure}

\vspace{0.5em}

\begin{subfigure}[t]{0.9\textwidth}
\centering
\includegraphics[width=0.4\textwidth,trim=2cm 12cm 6cm 2cm,clip]{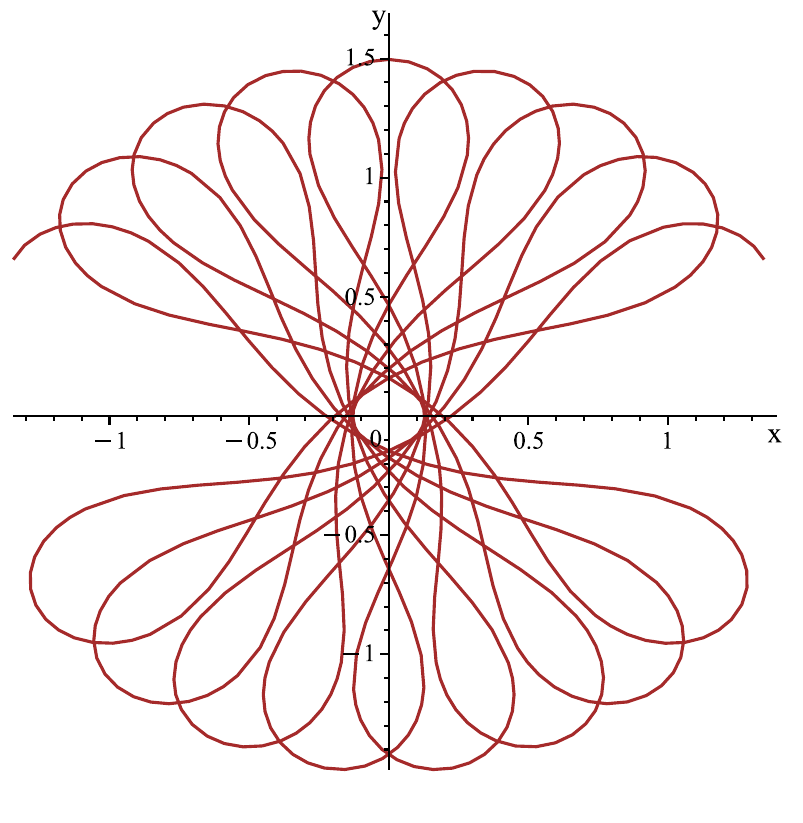}
\hfill
\includegraphics[width=0.4\textwidth,trim=2cm 12cm 6cm 2cm,clip]{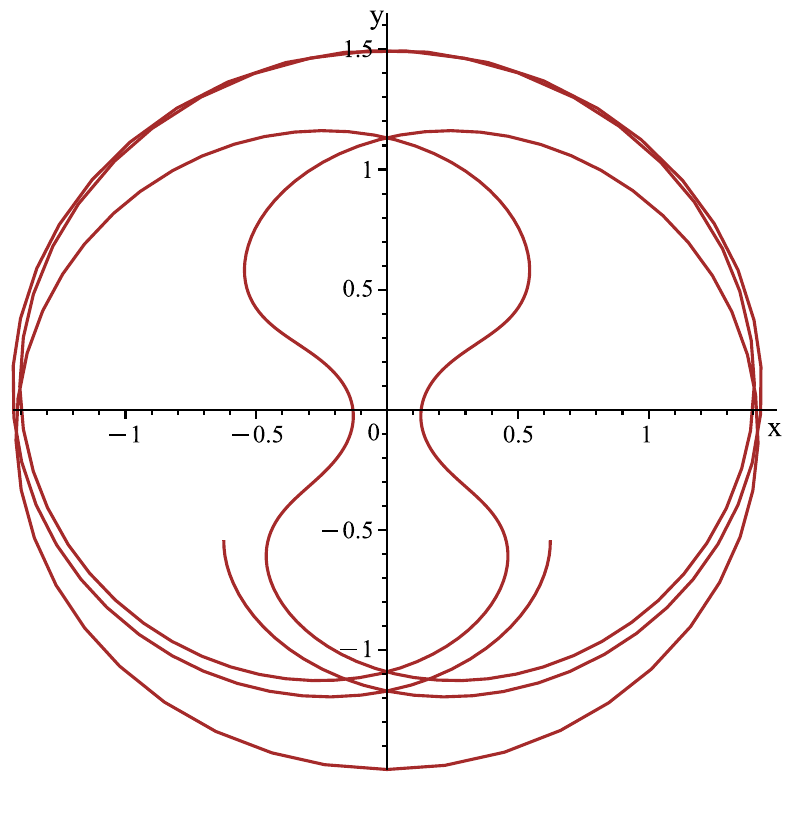}
\caption{$C = -0.4$. Left: $\sigma = 1$.  Right: $\sigma=-1$.}
\end{subfigure}
\caption{rational $\mathrm{CN}$ open loops with $c=1, q=0.2$.}
\label{fig:ratcn_loop_open_curve2}
\end{figure}

\begin{figure}[H]
\centering
\begin{subfigure}[t]{0.9\textwidth}
\centering
\includegraphics[width=0.4\textwidth,trim=2cm 12cm 6cm 2cm,clip]{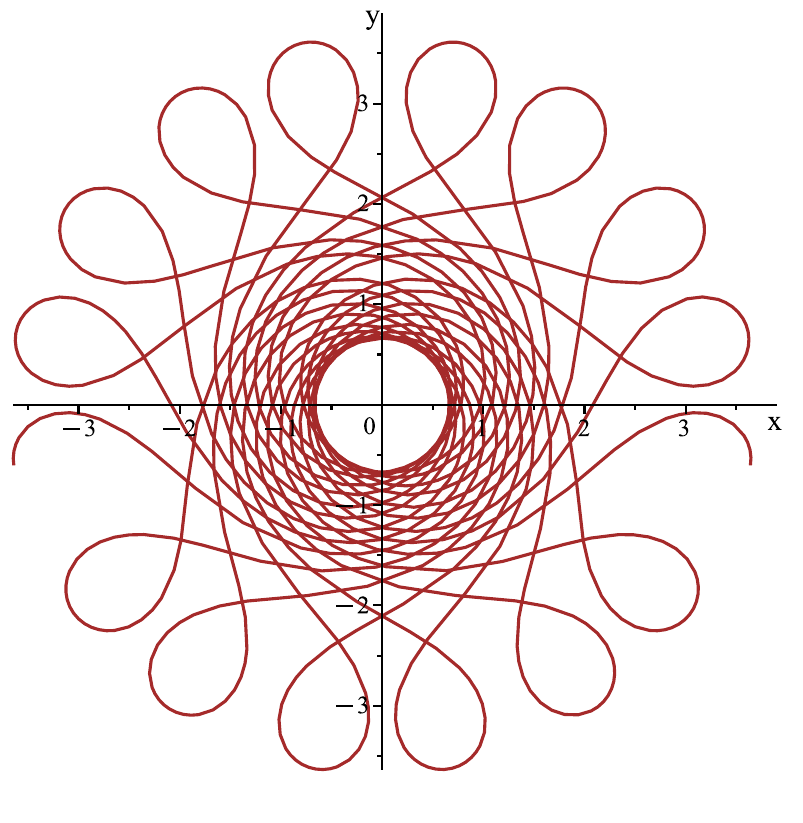}
\hfill
\includegraphics[width=0.4\textwidth,trim=2cm 12cm 6cm 2cm,clip]{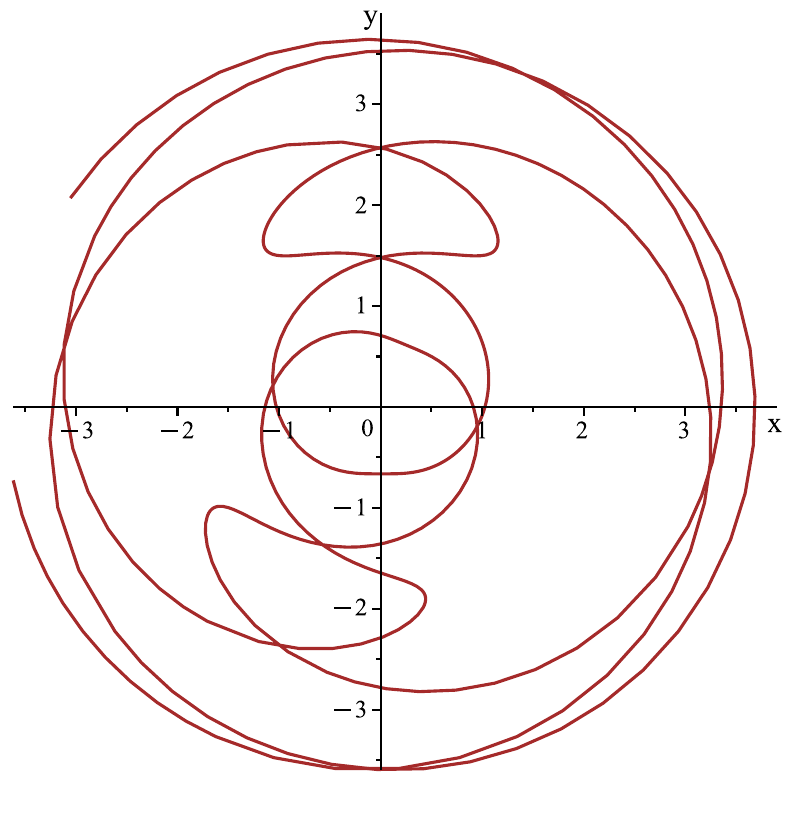}
\caption{$C = 0.7$. Left: $\sigma = 1$.  Right: $\sigma=-1$.}
\end{subfigure}

\vspace{0.5em}

\begin{subfigure}[t]{0.9\textwidth}
\centering
\includegraphics[width=0.4\textwidth,trim=2cm 12cm 6cm 2cm,clip]{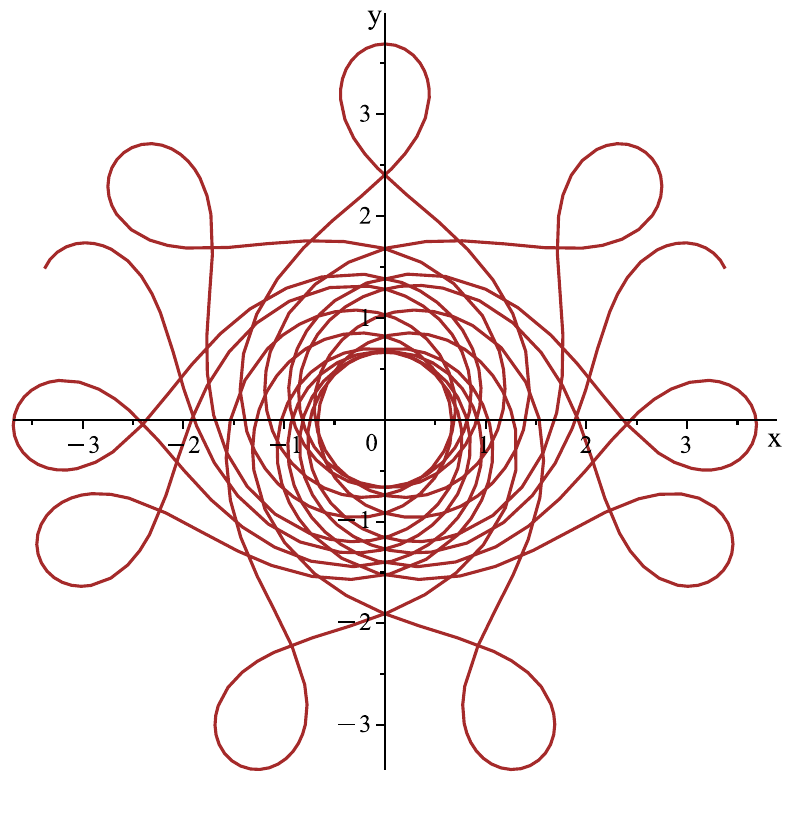}
\hfill
\includegraphics[width=0.4\textwidth,trim=2cm 12cm 6cm 2cm,clip]{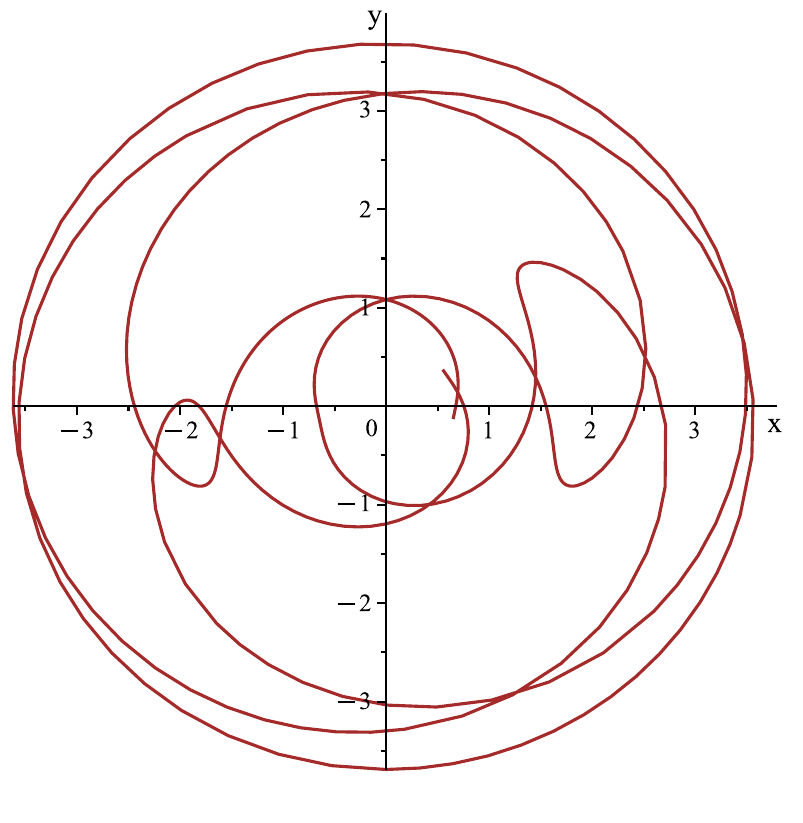}
\caption{$C = -0.7$. Left: $\sigma = 1$.  Right: $\sigma=-1$.}
\end{subfigure}
\caption{rational $\mathrm{CN}$ open loops with $c=1,q=0.9$.}
\label{fig:ratcn_loop_open_curve3}
\end{figure}

\begin{figure}[H]
\centering
\begin{subfigure}[t]{0.9\textwidth}
\centering
\includegraphics[width=0.4\textwidth,trim=2cm 12cm 6cm 2cm,clip]{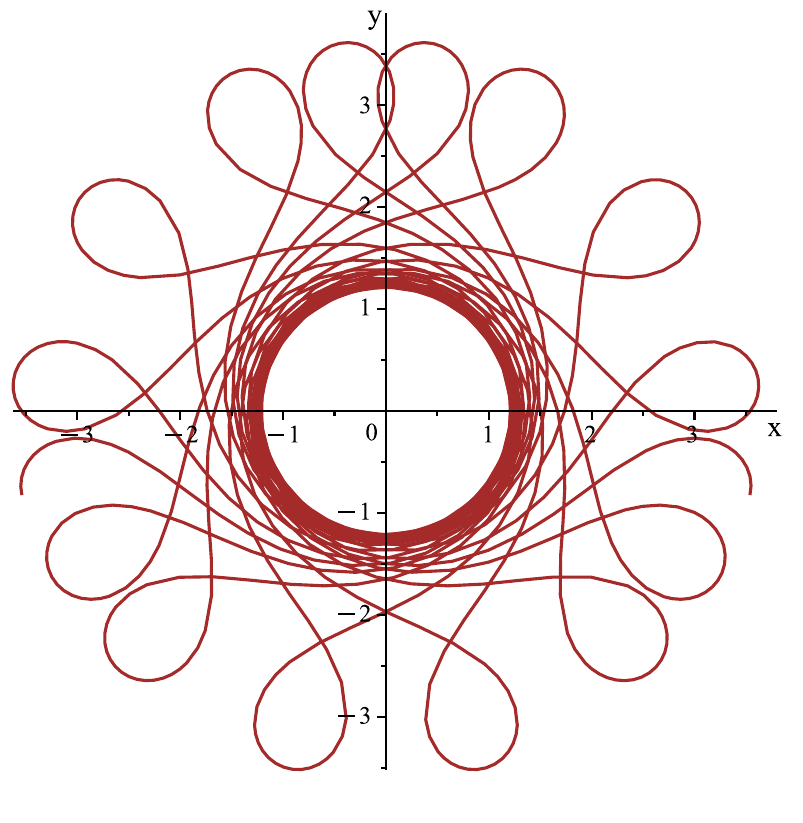}
\hfill
\includegraphics[width=0.4\textwidth,trim=2cm 12cm 6cm 2cm,clip]{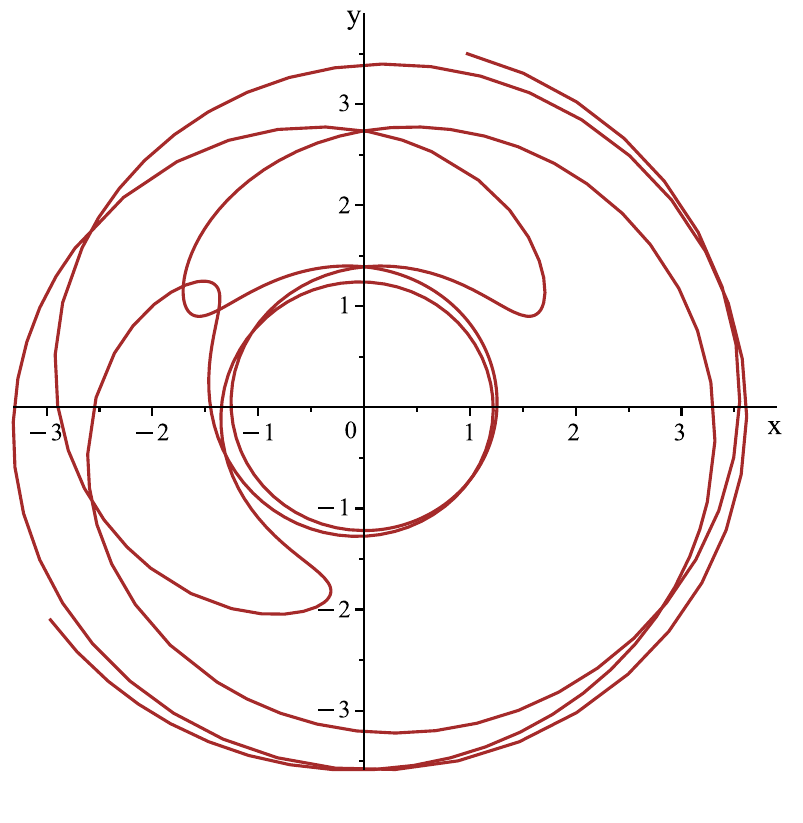}
\caption{$C = 0.9$. Left: $\sigma = 1$.  Right: $\sigma=-1$.}
\end{subfigure}

\vspace{0.5em}

\begin{subfigure}[t]{0.9\textwidth}
\centering
\includegraphics[width=0.4\textwidth,trim=2cm 12cm 6cm 2cm,clip]{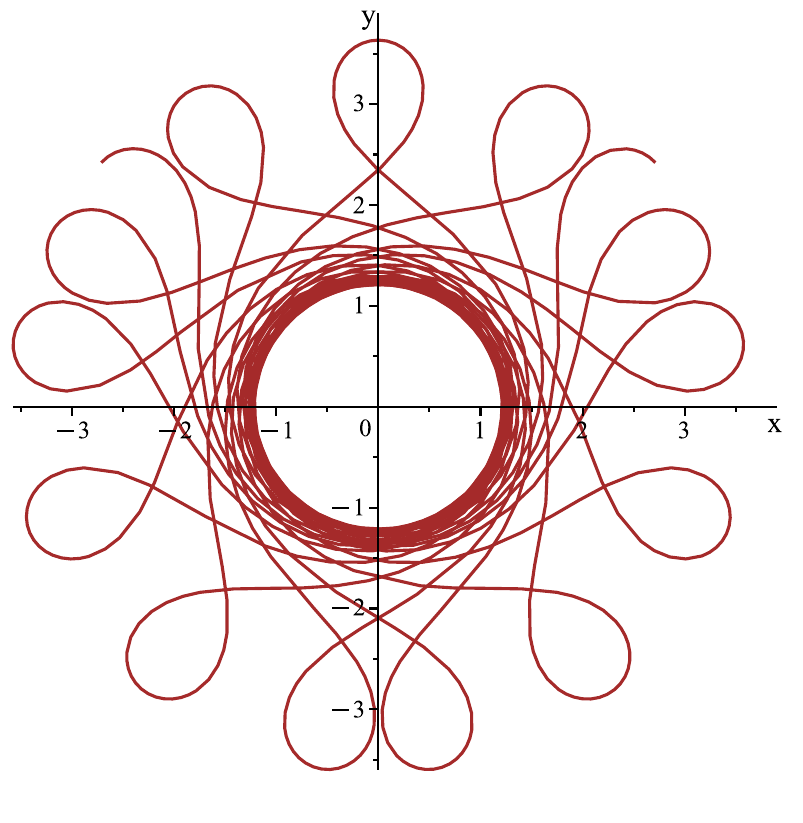}
\hfill
\includegraphics[width=0.4\textwidth,trim=2cm 12cm 6cm 2cm,clip]{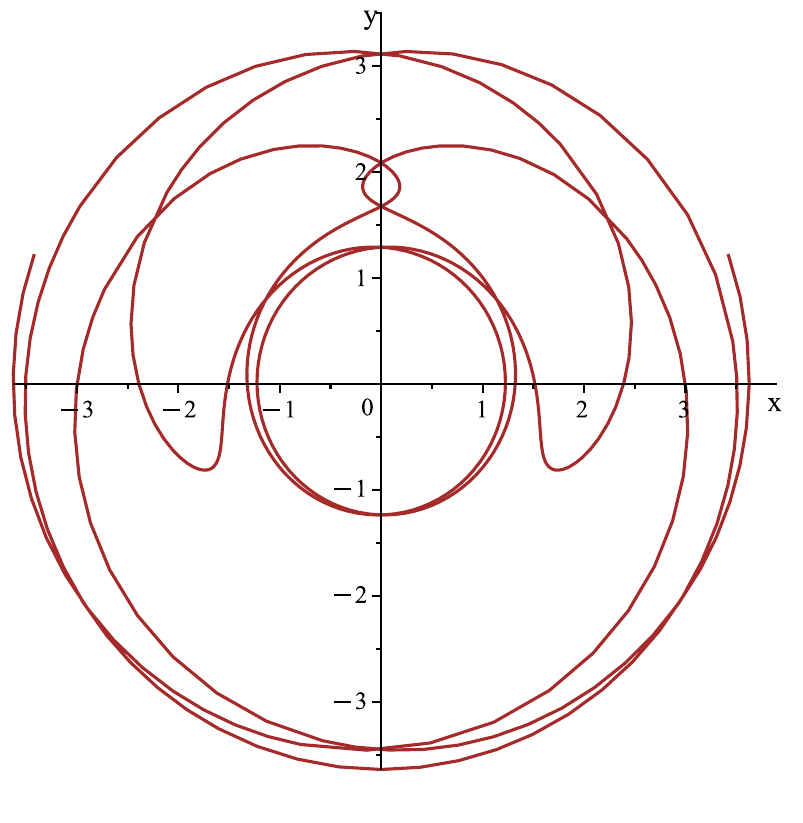}
\caption{$C = -0.9$. Left: $\sigma = 1$.  Right: $\sigma=-1$.}
\end{subfigure}
\caption{rational $\mathrm{CN}$ open loops with $c=1,q=0.4$.}
\label{fig:ratcn_loop_open_curve4}
\end{figure}

The static loop has the following basic features.
First, it is a bounded curve,
since 
\begin{equation}
  \begin{aligned}
  |\vec{r}_\pm(s,0)|^2 = |J(s)|^2 = & 
  \big(\tfrac{2c}{F C_1}\big)^2
  \Big(  (q +(1-q) C^4)^2 - 4q E
  \\&\qquad
  + 4(q + (1-q)C^4) E/(C\CN(\sqrt{cE/F},s,\sqrt{q}) + 1) \Big)
  \end{aligned}
\end{equation}
is a bounded function of $s$ with a respective maximum and minimum 
\begin{equation}\label{ratcn.annulus} 
  R_{\max}^2 = \big(\tfrac{2c}{F C_1}\big)^2 ((q+(1-q)C^4) + 2|C| G^2)^2,
  \quad
  R_{\min}^2 = \big(\tfrac{2c}{F C_1}\big)^2 ((q+(1-q)C^4) - 2|C| G^2)^2 . 
\end{equation}
Thus, the curve lies in an annulus with radii $R_{\max}$ and $R_{\min}$.
Second, its shape depends on the $\pm$ sign in the expression \eqref{ratcn.J}. 
In the ``$+$'' case,
the curve is comprised of petals that exhibit a double twist and face outwards,
with their tips lying on the outer annulus.
In the ``$-$'' case,
the petals are untwisted and face inwards, with their tips lying on the inner annulus. 
See the plots in Fig.~\ref{fig:ratcn_loop_open_curve1}--
~\ref{fig:ratcn_loop_open_curve4}. 

From the results in Table~\ref{table:conserved},
the winding number \eqref{winding.number} of the curve is 
\begin{equation}
  N= \mp \frac{4}{\pi}\frac{B}{C\sqrt{E}} \K(\sqrt{q})
\end{equation}
and the mean-winding rate \eqref{winding.rate} is 
\begin{equation}
  W = \frac{2}{\pi} \frac{(4(q - 1) E + C^2)}{\sqrt{E}} \sqrt{\frac{c}{F}} \K(\sqrt{q}) . 
\end{equation}

A main question now is whether the curve is open or closed.
Clearly, $|\vec{r}_\pm(s,0)|$ is periodic in $s$,
with the same periodic as the $\CN$ function,
which is given by expression \eqref{ratCN.period}.
Closure will hold whenever this period is commensurable
with the period of the turning angle of $\vec{r}_\pm(0,s)$.

The turning angle is defined by
\begin{equation}\label{ratcn.turningangle}
  \arg(\vec{r}_\pm(0,s)) = \vartheta(s) + \arg(J(s))
\end{equation}
where
\begin{equation}
\arg\big(J(s)\big) =
\pm\arctan\bigg( \frac{\tfrac{1}{2}F (C\CN(\sqrt{cE/F}\, s, \sqrt{q}) + 1)^2 -\big( B\CN(\sqrt{cE/F}\, s, \sqrt{q}) + LC \big)^2}{\sqrt{E}^3 \SN(\sqrt{cE/F}\, s, \sqrt{q}) \DN(\sqrt{cE/F}\, s, \sqrt{q})} \bigg) . 
\end{equation}
It is useful to decompose the turning angle \eqref{ratcn.turningangle}
into two distinct parts:
periodic terms, which have the same period as the $\CN$ function;
a secular term, which comprises all non-periodic terms. 
If $s$ is increased by the period of the $\CN$ function,
then the turning angle increases only by its secular change. 
From expression \eqref{vartheta.ratcn} for $\vartheta(s)$, 
this secular change is given by
\begin{equation}
  \Delta \vartheta =
  \pm \frac{8}{C}\Bigg( 
-\frac{B}{\sqrt{E}} \K(\sqrt{q})
+ G\, \Pi\big(C^2/(C^2 - 1), \sqrt{q} \big)
\Bigg)
\end{equation}
using expression \eqref{ratCN.period} for the period of the $\CN$ function,
where $\K$ is the complete elliptic integral of the first kind
and $\Pi$ is the complete elliptic integral of the third kind. 
Commensurability between $\Delta \vartheta$ and the period of the $\CN$ function
requires that some integer multiple of $\Delta \vartheta$ vanishes modulo $2\pi$. 
This establishes the following closure result.

\begin{prop}
(i) 
The bounded curve $\vec{r}(s,0)$ is closed and periodic iff
\begin{equation}\label{ratcn.closed.cond}
\Delta \vartheta/(2\pi) = n/m
\end{equation}
holds for some non-zero integers $n$, $m$ (with no common factor).
(ii)
The period of a closed curve is 
\begin{equation}
\Delta s =   4 m \sqrt{F/(cE)}\,\K(\sqrt{q}) . 
\end{equation}
\end{prop}

\begin{figure}[h]
\centering
\includegraphics[width=0.3\textwidth,trim=2cm 12cm 6cm 2cm,clip]{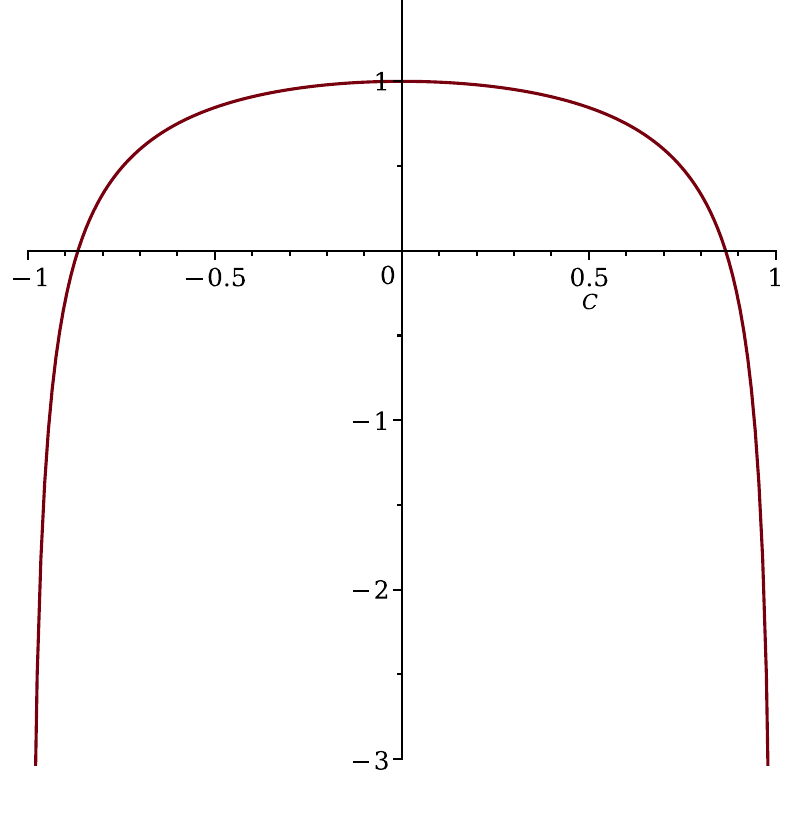}
\caption{Largest and smallest ratio for $n/m$ in the case $\sigma\sgn(C)=1$. }
\label{fig:ratcn_closed_ratio}
\end{figure}

For a fixed value of $C$,
condition \eqref{ratcn.closed.cond} is graphically equivalent to
finding the intersection points of the left-hand side curve and the right-hand side horizontal line.
The expression $\Delta\vartheta/(2\pi)$ has overall sign $\sigma\,\sgn(C)$.
Modulo this sign, it goes to the finite value $(2 - 1/\sqrt{1-C^2})<1$ at $q=0$
and approaches $-\infty$ as $q\to 1$. 
This implies that when $C^2< 3/4$ there will be 
a largest positive ratio allowed for $n/m$ in the case $\sigma\,\sgn(C)=1$,
and a largest negative ratio in the case $\sigma\,\sgn(C)=-1$.
These largest ratios obey $|n/m|<1$. 
When $3/4\leq C^2 < 1$, there will be a smallest negative ratio in the former case,
and a smallest positive ratio in the latter case,
which are unbounded. 
See Fig.~\ref{fig:ratcn_closed_ratio}. 

\begin{figure}[H]
\centering
\begin{subfigure}[t]{0.9\textwidth}
\centering
\includegraphics[width=0.325\textwidth,trim=2cm 12cm 6cm 2cm,clip]{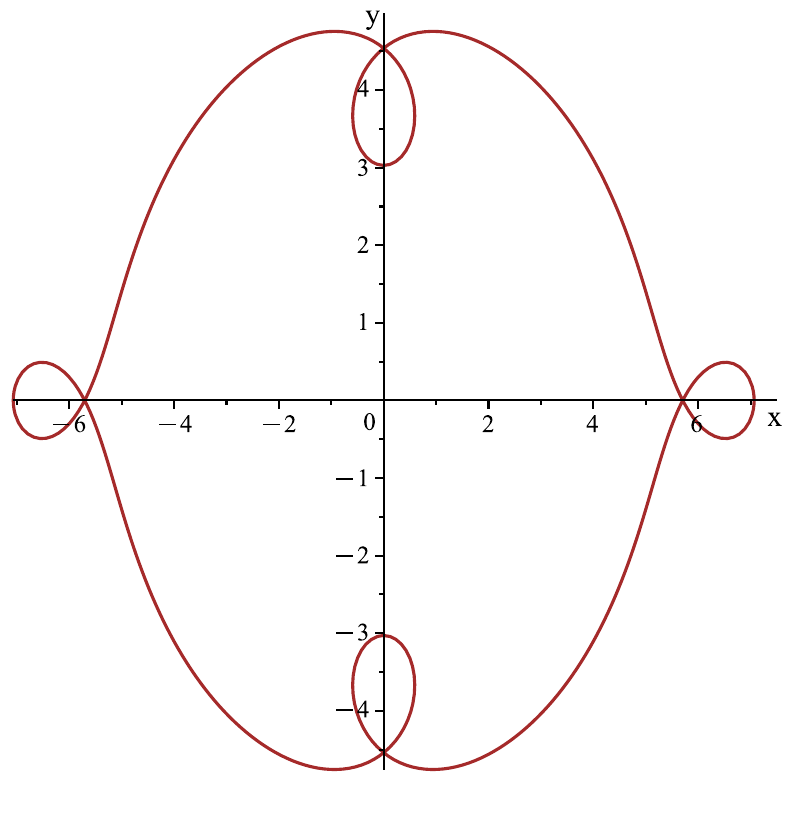}
\hfill
\includegraphics[width=0.325\textwidth,trim=2cm 12cm 6cm 2cm,clip]{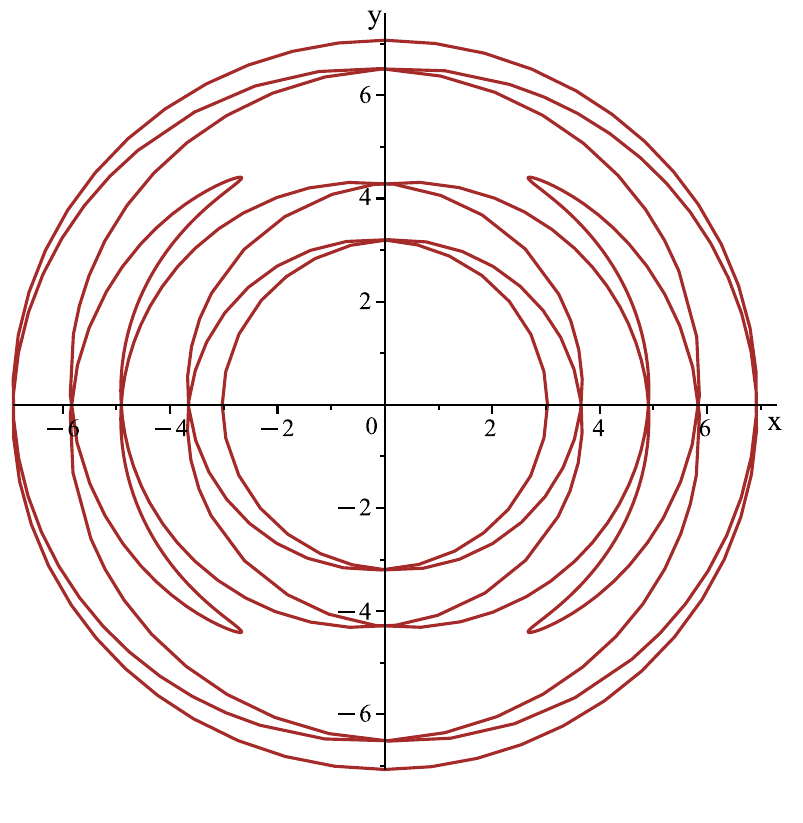}
\caption{$m=2,n=-1,q=0.9999$. Left: $\sigma = 1,C=0.2$.  Right: $\sigma=-1,C=-0.2$.}
\end{subfigure}

\vspace{0.5em}

\begin{subfigure}[t]{0.9\textwidth}
\centering
\includegraphics[width=0.325\textwidth,trim=2cm 12cm 6cm 2cm,clip]{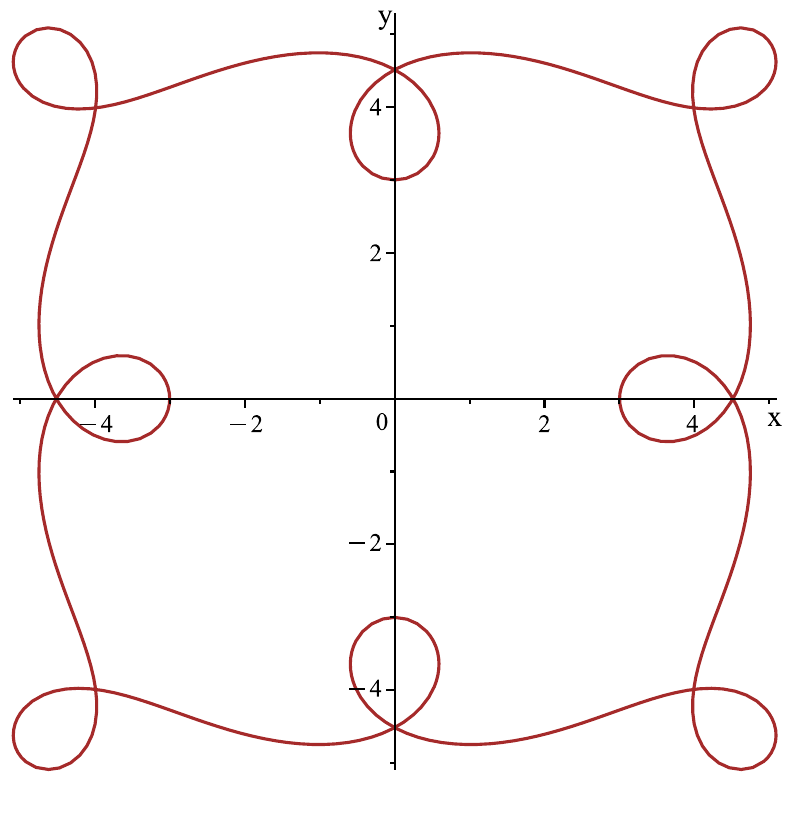}
\hfill
\includegraphics[width=0.325\textwidth,trim=2cm 12cm 6cm 2cm,clip]{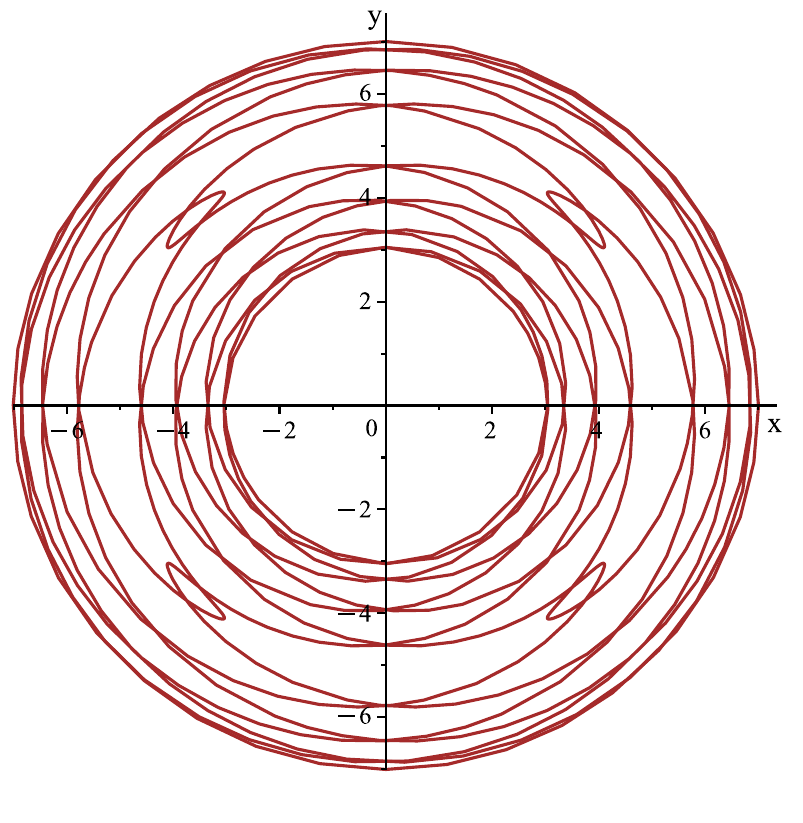}
\hfill
\caption{$m=4,n=-1,q=0.9935$. Left: $\sigma = 1,C=0.2$.  Right: $\sigma=-1,C=-0.2$.}
\end{subfigure}

\vspace{0.5em}

\begin{subfigure}[t]{0.9\textwidth}
\centering
\includegraphics[width=0.325\textwidth,trim=2cm 12cm 6cm 2cm,clip]{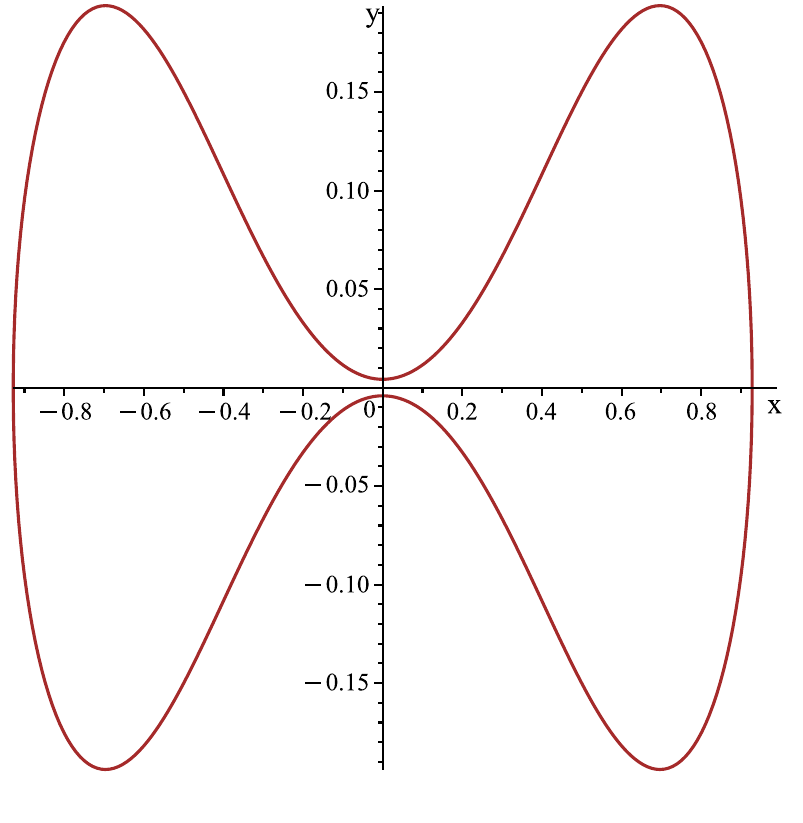}
\hfill
\includegraphics[width=0.325\textwidth,trim=2cm 12cm 6cm 2cm,clip]{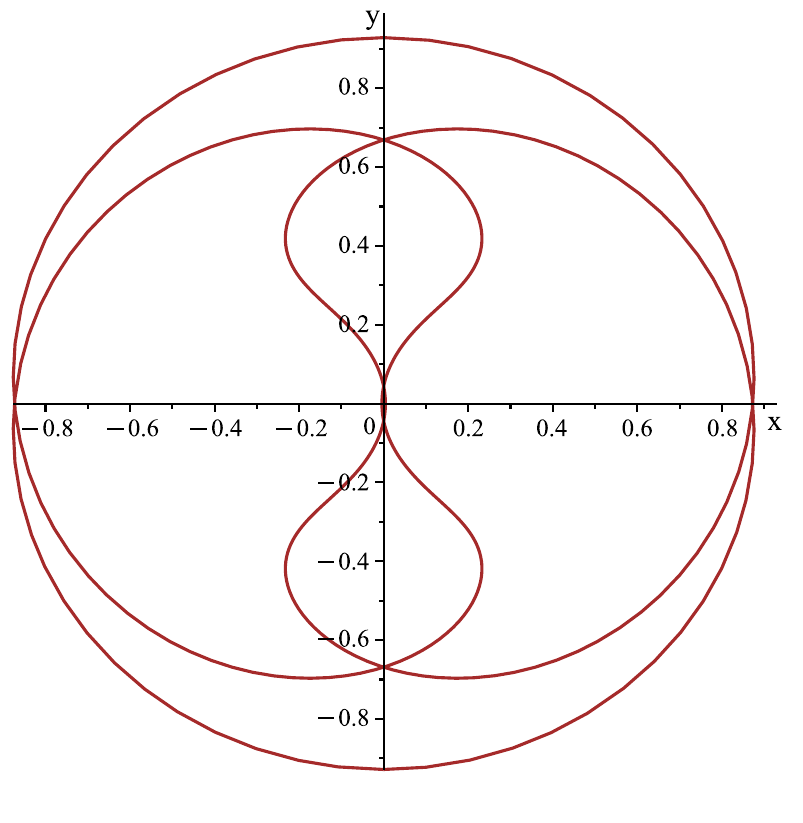}
\hfill
\caption{$m=2,n=1,q=0.1958$. Left: $\sigma = 1,C=0.35$.  Right: $\sigma=-1,C=-0.35$.}
\end{subfigure}
\caption{rational $\mathrm{CN}$ closed loops with $c=1$.}
\label{fig:ratcn_loop_closed_curve1}
\end{figure}

When a curve is closed, note that it has $m$ maxima points,
since this is the number of periods of the $\CN$ function that will occur
as $\vec{r}(s,0)$ traces out the curve once.
Plots of closed loops are shown in
Figs.~\ref{fig:ratcn_loop_closed_curve1}, ~\ref{fig:ratcn_loop_closed_curve2}, ~\ref{fig:ratcn_loop_closed_curve3}, ~\ref{fig:ratcn_loop_closed_curve4}, ~\ref{fig:ratcn_loop_closed_curve5} and~\ref{fig:ratcn_loop_closed_curve6}.

\begin{figure}[H]
\centering
\includegraphics[width=0.325\textwidth,trim=2cm 12cm 6cm 2cm,clip]{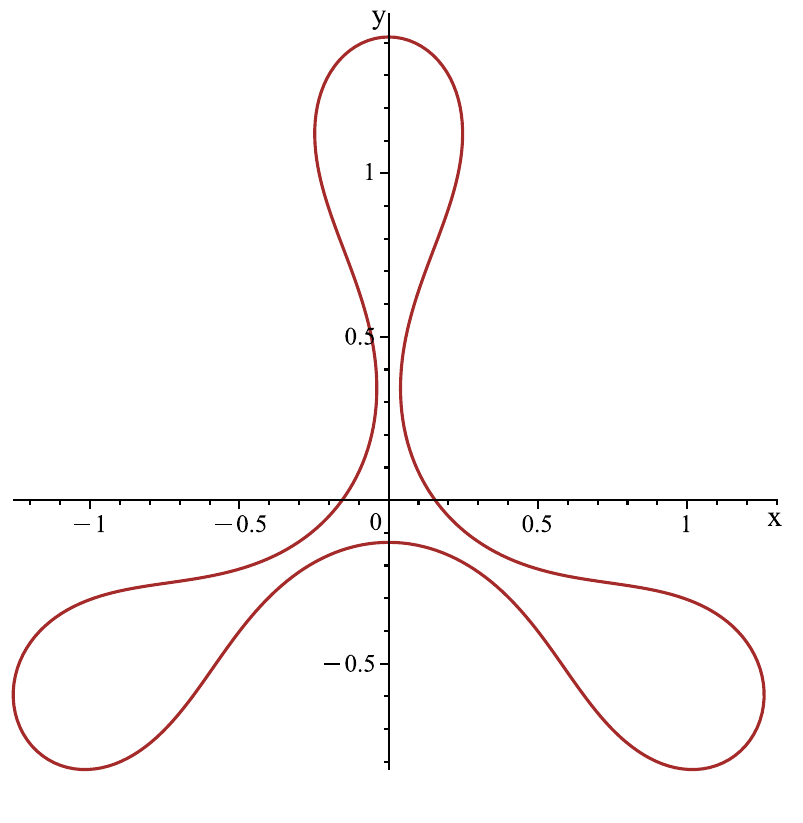}
\hfill
\includegraphics[width=0.325\textwidth,trim=2cm 12cm 6cm 2cm,clip]{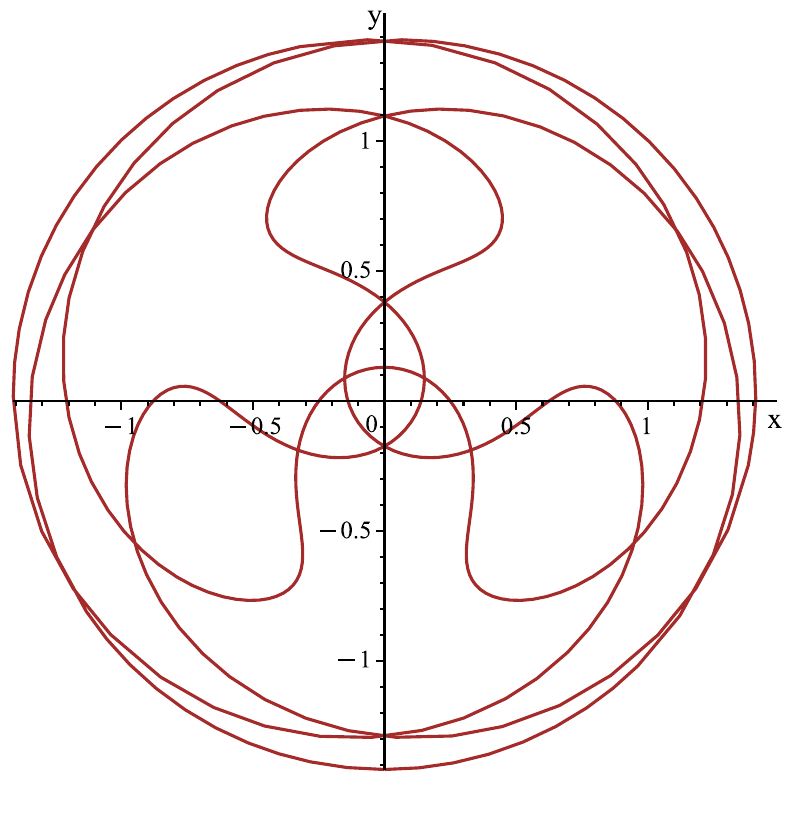}
\caption{$m=3,n=1,q=0.3549$. Left: $\sigma = 1,C=0.35$.  Right: $\sigma=-1,C=-0.35$.}
\end{figure}

\begin{figure}[H]
\centering
\begin{subfigure}[t]{0.9\textwidth}
\centering
\includegraphics[width=0.325\textwidth,trim=2cm 12cm 6cm 2cm,clip]{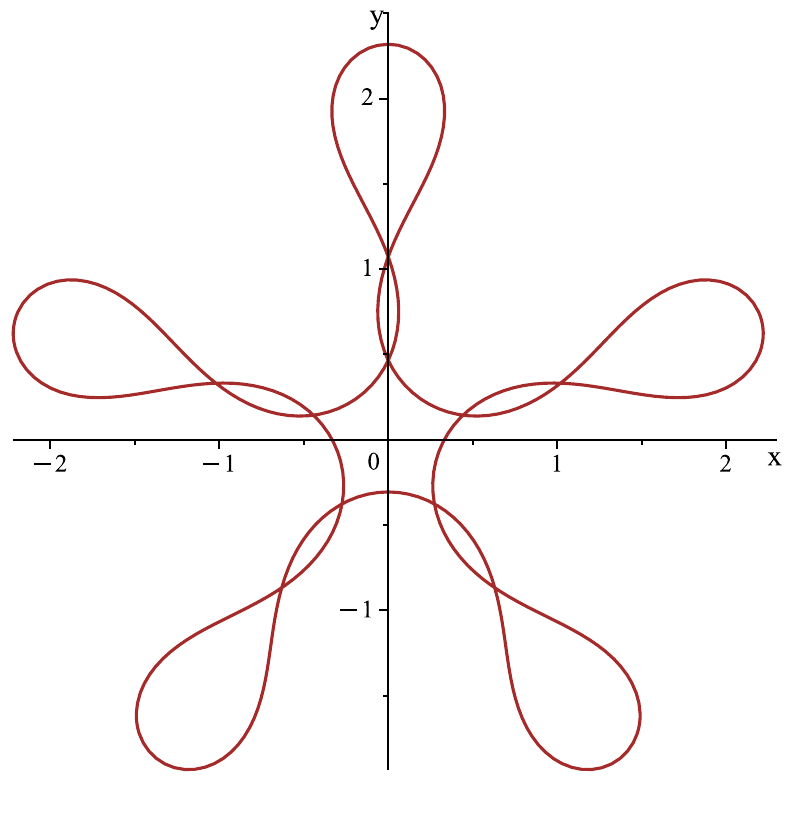}
\hfill
\includegraphics[width=0.325\textwidth,trim=2cm 12cm 6cm 2cm,clip]{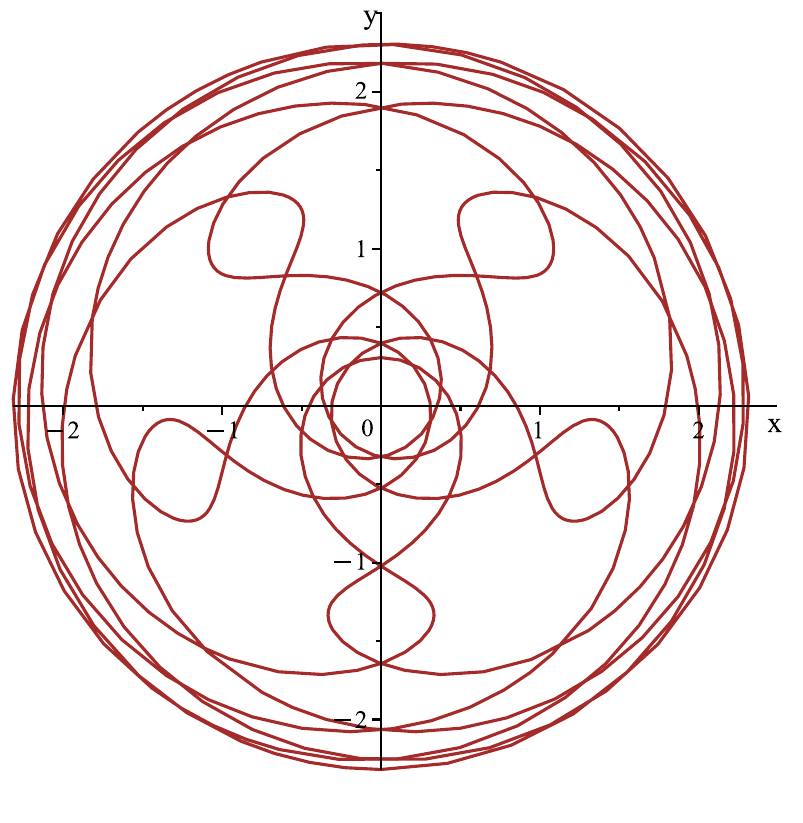}
\caption{$m=5,n=1,q=0.5265$. Left: $\sigma = 1,C=0.35$.  Right: $\sigma=-1,C=-0.35$.}
\end{subfigure}

\vspace{0.5em}

\begin{subfigure}[t]{0.9\textwidth}
\centering
\includegraphics[width=0.325\textwidth,trim=2cm 12cm 6cm 2cm,clip]{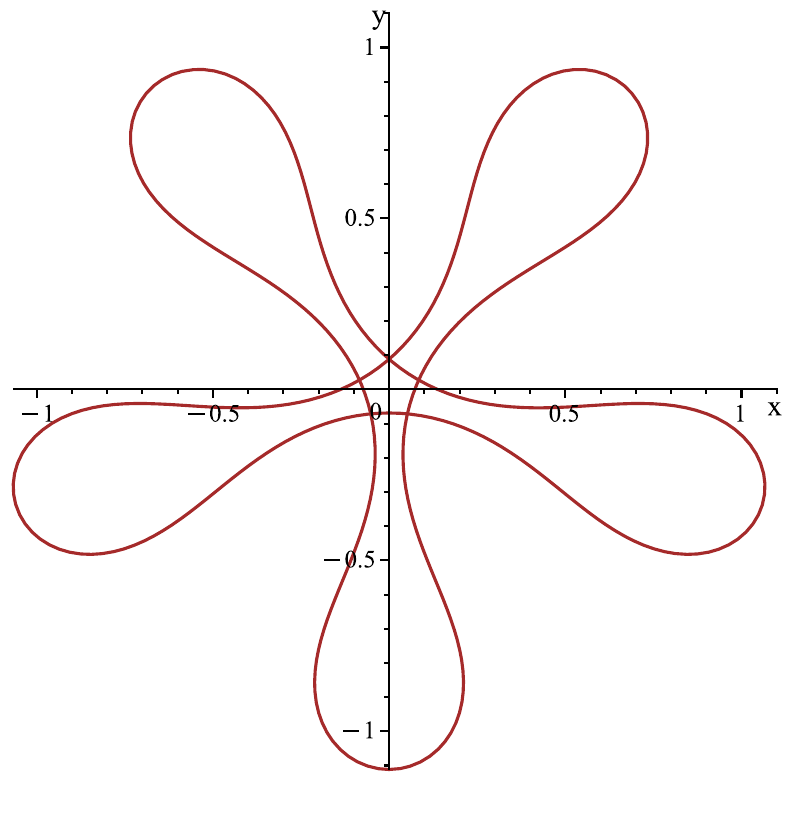}
\hfill
\includegraphics[width=0.325\textwidth,trim=2cm 12cm 6cm 2cm,clip]{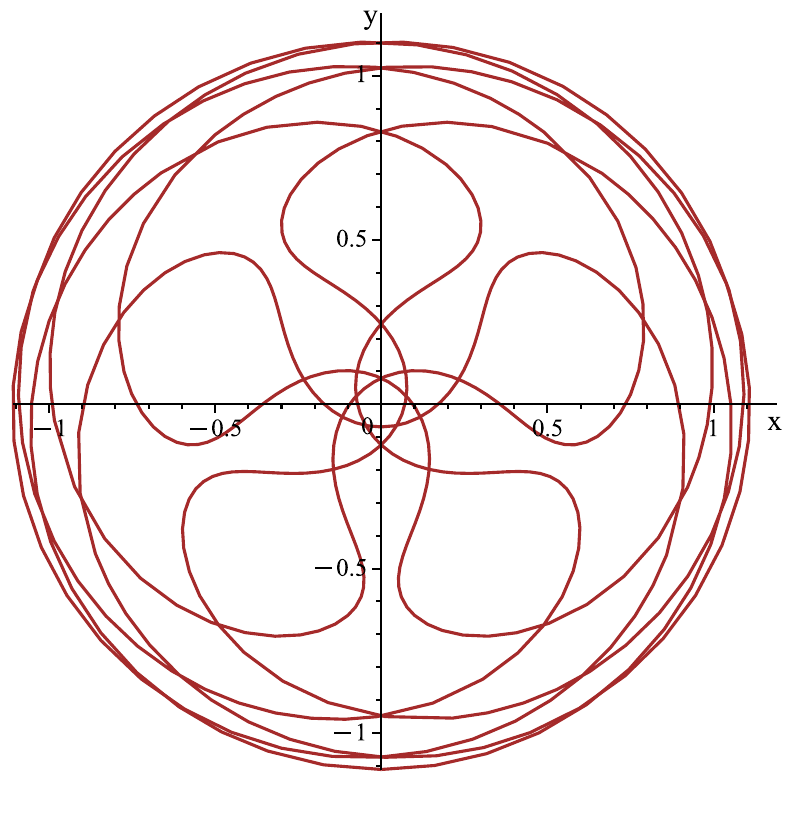}
\hfill
\caption{$m=5,n=2,q=0.2835$. Left: $\sigma = 1,C=0.35$.  Right: $\sigma=-1,C=-0.35$.}
\end{subfigure}

\vspace{0.5em}

\begin{subfigure}[t]{0.9\textwidth}
\centering
\includegraphics[width=0.325\textwidth,trim=2cm 12cm 6cm 2cm,clip]{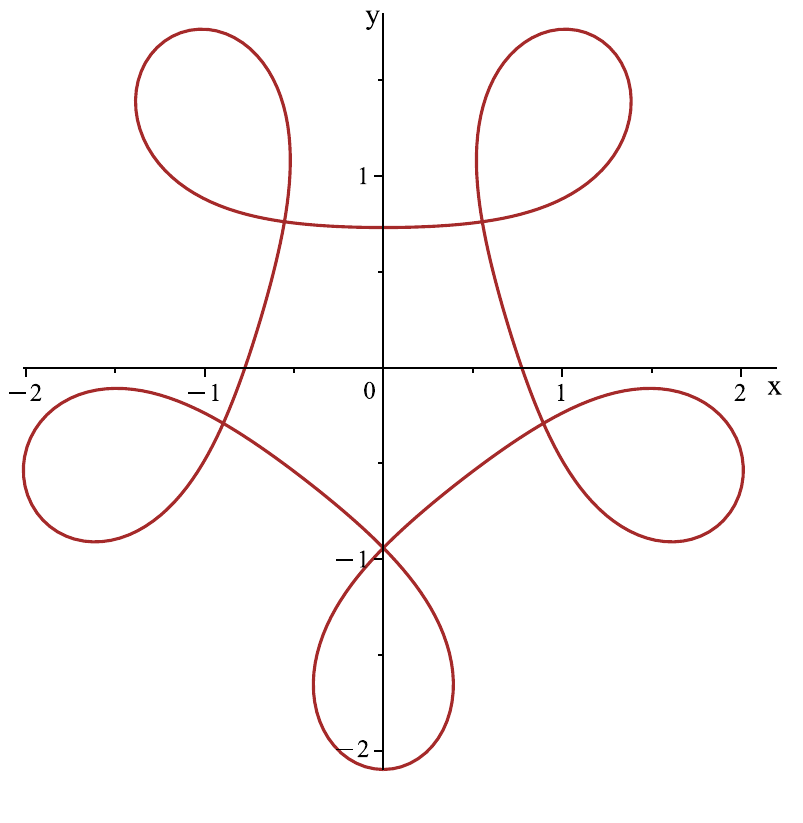}
\hfill
\includegraphics[width=0.325\textwidth,trim=2cm 12cm 6cm 2cm,clip]{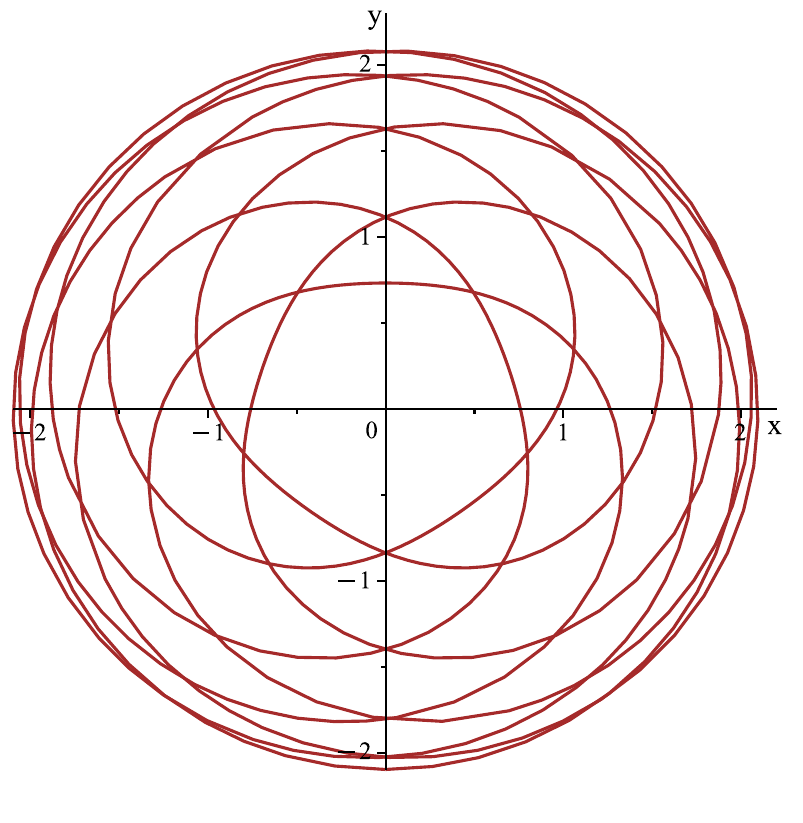}
\hfill
\caption{$m=5,n=4,q=0.0384$. Left: $\sigma = 1,C=0.35$.  Right: $\sigma=-1,C=-0.35$.}
\end{subfigure}
\caption{rational $\mathrm{CN}$ closed loops with $c=1$.}
\label{fig:ratcn_loop_closed_curve2}
\end{figure}

\begin{figure}[H]
\centering
\begin{subfigure}[t]{0.9\textwidth}
\centering
\includegraphics[width=0.325\textwidth,trim=2cm 12cm 6cm 2cm,clip]{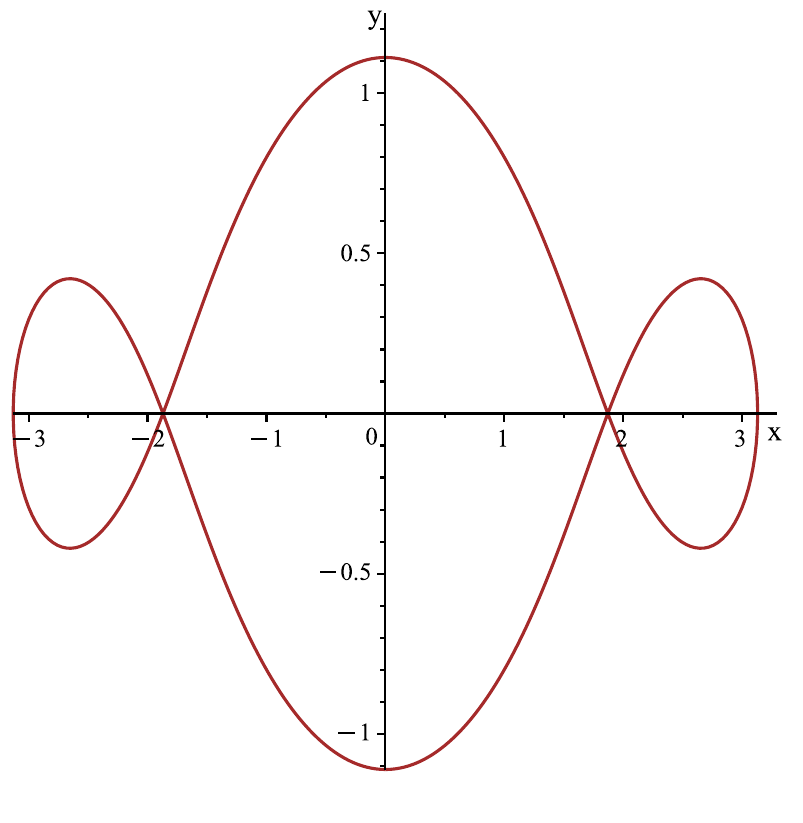}
\hfill
\includegraphics[width=0.325\textwidth,trim=2cm 12cm 6cm 2cm,clip]{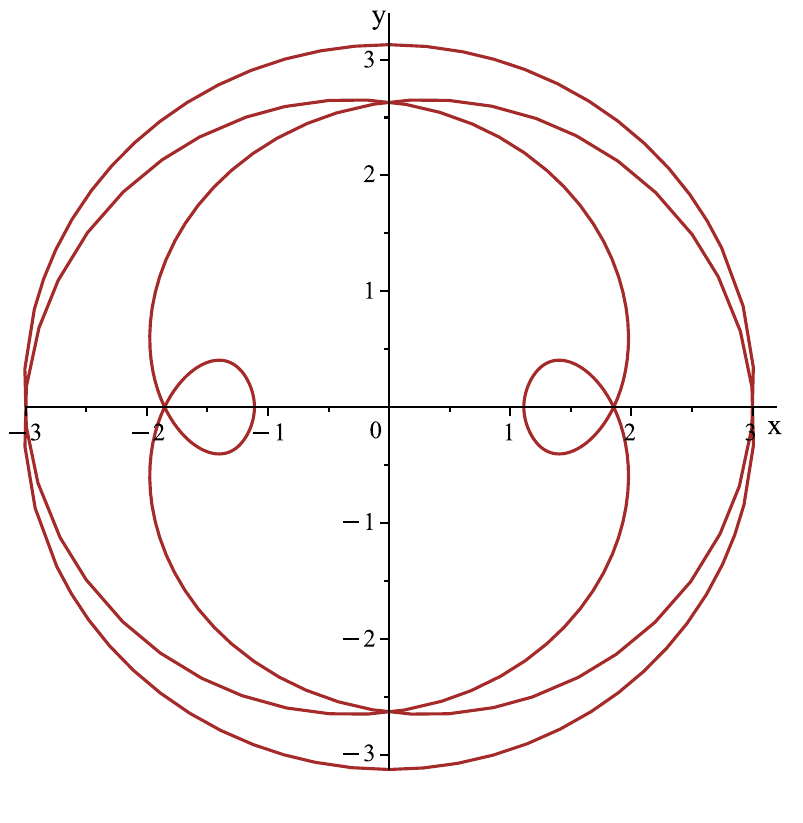}
\hfill
\caption{$m=2,n=1,q=0.1726$. Left: $\sigma = 1,C=0.64$.  Right: $\sigma=-1,C=-0.64$.}
\end{subfigure}

\vspace{0.5em}

\begin{subfigure}[t]{0.9\textwidth}
\centering
\includegraphics[width=0.325\textwidth,trim=2cm 12cm 6cm 2cm,clip]{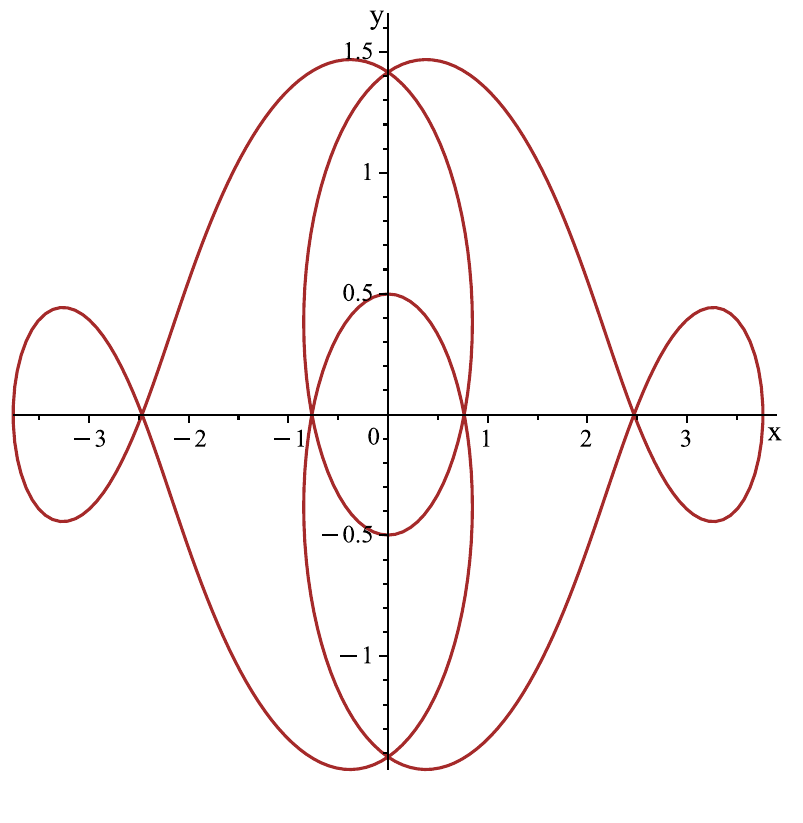}
\hfill
\includegraphics[width=0.325\textwidth,trim=2cm 12cm 6cm 2cm,clip]{fig-rat_cn_C_0.64_Csq_0.4096_c_1_m_2_n_-1_q_0.9222839953_sigma_1.pdf}
\hfill
\caption{$m=2,n=-1,q=0.9223$. Left: $\sigma = 1,C=0.64$.  Right: $\sigma=-1,C=-0.64$.}
\end{subfigure}

\vspace{0.5em}

\begin{subfigure}[t]{0.9\textwidth}
\centering
\includegraphics[width=0.325\textwidth,trim=2cm 12cm 6cm 2cm,clip]{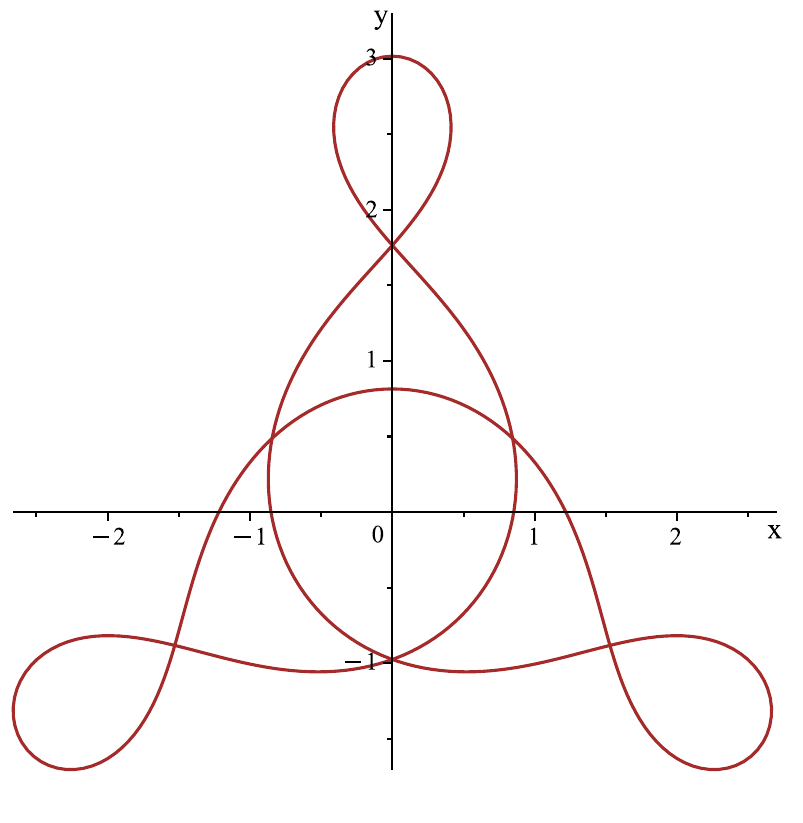}
\hfill
\includegraphics[width=0.325\textwidth,trim=2cm 12cm 6cm 2cm,clip]{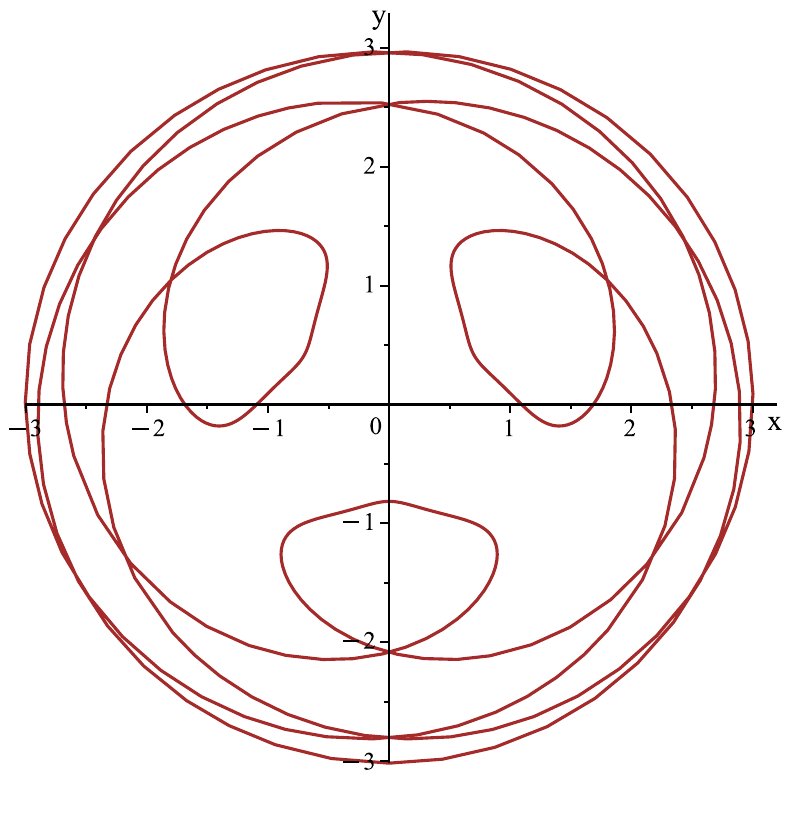}
\hfill
\caption{$m=3,n=-1,q=0.3357$. Left: $\sigma = 1,C=0.64$.  Right: $\sigma=-1,C=-0.64$.}
\end{subfigure}
\caption{rational $\mathrm{CN}$ closed loops with $c=1$.}
\label{fig:ratcn_loop_closed_curve3}
\end{figure}

\begin{figure}[H]
\centering
\begin{subfigure}[t]{0.9\textwidth}
\centering
\includegraphics[width=0.325\textwidth,trim=2cm 12cm 6cm 2cm,clip]{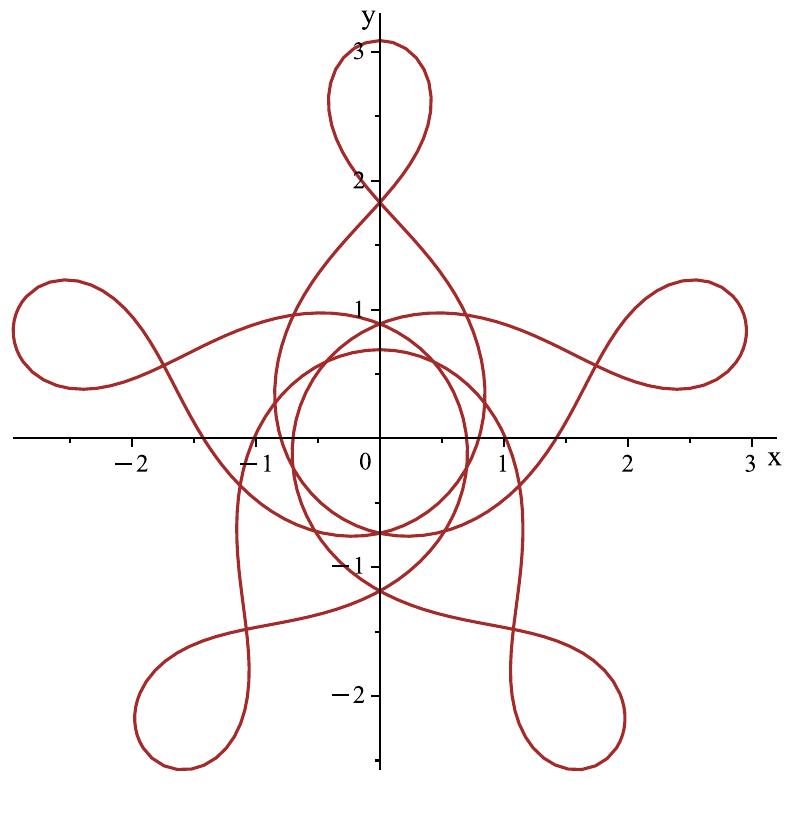}
\hfill
\includegraphics[width=0.325\textwidth,trim=2cm 12cm 6cm 2cm,clip]{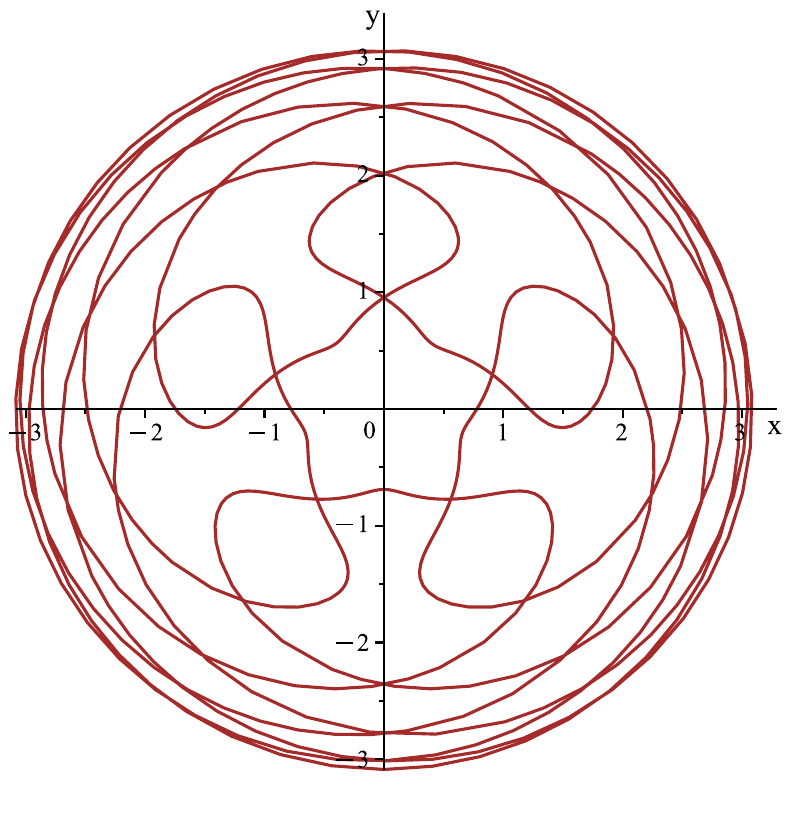}
\hfill
\caption{$m=5,n=1,q=0.4694$. Left: $\sigma=1,C=0.64$. Right: $\sigma=-1,C=-0.64$.}
\end{subfigure}

\vspace{0.5em}

\begin{subfigure}[t]{0.9\textwidth}
\centering
\includegraphics[width=0.325\textwidth,trim=2cm 12cm 6cm 2cm,clip]{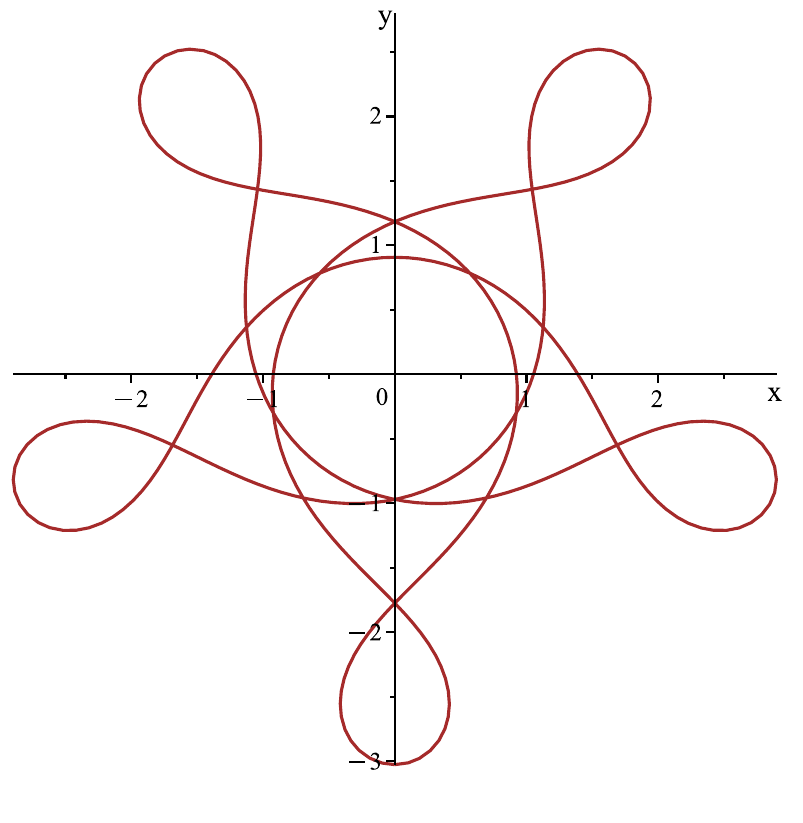}
\hfill
\includegraphics[width=0.325\textwidth,trim=2cm 12cm 6cm 2cm,clip]{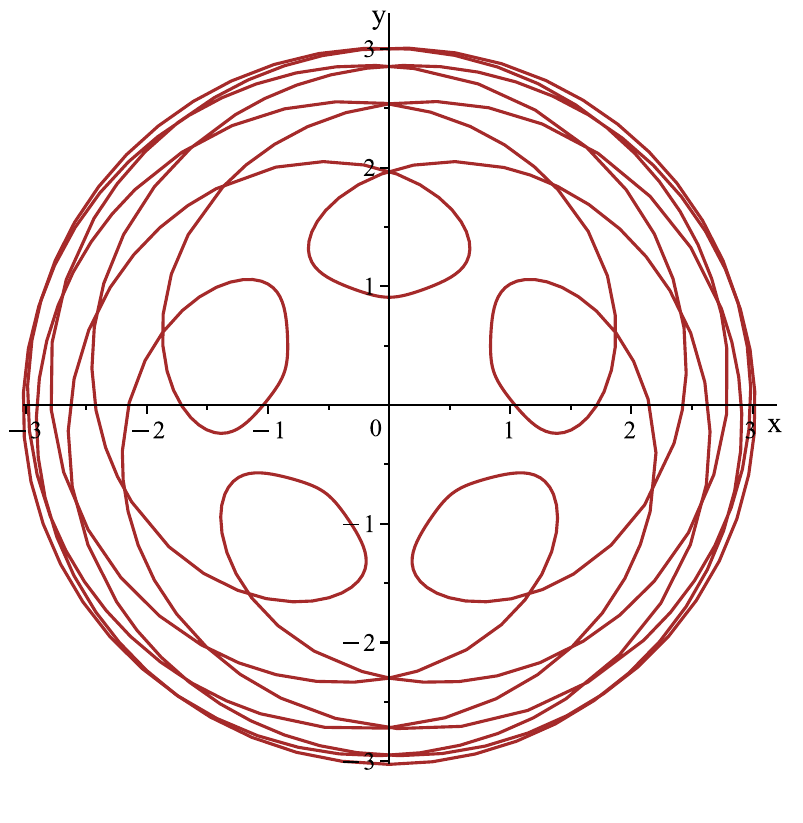}
\hfill
\caption{$m=5,n=2,q=0.2692$. Left: $\sigma = 1,C=0.64$.  Right: $\sigma=-1,C=-0.64$.}
\end{subfigure}

\vspace{0.5em}

\begin{subfigure}[t]{0.9\textwidth}
\centering
\includegraphics[width=0.325\textwidth,trim=2cm 12cm 6cm 2cm,clip]{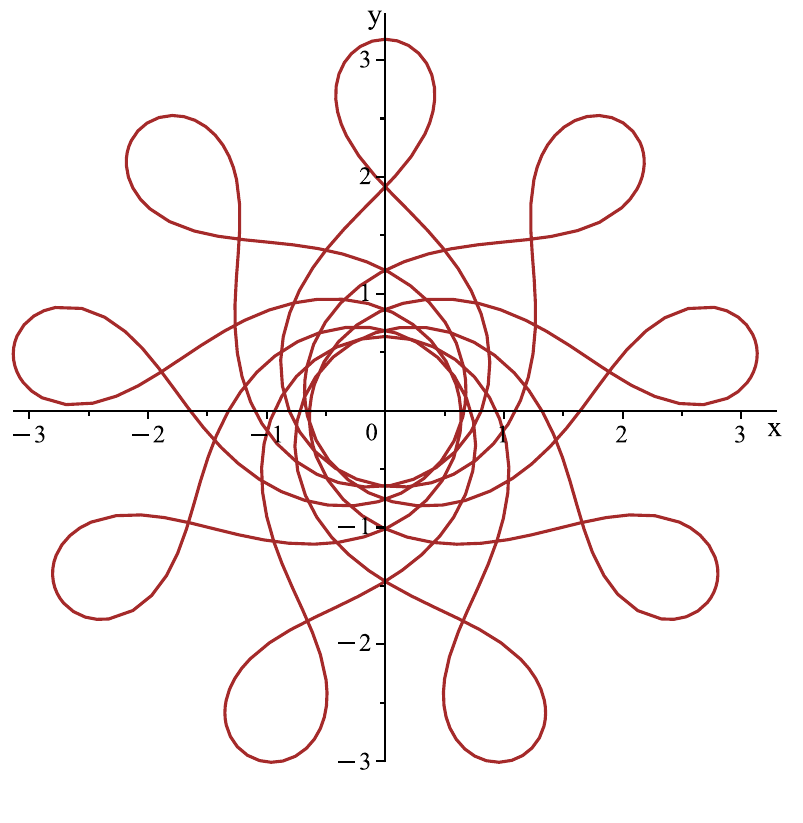}
\hfill
\includegraphics[width=0.325\textwidth,trim=2cm 12cm 6cm 2cm,clip]{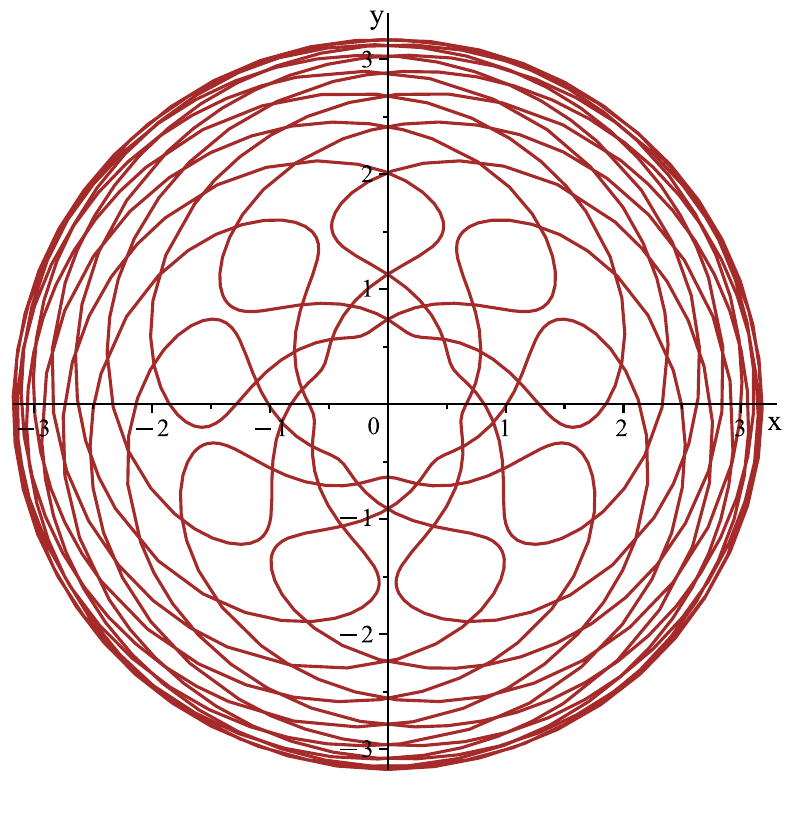}
\hfill
\caption{$m=9,n=1,q=0.5554$. Left: $\sigma = 1,C=0.64$.  Right: $\sigma=-1,C=-0.64$.}
\end{subfigure}
\caption{rational $\mathrm{CN}$ closed loops with $c=1$.}
\label{fig:ratcn_loop_closed_curve4}
\end{figure}

\begin{figure}[H]
\centering
\begin{subfigure}[t]{0.9\textwidth}
\centering
\includegraphics[width=0.325\textwidth,trim=2cm 12cm 6cm 2cm,clip]{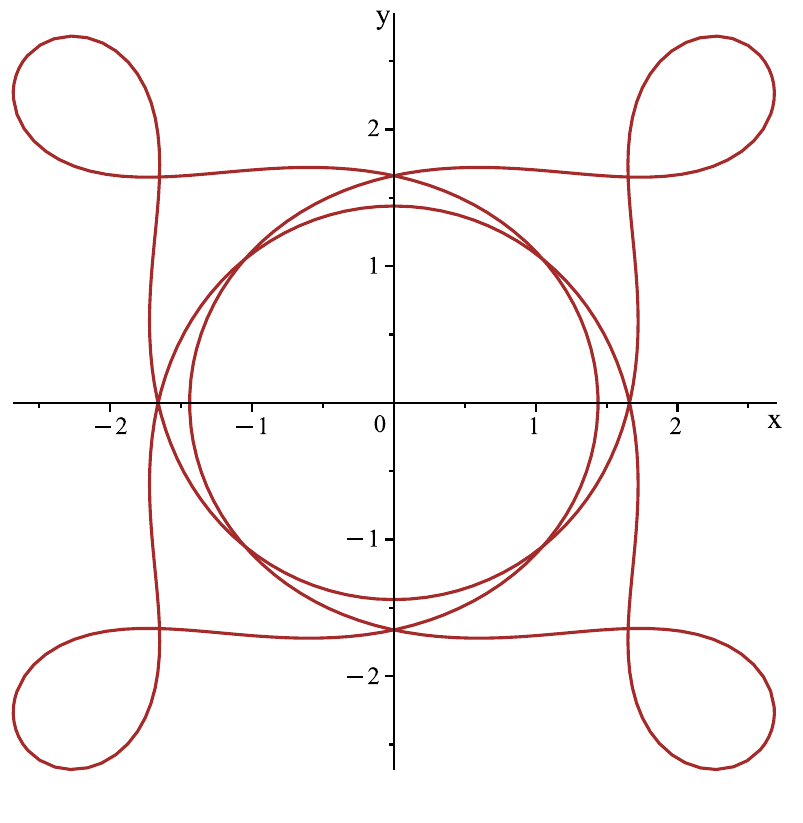}
\hfill
\includegraphics[width=0.325\textwidth,trim=2cm 12cm 6cm 2cm,clip]{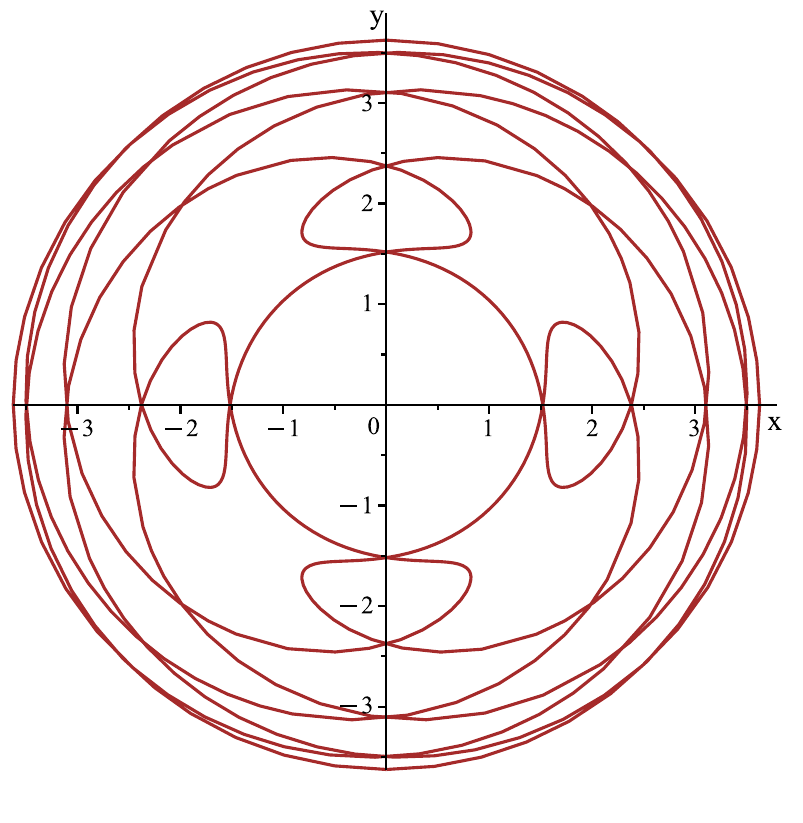}
\hfill
\caption{$m=4,n=1,q=0.0935$. Left: $\sigma = 1,C=0.8$.  Right: $\sigma=-1,C=-0.8$.}
\end{subfigure}

\vspace{0.5em}

\begin{subfigure}[t]{0.9\textwidth}
\centering
\includegraphics[width=0.325\textwidth,trim=2cm 12cm 6cm 2cm,clip]{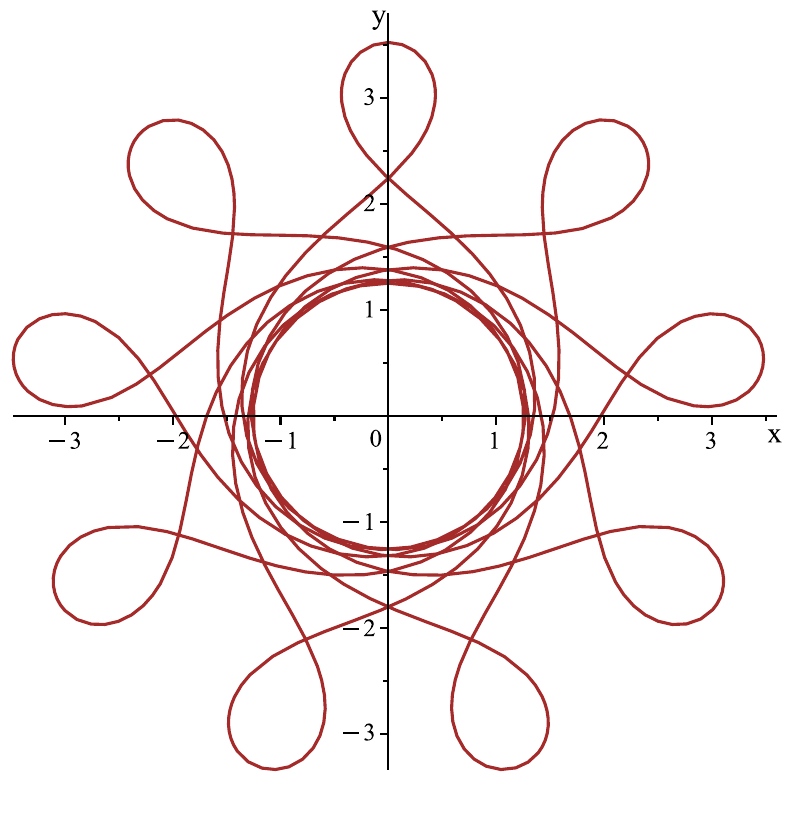}
\hfill
\includegraphics[width=0.325\textwidth,trim=2cm 12cm 6cm 2cm,clip]{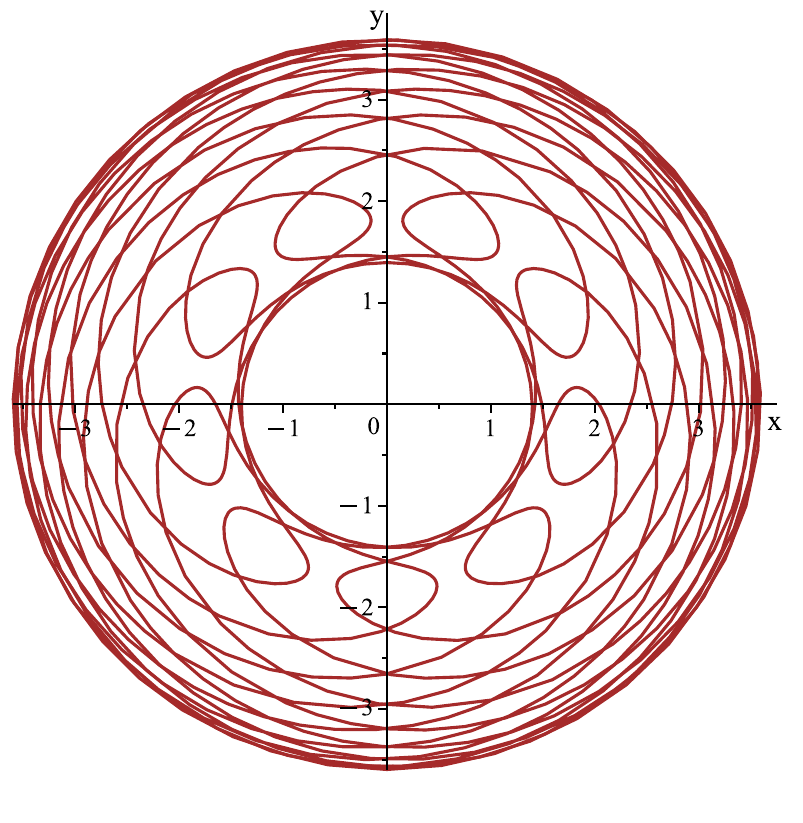}
\hfill
\caption{$m=9,n=1,q=0.2445$. Left: $\sigma = 1,C=0.8$.  Right: $\sigma=-1,C=-0.8$.}
\end{subfigure}

\vspace{0.5em}

\begin{subfigure}[t]{0.9\textwidth}
\centering
\includegraphics[width=0.325\textwidth,trim=2cm 12cm 6cm 2cm,clip]{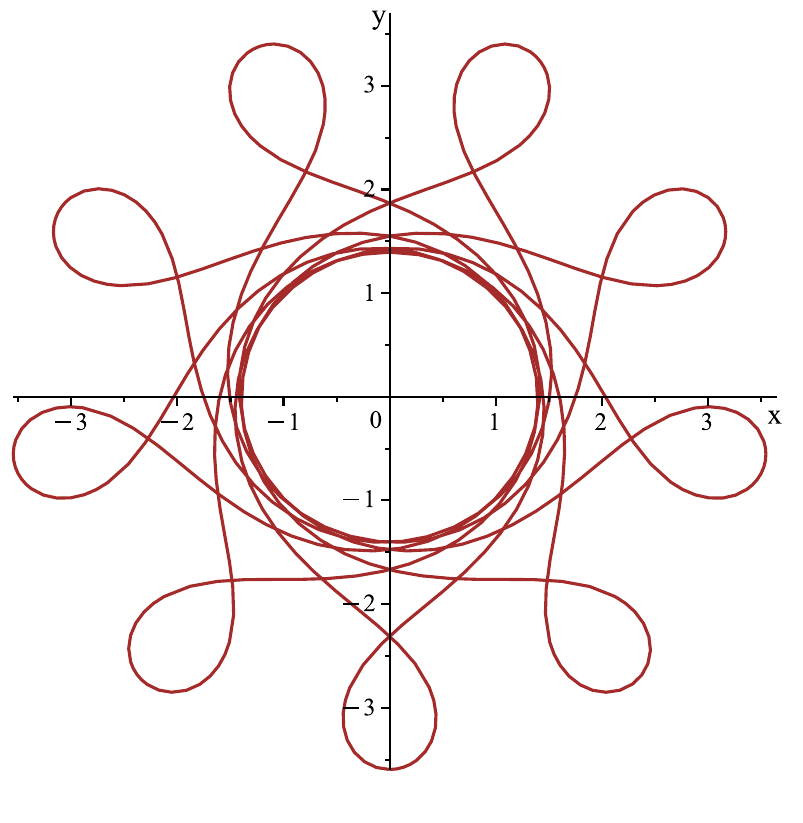}
\hfill
\includegraphics[width=0.325\textwidth,trim=2cm 12cm 6cm 2cm,clip]{fig-rat_cn_C_-0.8_Csq_0.64_c_1_m_9_n_2_q_0.1243572082_sigma_-1.pdf}
\hfill
\caption{$m=9,n=2,q=0.1244$. Left: $\sigma = 1,C=0.8$.  Right: $\sigma=-1,C=-0.8$.}
\end{subfigure}
\caption{rational $\mathrm{CN}$ closed loops with $c=1, m=2, n=-1$.}
\label{fig:ratcn_loop_closed_curve5}
\end{figure}

\begin{figure}[H]
\centering
\includegraphics[width=0.325\textwidth,trim=2cm 12cm 6cm 2cm,clip]{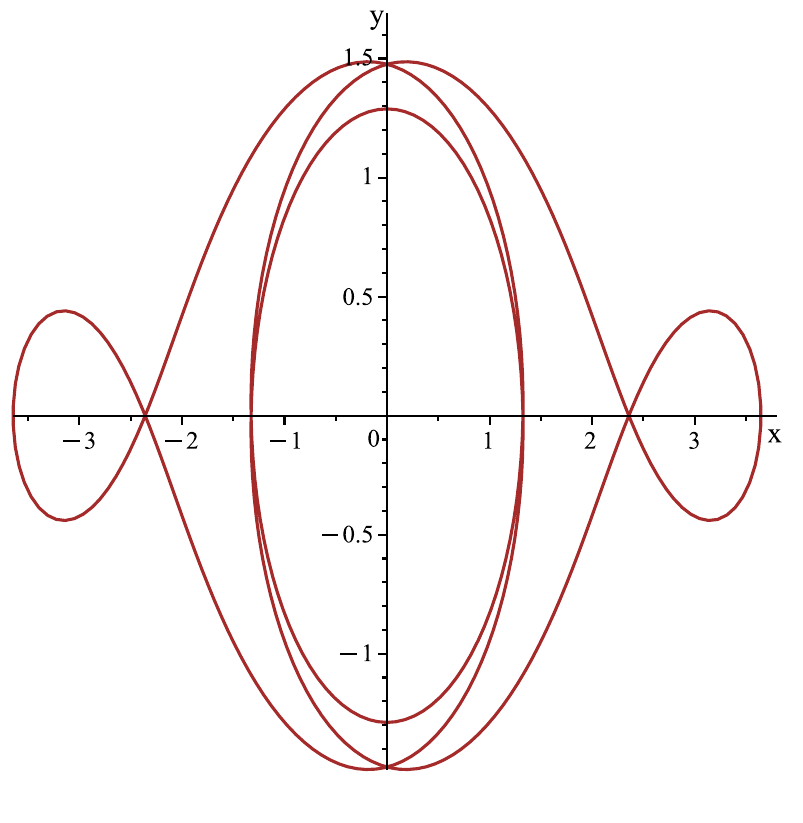}
\hfill
\includegraphics[width=0.325\textwidth,trim=2cm 12cm 6cm 2cm,clip]{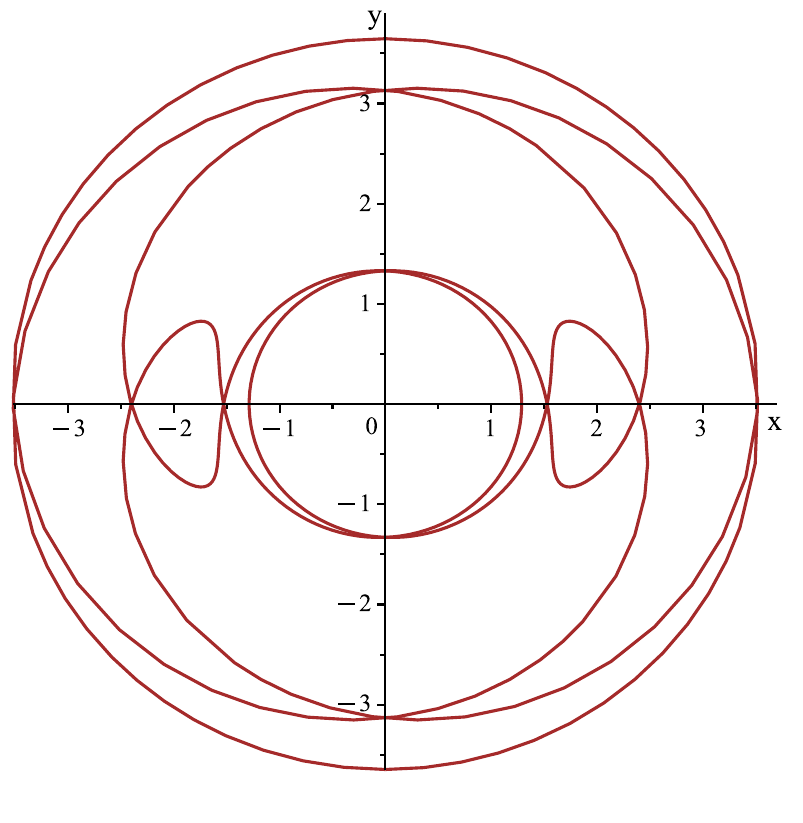}
\caption{rational $\mathrm{CN}$ closed loops with $c=1, m=2, n=-1, q=0.2244$. Left: $\sigma=1,C=0.9$. Right: $\sigma=-1, C=-0.9$.}
\label{fig:ratcn_loop_closed_curve6}
\end{figure}

\subsection{Rational dn loops}
\label{sec:ratdn.loops}

Periodic rational-dn mKdV waves with $q>1$
are bright/dark corresponding to the $\pm$ sign in expression \eqref{ratdn}. 
The resulting loop solutions are obtained similarly to the rational cn loops 
by evaluation of the parametric integral \eqref{vartheta}
and the algebraic formula \eqref{z.loop.eqn.C1not0}, 
in addition to expression \eqref{J}.
Note that $q$, $C$, and $c$ obey the inequalities
\begin{equation}
     0 < C^2 < 1,
     \quad
     q>1, 
     \quad
     c>0 . 
\end{equation}

The parametric integral \eqref{vartheta} is given by 
\begin{equation}\label{vartheta.ratdn}
\begin{aligned}
\vartheta(s) = 
\pm 2\Bigg( &
-\frac{B}{C} \sqrt{c/F}\, s
- \arctan\bigg(\frac{G \SN(\sqrt{cqE/F}\, s, 1/\sqrt{q})}{\sqrt{q}\, \CN(\sqrt{cqE/F}\, s, 1/\sqrt{q})}\bigg)
\\&\qquad
+ \frac{G}{\sqrt{q} C}\, \Pi\big( \SN(\sqrt{cqE/F}\, s, 1/\sqrt{q}), C^2/(q(C^2 - 1)), 1/\sqrt{q} \big)
    \Bigg) 
\end{aligned}
\end{equation}
where $\Pi$ denotes the elliptic integral of the third kind. 
This yields 
\begin{equation}
  \begin{aligned}
    e^{i\vartheta(s)} = & 
    \exp\bigg( \pm i   \frac{2}{\sqrt{q}\,C}\Big(
    -B\sqrt{cq/F}\, s + G \Pi\big( \SN(\sqrt{cqE/F}\, s, 1/\sqrt{q}), C^2/(q(C^2 - 1)), 1/\sqrt{q} \big) \Big) 
\bigg)
    \\&\qquad \times
    \bigg( \frac{\mp G\, \SN(\sqrt{cqE/F}\,s, 1/\sqrt{q}) + i\,\sqrt{q}\, \CN(\sqrt{cqE/F}\, s, 1/\sqrt{q})}{\pm G\, \SN(\sqrt{cqE/F}\,s, 1/\sqrt{q}) + i\,\sqrt{q}\, \CN(\sqrt{cqE/F}\, s, 1/\sqrt{q})} \bigg) . 
  \end{aligned}
\end{equation}
Expression \eqref{J} is given by 
\begin{equation}\label{ratdn.J}
  \begin{aligned}
J(s) =
\frac{2\sqrt{F/c}}{(q-(q - 1)C^4)C} \bigg( &
\sqrt{E}^3\, \frac{\CN(\sqrt{cqE/F}\, s, 1/\sqrt{q})\, \SN(\sqrt{cqE/F}\, s, 1/\sqrt{q})}{\big( C\, \DN(\sqrt{cqE/F}\, s, 1/\sqrt{q}) + 1\big)^2}
\\&\quad
\pm i \bigg(
\tfrac{1}{2} F -\sqrt{q}\bigg( \frac{B\, \DN(\sqrt{cqE/F}\, s, 1/\sqrt{q}) +LC}{C\, \DN(\sqrt{cqE/F}\, s, 1/\sqrt{q}) + 1}\bigg)^2 \bigg) 
\bigg)
  \end{aligned}
\end{equation}
with $L$ given by expression \eqref{ratcn.L}, 
using $C_1 = \pm (q + (1-q)C^4)C \sqrt{c/F}^3$
from expression \eqref{C1.ratcn}.

The loop equation \eqref{z.loop.eqn.C1not0} is thus 
\begin{equation}\label{ratcn.loop}
z(s,t) = J(s)  e^{i\vartheta(s)} e^{\pm i (q+ (1-q)C^4)C \sqrt{c/F}^3 t}  + z_0
\end{equation}
which describes a rotating loop whose angular speed is 
$\pm (q + (1-q)C^4)C \sqrt{c/F}^3$.
This gives a geometric curve flow \eqref{r.z} from 
the real and imaginary parts of the loop solution \eqref{ratcn.loop},
$\vec{r}_\pm(s,t)  =\big(\mathrm{Re}\,z(s,t),\mathrm{Im}\,z(s,t)\big)$, 
which is the composition of a static loop $\vec{r}_\pm(s,0)$ plus a rotational motion. 

\begin{figure}[H]
\centering
 \begin{subfigure}[t]{1\textwidth}
\centering
\includegraphics[width=0.4\textwidth,trim=2cm 12cm 6cm 2cm,clip]{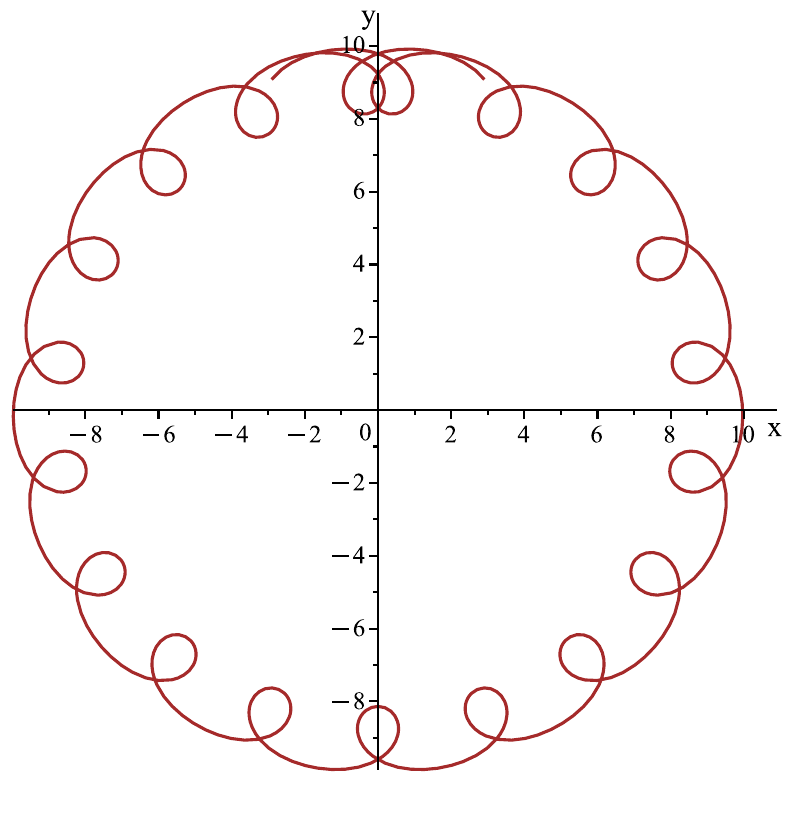}
\hfill
\includegraphics[width=0.4\textwidth,trim=2cm 12cm 6cm 2cm,clip]{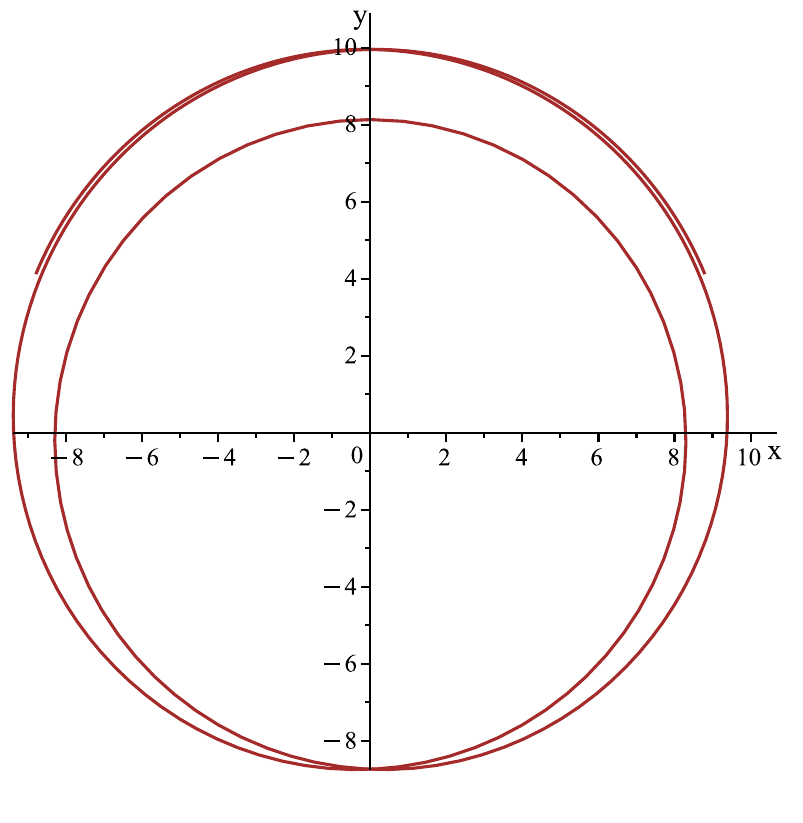}
\hfill
\includegraphics[width=0.4\textwidth,trim=2cm 12cm 6cm 2cm,clip]{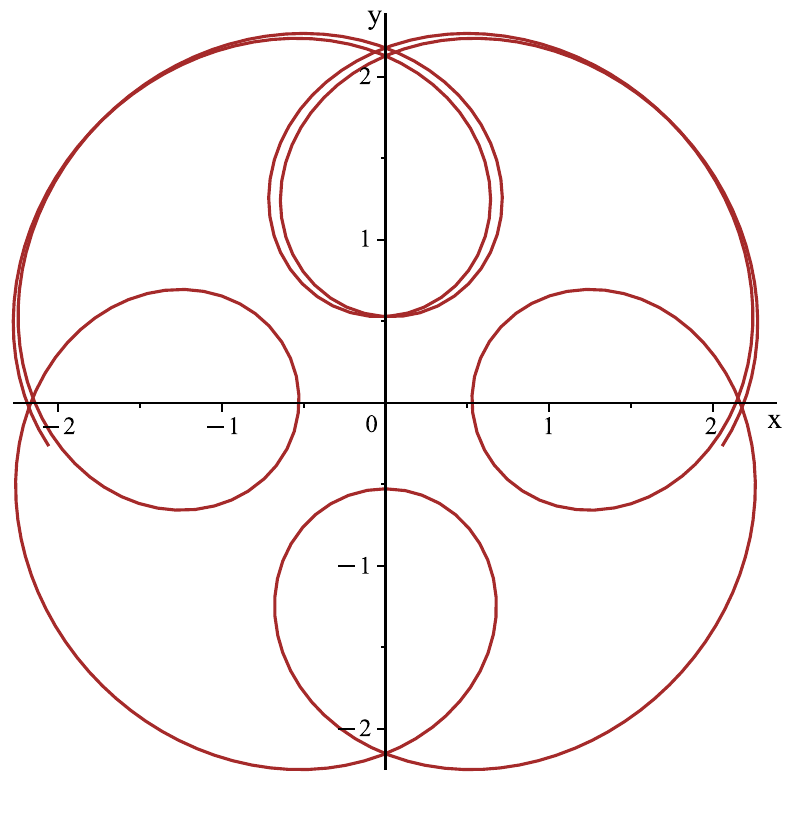}
\hfill
\includegraphics[width=0.4\textwidth,trim=2cm 12cm 6cm 2cm,clip]{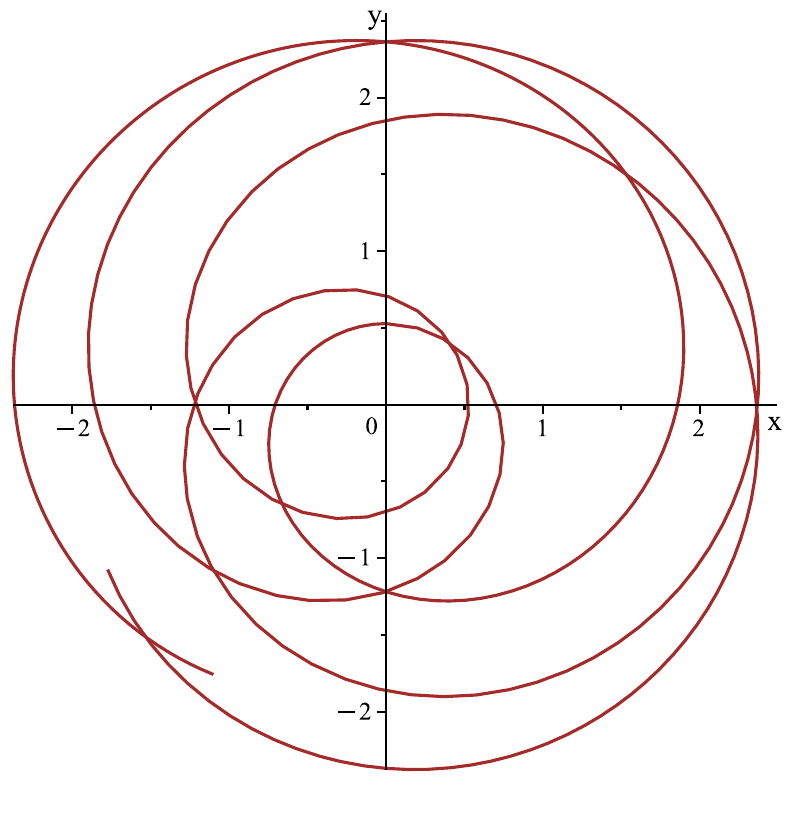}
\hfill
\caption{Top Left: $C=0.1,q=1.01,\sigma=1$. Top Right: $C=0.1,q=1.01,\sigma=-1$. Bottom Left: $C=0.4,q=1.01,\sigma=1$. Bottom Right: $C=0.4,q=1.01,\sigma=-1$.}
\end{subfigure}
\caption{rational DN open loops $c=1$}
\label{fig:ratdn_loop_open_curve_row1}
\end{figure}

\begin{figure}[H]
\begin{subfigure}[t]{1\textwidth}
\centering
\includegraphics[width=0.4\textwidth,trim=2cm 12cm 6cm 2cm,clip]{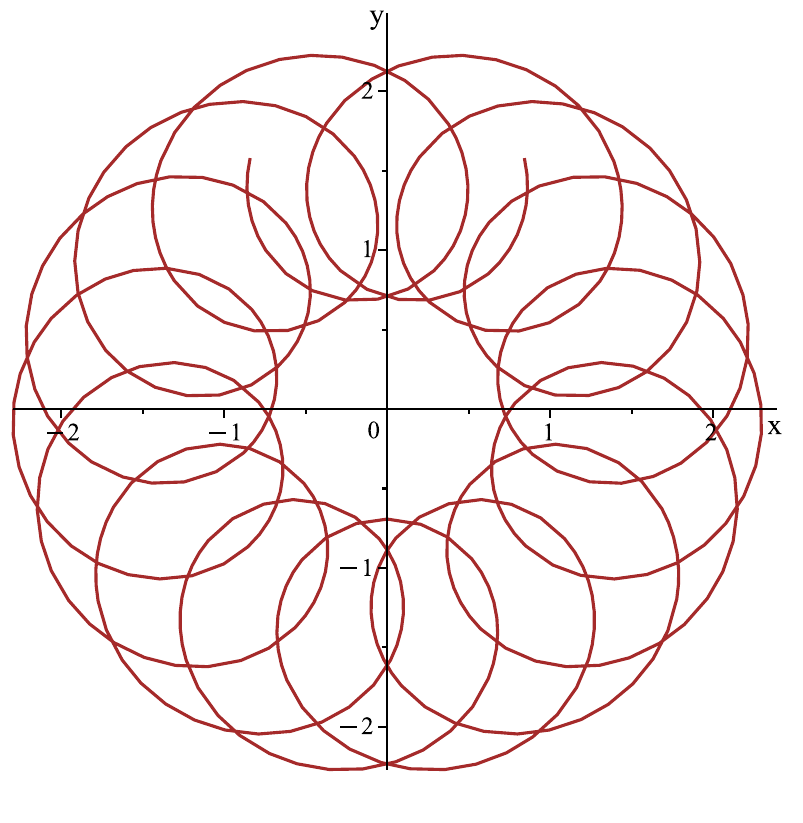}
\hfill
\includegraphics[width=0.4\textwidth,trim=2cm 12cm 6cm 2cm,clip]{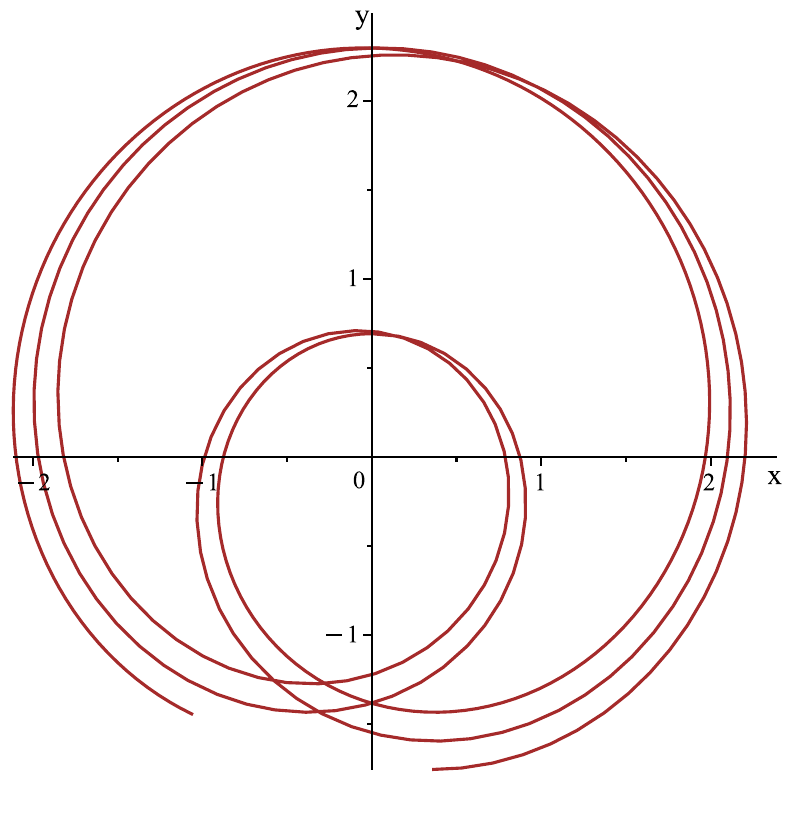}
\hfill
\includegraphics[width=0.4\textwidth,trim=2cm 12cm 6cm 2cm,clip]{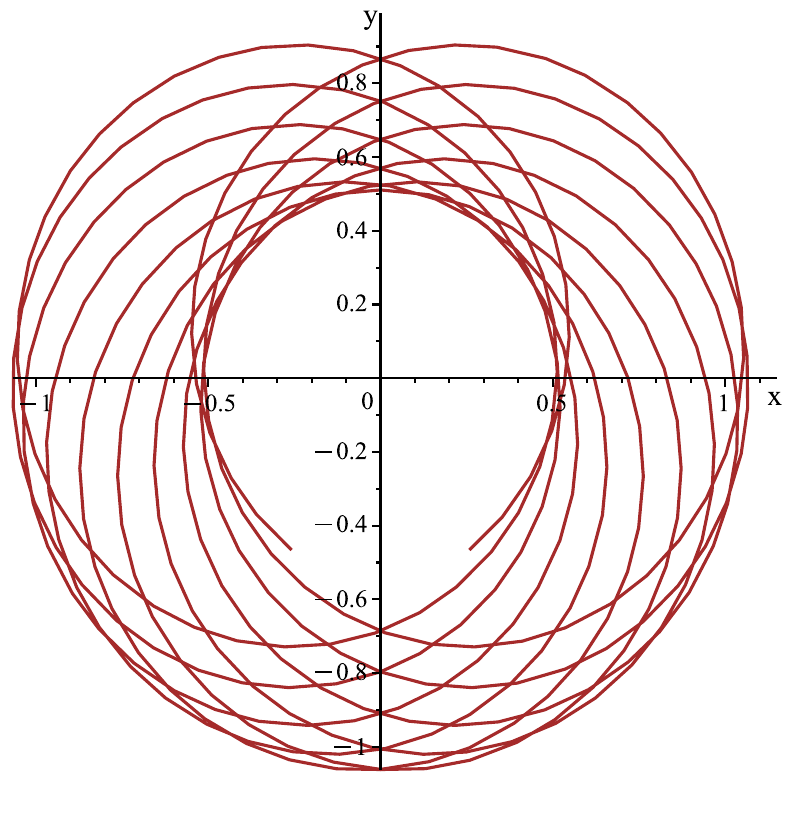}
\hfill
\includegraphics[width=0.4\textwidth,trim=2cm 12cm 6cm 2cm,clip]{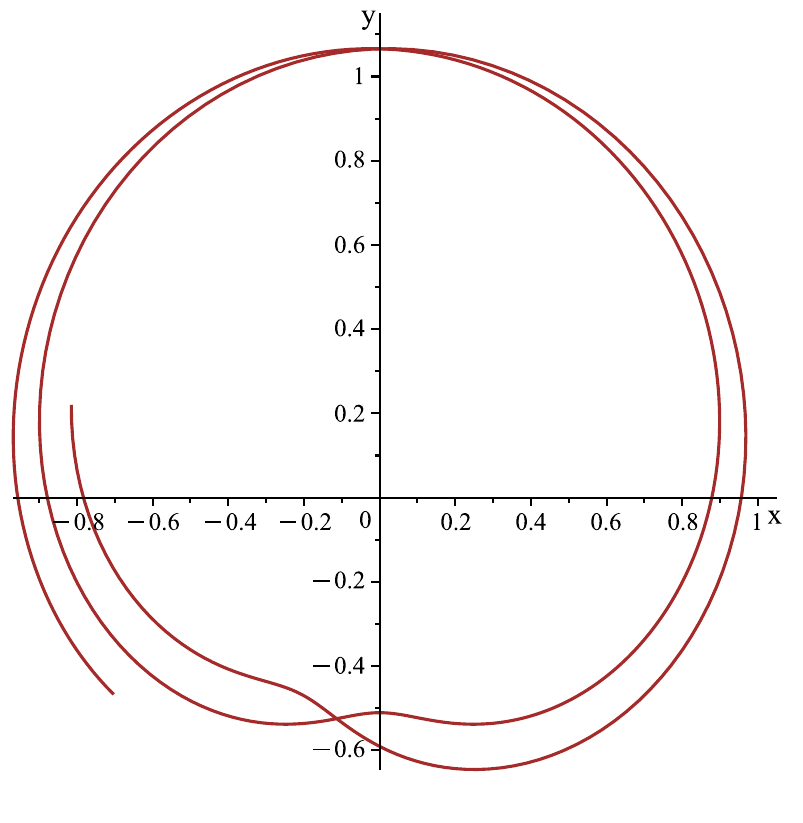}
\caption{Top Left: $C=0.4,q=1.2,\sigma=1$. Top Right: $C=0.4,q=1.2,\sigma=-1$. Bottom Left: $C=0.7,q=1.9,\sigma=1$. Bottom Right: $C=0.7,q=1.9,\sigma=-1$.}
\end{subfigure}
\caption{rational DN open loops $c=1$.}
\label{fig:ratdn_loop_open_curve_row2}
\end{figure}

\begin{figure}[H]
\centering
\includegraphics[width=0.4\textwidth,trim=2cm 12cm 6cm 2cm,clip]{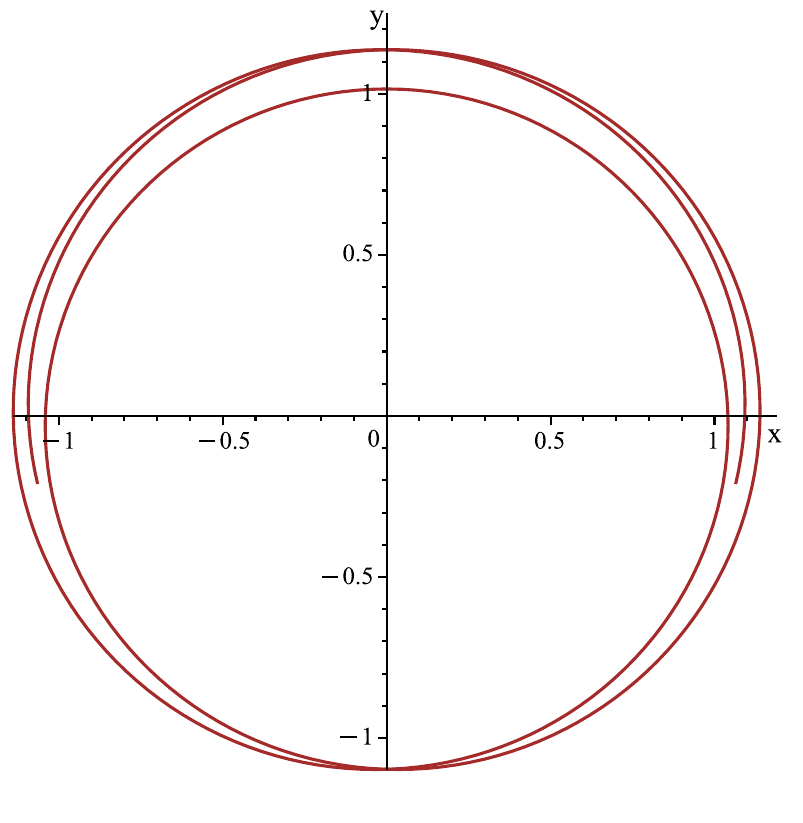}
\hfill
\includegraphics[width=0.4\textwidth,trim=2cm 12cm 6cm 2cm,clip]{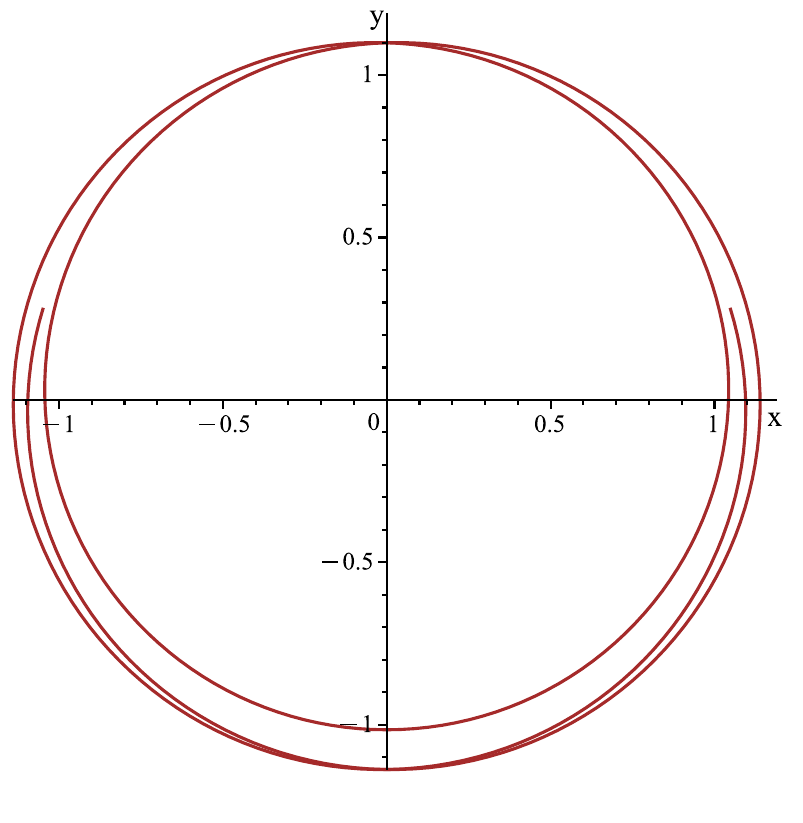}

\caption{rational DN open loops $C=0.9,q=1.2,c=1$. Left: $\sigma=1$. Right:$\sigma=-1$.}
\label{fig:ratdn_loop_open_curve_row3}
\end{figure}

The static loop is, firstly, a bounded curve,
since 
\begin{equation}
  \begin{aligned}
  |\vec{r}_\pm(s,0)|^2 = |J(s)|^2 = & 
  \big(\tfrac{2c}{F C_1}\big)^2
  \Big(  (q +(1-q) C^4)^2 - 4q E
  \\&\qquad
  + 4(q + (1-q)C^4) E/(C\DN(\sqrt{cqE/F},s,1/\sqrt{q}) + 1) \Big)
  \end{aligned}
\end{equation}
is a bounded function of $s$.
Secondly, this curve lies in an annulus whose maximum and minimum radius
\eqref{ratcn.annulus} 
is the same as for rational cn loops. 
Thirdly, the shape of the curve is comprised of inward-facing petals. 
See the plots in Fig.~\ref{fig:ratdn_loop_open_curve_row1}, ~\ref{fig:ratdn_loop_open_curve_row2} and ~\ref{fig:ratdn_loop_open_curve_row3}

These curves have complicated expressions for
the winding number \eqref{winding.number}
and the mean-winding rate \eqref{winding.rate},
which arise from the results in Table~\ref{table:conserved}. 

Similarly to the rational cn loops,
the curve will be closed and periodic
when its turning angle satisfies the following condition.

\begin{prop}
(i) 
A closed curve is periodic iff it satisfies condition \eqref{ratcn.closed.cond}
where the secular change in the turning angle is given by 
\begin{equation}
  \Delta\vartheta = \frac{4}{C\sqrt{q}}\bigg( \frac{-B}{\sqrt{E}} \K(1/\sqrt{q})
  + G\, \Pi(C^2/(q(C^2 - 1)), 1/\sqrt{q})
  \bigg) . 
\end{equation}
(ii) Its period is
\begin{equation}
  \Delta s =
  2 \sqrt{F/(cqE)}\, \K(1/\sqrt{q}) . 
\end{equation}  
\end{prop}

For a fixed value of $C$,
the periodicity condition is graphically equivalent to
finding the intersection points of the curve $\Delta\vartheta/(2\pi)$
\and the horizontal line $n/m$.
Modulo the overall sign $\sigma\,\sgn(C)$,
$\Delta\vartheta/(2\pi)$ goes to $0$ as $q\to\infty$ 
and approaches $-\infty$ as $q\to 1$.
Thus, in the case $\sigma\,\sgn(C)=1$,
any negative ratio is allowed for $n/m$, 
and in the case $\sigma\,\sgn(C)=-1$,
any positive ratio is allowed. 

\begin{figure}[H]
\begin{subfigure}[t]{1\textwidth}
\centering
\includegraphics[width=0.4\textwidth,trim=2cm 12cm 6cm 2cm,clip]{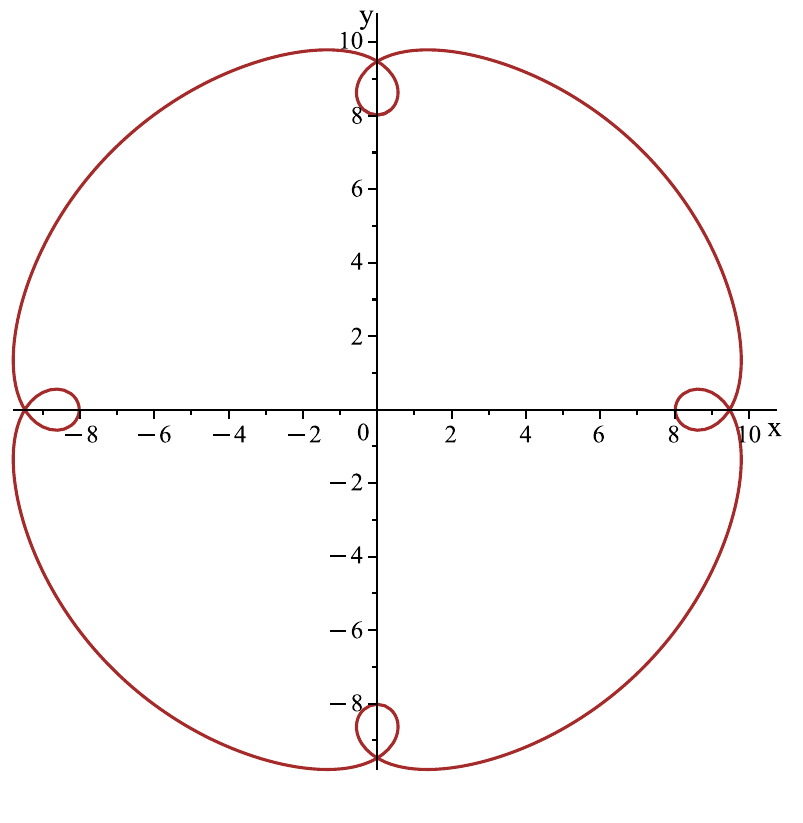}
\hfill
\includegraphics[width=0.4\textwidth,trim=2cm 12cm 6cm 2cm,clip]{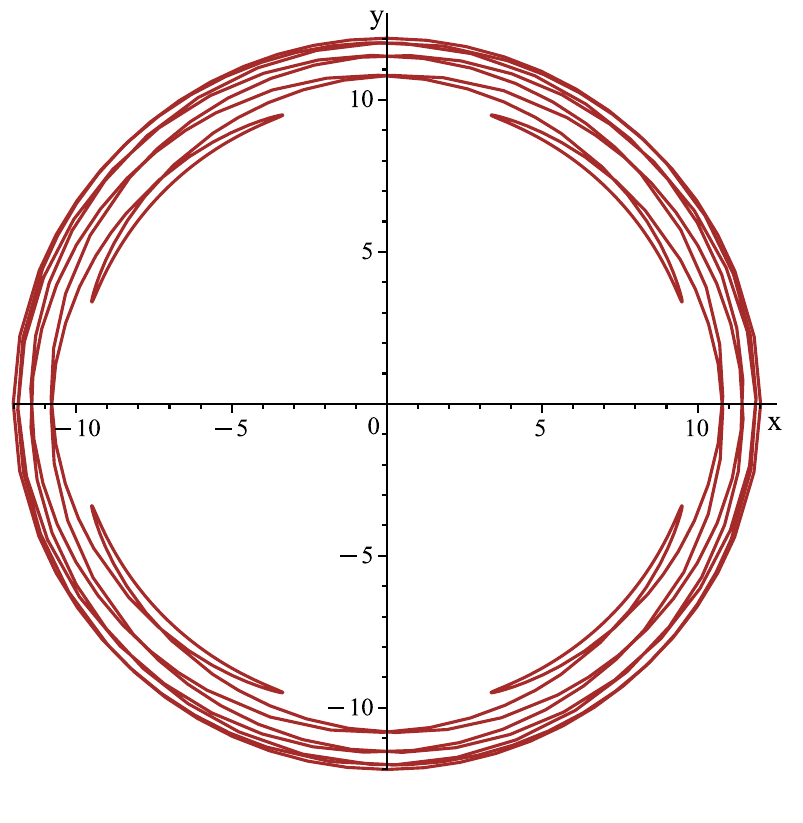}
\caption{$m=4,n=-1, q=1.0000000484$. Left: $C= 0.1,\sigma=1$. Right: $C= -0.1,\sigma=-1$.}
\end{subfigure}

\vspace{0.5em}

\begin{subfigure}[t]{0.9\textwidth}
\centering
\includegraphics[width=0.4\textwidth,trim=2cm 12cm 6cm 2cm,clip]{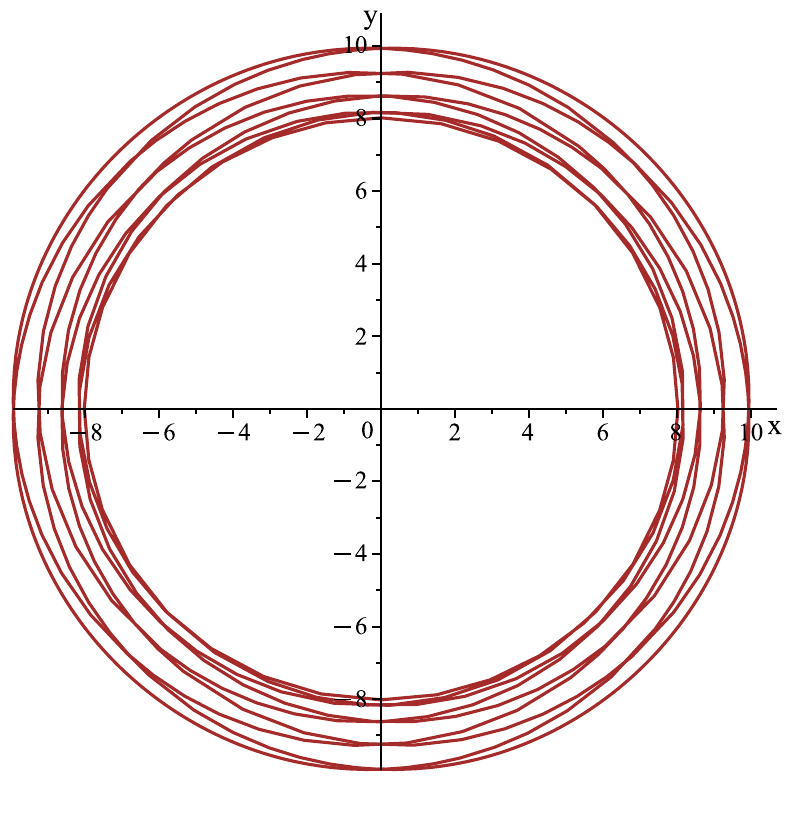}
\hfill
\includegraphics[width=0.4\textwidth,trim=2cm 12cm 6cm 2cm,clip]{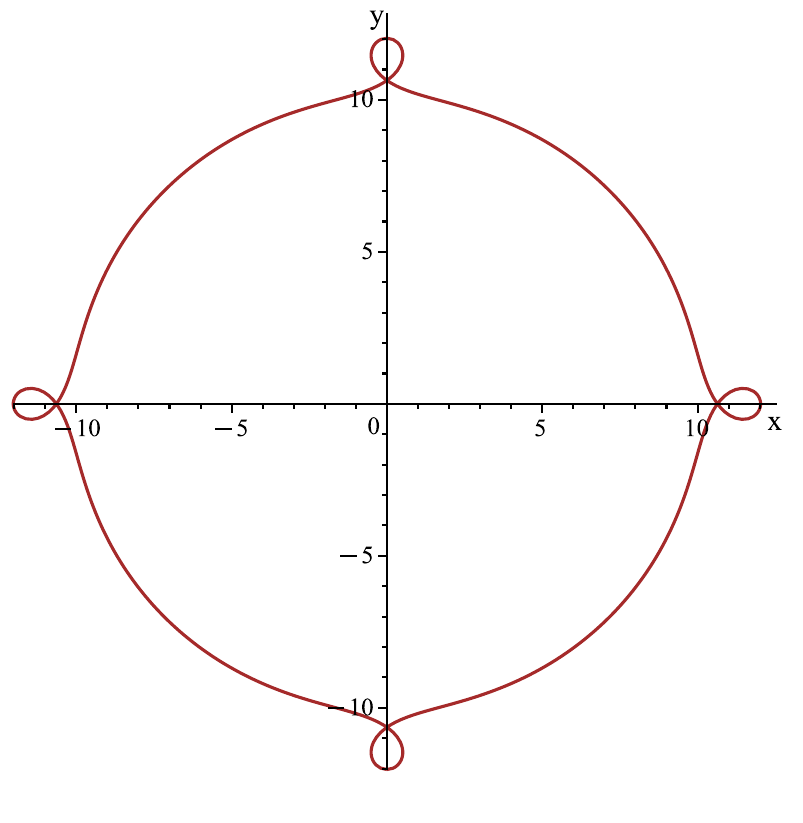}
\caption{$m=4,n=1, q=1.0000000484$. Left: $C= 0.1,\sigma=-1$. Right: $C= -0.1,\sigma=1$.}
\end{subfigure}
\caption{rational DN closed loops $c=1$.}
\label{fig:ratdn_loop_closed_curve1}
\end{figure}

A closed curve will have $m$ maxima points,
since this is the number of periods of the $\DN$ function that will occur
as $\vec{r}(s,0)$ traces out the curve once.
Plots of closed loops  are shown in Figs.~\ref{fig:ratdn_loop_closed_curve1}, ~\ref{fig:ratdn_loop_closed_curve3} and ~\ref{fig:ratdn_loop_closed_curve3}.

\begin{figure}[H]
\begin{subfigure}[t]{1\textwidth}
\centering
\includegraphics[width=0.4\textwidth,trim=2cm 12cm 6cm 2cm,clip]{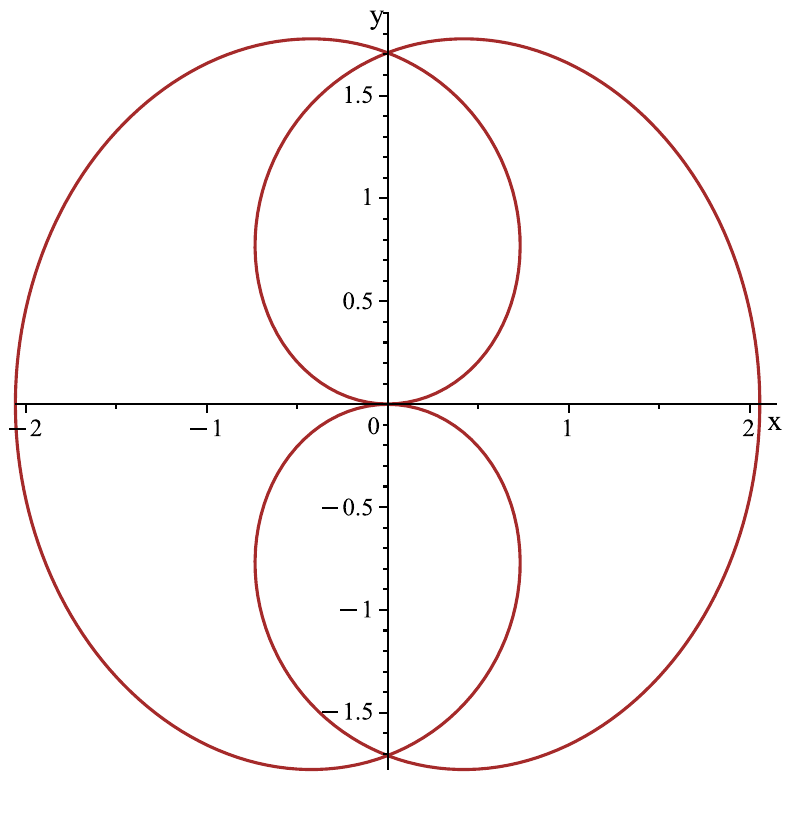}
\hfill
\includegraphics[width=0.4\textwidth,trim=2cm 12cm 6cm 2cm,clip]{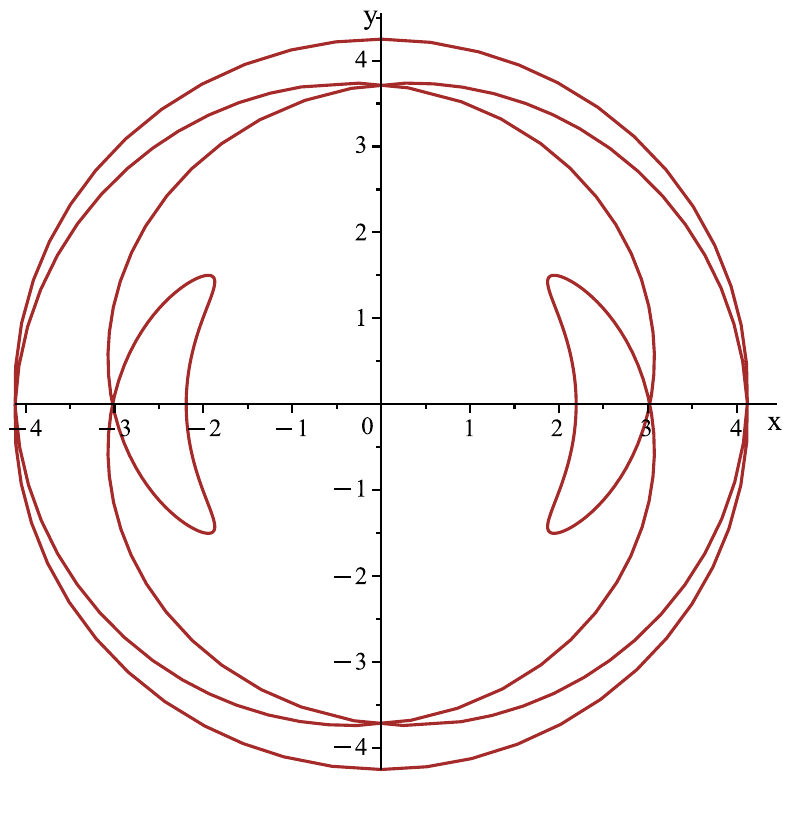}
\caption{$m=2,n=-1, q=1.00185$. Left: $C= 0.5,\sigma=1$. Right: $C= -0.5,\sigma=-1$.}
\end{subfigure}

\vspace{0.5em}

\begin{subfigure}[t]{0.9\textwidth}
\centering
\includegraphics[width=0.4\textwidth,trim=2cm 12cm 6cm 2cm,clip]{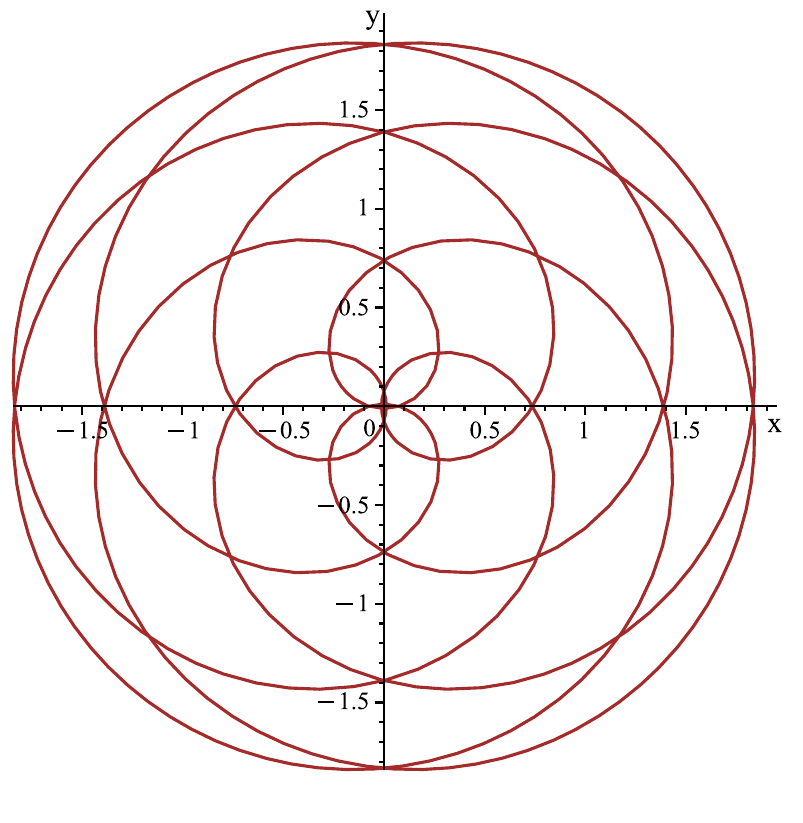}
\hfill
\includegraphics[width=0.4\textwidth,trim=2cm 12cm 6cm 2cm,clip]{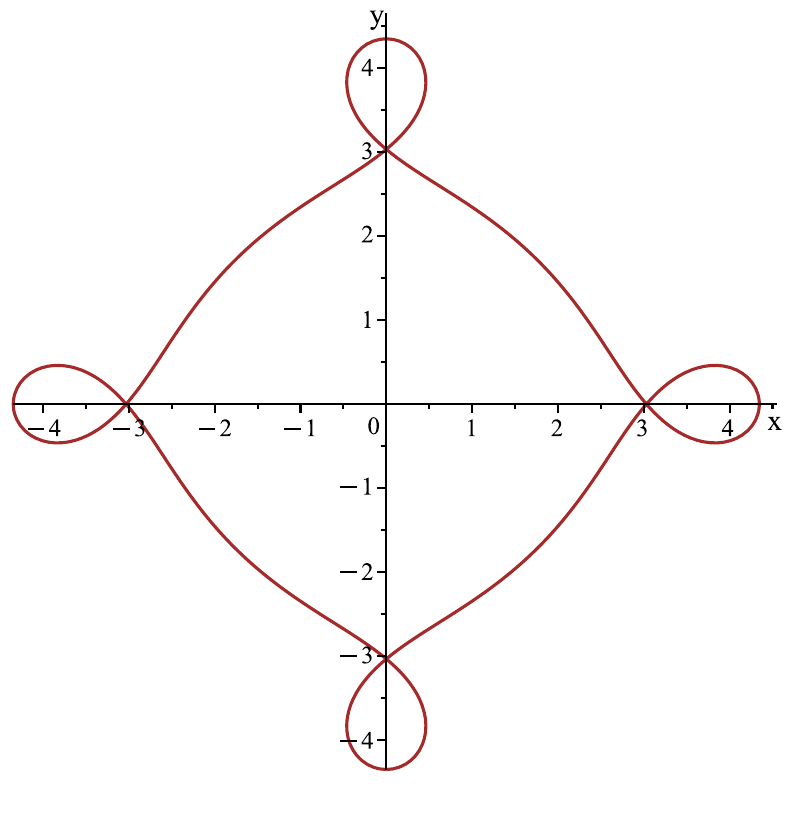}
\caption{$m=4,n=1, q=1.0297151$. Left: $C=0.5,\sigma=-1$. Right: $C=-0.5,\sigma=1$.}
\end{subfigure}
\caption{rational DN closed loops $c=1$}
\label{fig:ratdn_loop_closed_curve2}
\end{figure}

\begin{figure}[H]
\begin{subfigure}[t]{1\textwidth}
\centering
\includegraphics[width=0.4\textwidth,trim=2cm 12cm 6cm 2cm,clip]{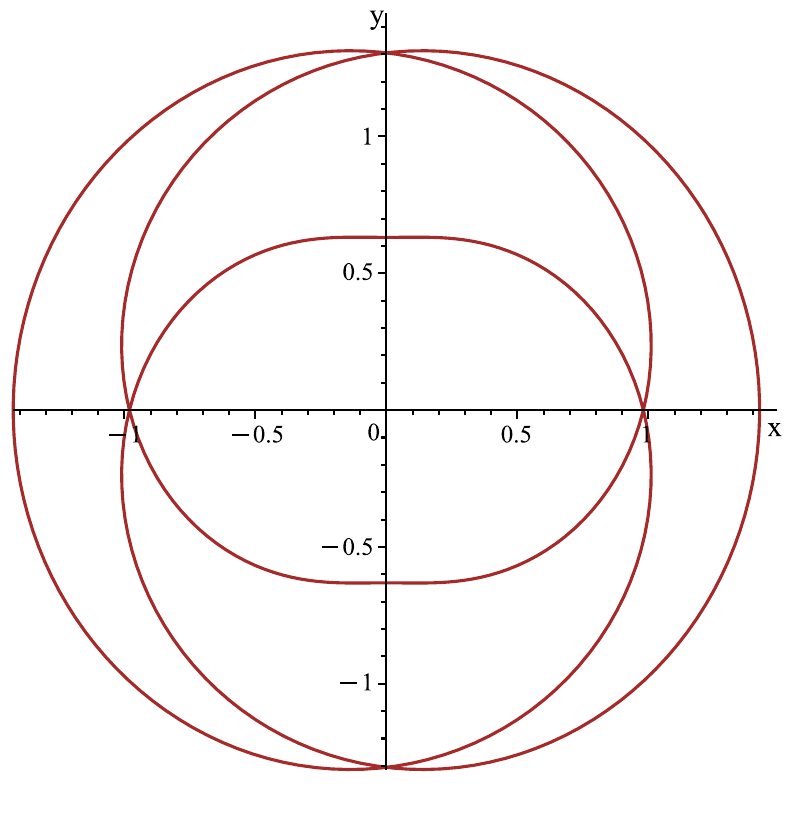}
\hfill
\includegraphics[width=0.4\textwidth,trim=2cm 12cm 6cm 2cm,clip]{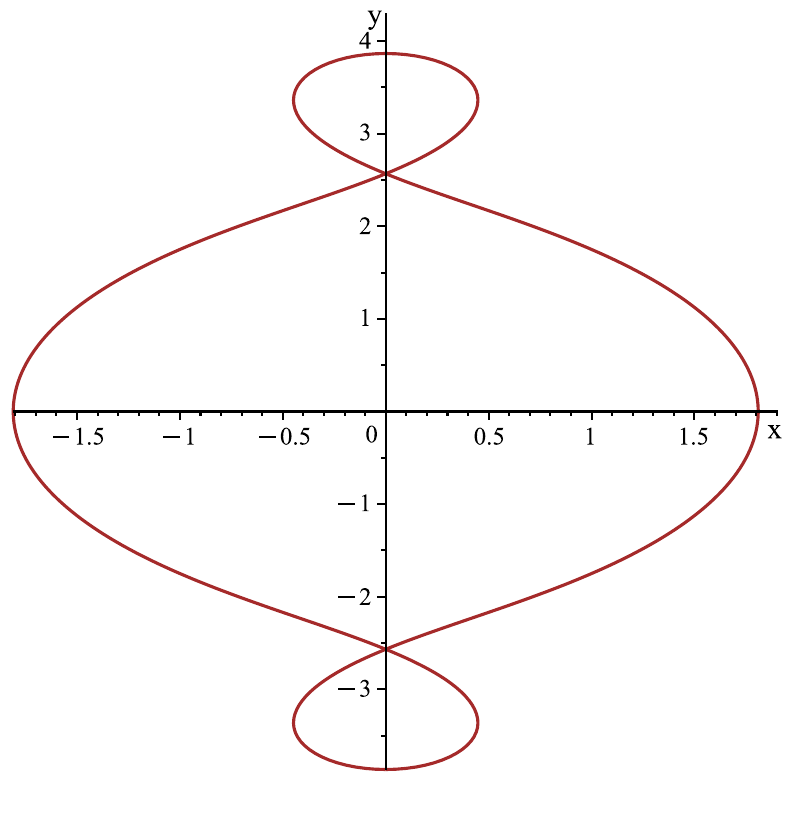}
\caption{$m=2,n=1,q=1.0282$. Left: $C=0.7.\sigma=-1$. Right: $C=-0.7.\sigma=1$.}
\end{subfigure}
\caption{rational DN closed loops $c=1$}
\label{fig:ratdn_loop_closed_curve3}
\end{figure}

\subsection{Rational cosine loops}\label{sec:ratcos.loops}

For the rational-cosine mKdV waves \eqref{ratcos}, 
the corresponding loop solutions are derived by
specializing the rational cn loops to the case $q=0$.
Note that $c$ and $C$ obey the inequalities 
$c>0$ and $0<C^2<1$.
It will be convenient now to let
\begin{equation} 
  F = (1-C)/(1+C),
  \quad
  G = (1-C^2)/(1 + 2C^2) . 
\end{equation}

The parametric integral \eqref{vartheta} is given by 
\begin{equation}\label{vartheta.ratcos}
  \vartheta(s) =
\pm\Big(  \sqrt{2c/(1+2C^2)}\,s
  - 4\arctan\Big( \sqrt{F} \tan\big(\tfrac{1}{2}\sqrt{c G}\, s\big) \Big) , 
\end{equation}
which yields 
\begin{equation}
  e^{i\vartheta(s)} =
  \bigg( \frac{\mp\sqrt{F}\, \tan\big(\tfrac{1}{2}\sqrt{c G}\, s\big) + i}{\pm \sqrt{F}\, \tan\big(\tfrac{1}{2}\sqrt{c G}\, s\big) + i}\bigg)^2 e^{\pm i \sqrt{2c/(1+2C^2)}\,s} . 
\end{equation}
Expression \eqref{J} is given by 
\begin{equation}
  \begin{aligned}
    J(s) = 
    \frac{\sqrt{F/c}}{C} \Bigg( &
    \frac{2 \sqrt{1-C^2}^3 \sin\big(\tfrac{1}{2}\sqrt{c G}\, s\big)}{\big( C \cos\big(\tfrac{1}{2}\sqrt{c G}\, s\big) + 1 \big)^2}
\\&\quad
\mp i \frac{1}{C}
\bigg( F - \tfrac{1}{2}\Big( \frac{C \cos\big(\tfrac{1}{2}\sqrt{c G}\, s\big) + 2C^2 - 1}{C \cos\big(\tfrac{1}{2}\sqrt{c G}\, s\big) + 1} \Big)^2 \bigg)
\Bigg) 
  \end{aligned}
\end{equation}
using $C_1 = \mp \sqrt{2c/(1+2C^2)}^3 C^2$
from expression \eqref{C1.ratcn}.

The loop equation \eqref{z.loop.eqn.C1not0} is 
\begin{equation}\label{ratcos.loop}
  z(s,t) = J(s)  e^{i\vartheta(s)} e^{\mp \sqrt{2c/(1+2C^2)}^3 C^2  t}  + z_0 , 
\end{equation}
which describes a rotating loop with angular speed 
$\mp \sqrt{2c/(1+2C^2)}^3 C^2$. 
The resulting geometric curve flow \eqref{r.z} consists of 
the real and imaginary parts of the loop solution \eqref{ratcn.loop},
$\vec{r}_\pm(s,t)  =\big(\mathrm{Re}\,z(s,t),\mathrm{Im}\,z(s,t)\big)$.
This is the composition of a static loop $\vec{r}_\pm(s,0)$ plus a rotational motion.

\begin{figure}[h]
\centering
 \begin{subfigure}[t]{0.9\textwidth}
\centering
\includegraphics[width=0.3\textwidth,trim=2cm 12cm 6cm 2cm,clip]{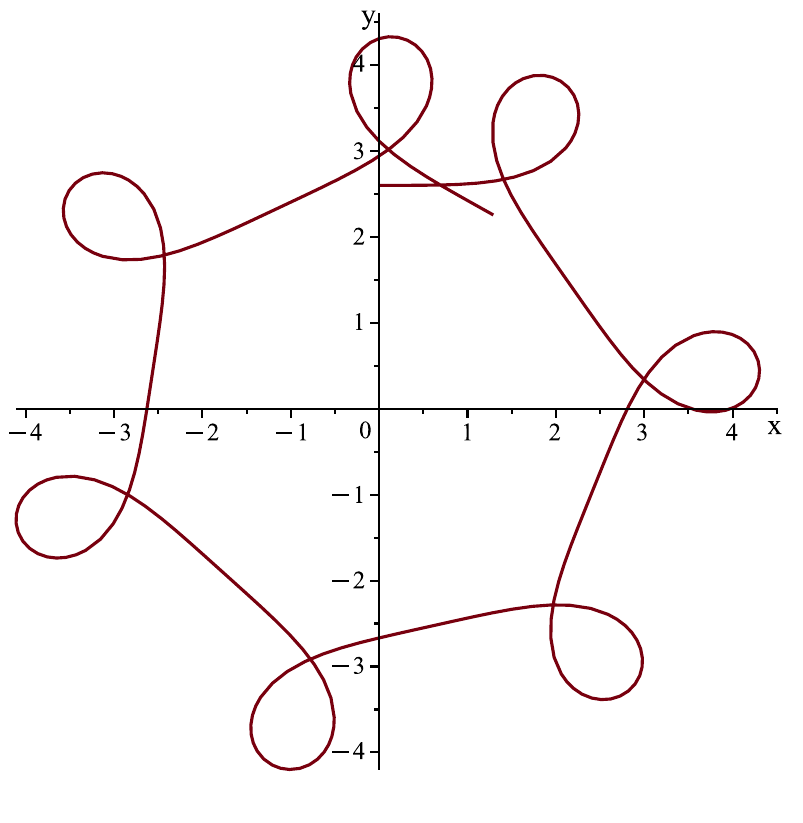}
\hfill
\includegraphics[width=0.3\textwidth,trim=2cm 12cm 6cm 2cm,clip]{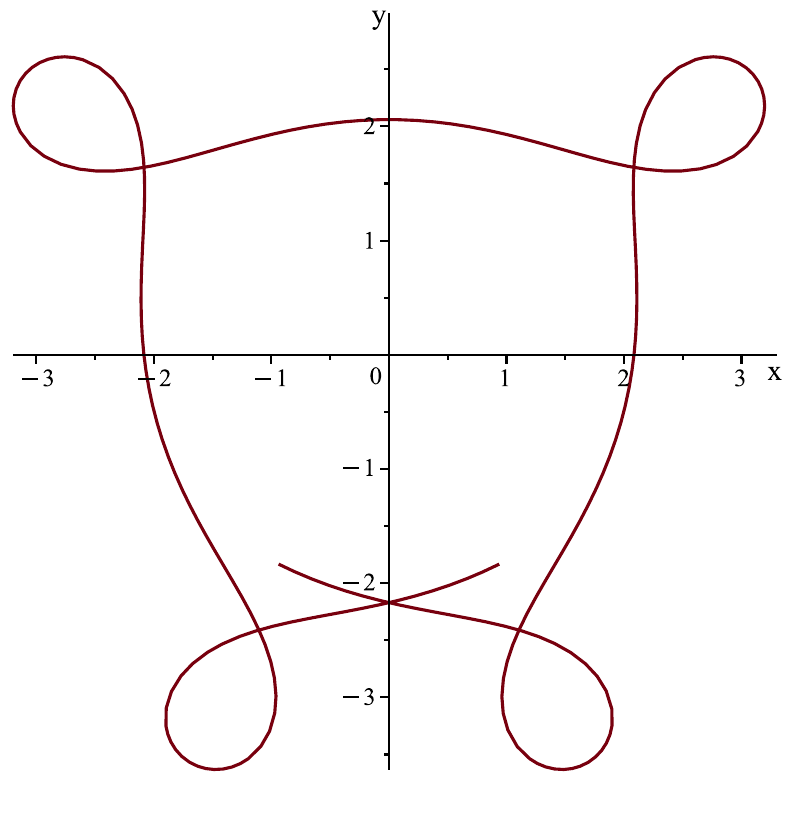}
\caption{Left: $C= 0.5$. Right: $C= 0.63$.}
\end{subfigure}

\begin{subfigure}[t]{0.9\textwidth}
\centering
\includegraphics[width=0.3\textwidth,trim=2cm 12cm 6cm 2cm,clip]{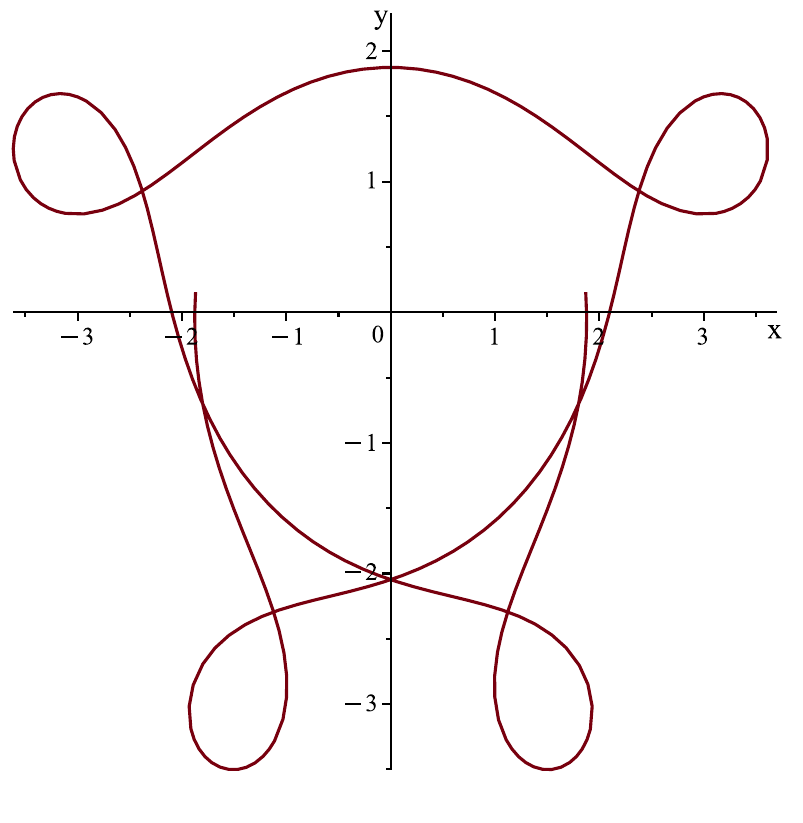}
\hfill
\includegraphics[width=0.3\textwidth,trim=2cm 12cm 6cm 2cm,clip]{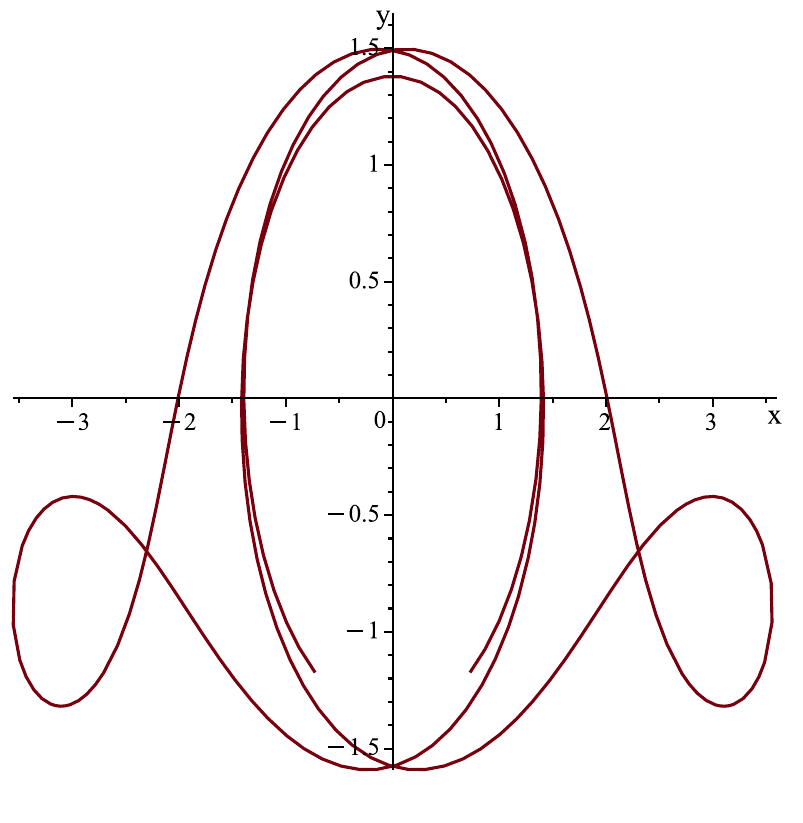}
\caption{Left: $C= 0.69$. Right: $C= 0.91$.}
\end{subfigure}

\begin{subfigure}[t]{0.9\textwidth}
\centering
\includegraphics[width=0.3\textwidth,trim=2cm 12cm 6cm 2cm,clip]{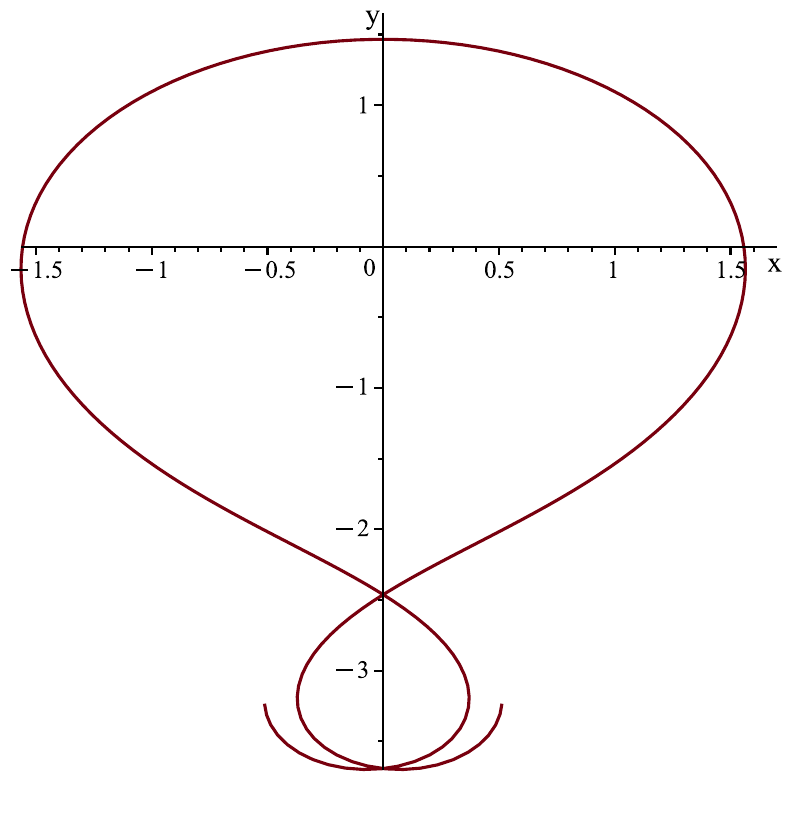}
\hfill
\includegraphics[width=0.3\textwidth,trim=2cm 12cm 6cm 2cm,clip]{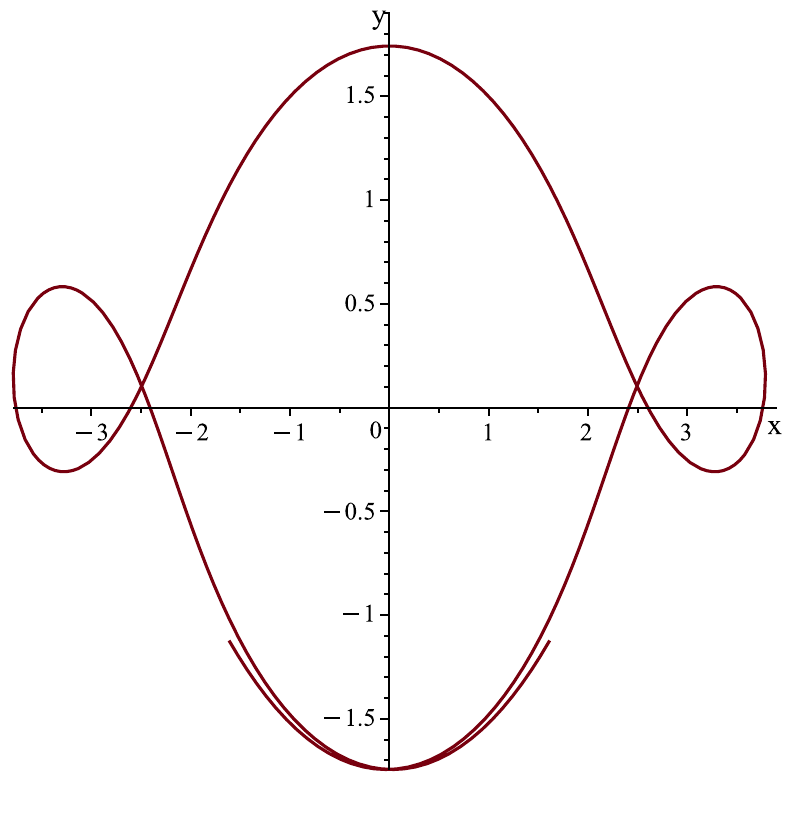}
\caption{Left: $C= 0.86$. Right: $C= 0.74$.}
\end{subfigure}
\caption{rational cosine open loops $c=1$.}
\label{fig:ratcos_loop_open_curve}
\end{figure}

The static loop has the following basic features,
which are specializations of those of the static rational cn loops. 
First,
it is a bounded curve lying in an annulus with radii $R_+$ and $R_-$ given by
\begin{equation}
R_\pm^2 = \frac{F(1 \pm 2/C)^2}{c}  . 
\end{equation}
This follows from the property that 
\begin{equation}
  \begin{aligned}
    |\vec{r}(s,0)|^2 = |J(s)|^2 = &
    \frac{F}{c}
    \Big( 1    +\frac{4(1-C^2)}{C^2(C \cos\big(\tfrac{1}{2}\sqrt{c G}\, s\big) + 1)} \Big)
  \end{aligned}
\end{equation}
is a bounded function of $s$. 
See the plots in Fig.~\ref{fig:ratcos_loop_open_curve}

Second, the curve has winding number 
\begin{equation} 
  N = \mp \big(2 - \frac{1}{\sqrt{1-C^2}}\big)
\end{equation}
and winding rate
\begin{equation} 
  W = \frac{\sqrt{2c}}{\sqrt{(1 +2C^2)(1-C^2)}}
\end{equation}
using the results in Table~\ref{table:conserved}.

\begin{figure}[h]
\centering
 \begin{subfigure}[t]{1\textwidth}
\centering
\includegraphics[width=0.3\textwidth,trim=2cm 12cm 6cm 2cm,clip]{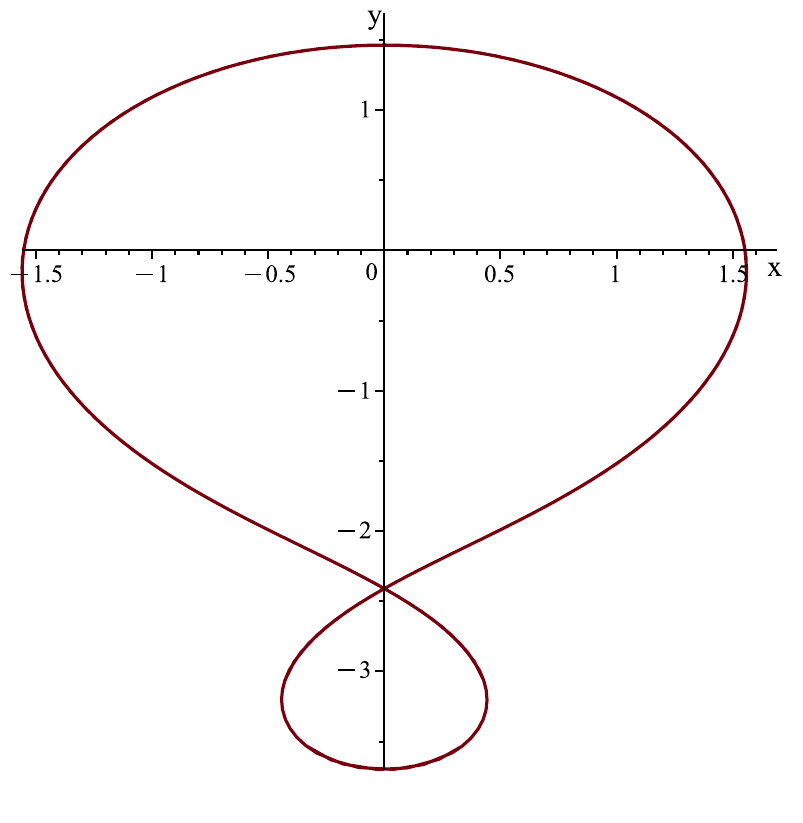}
\hfill
\includegraphics[width=0.3\textwidth,trim=2cm 12cm 6cm 2cm,clip]{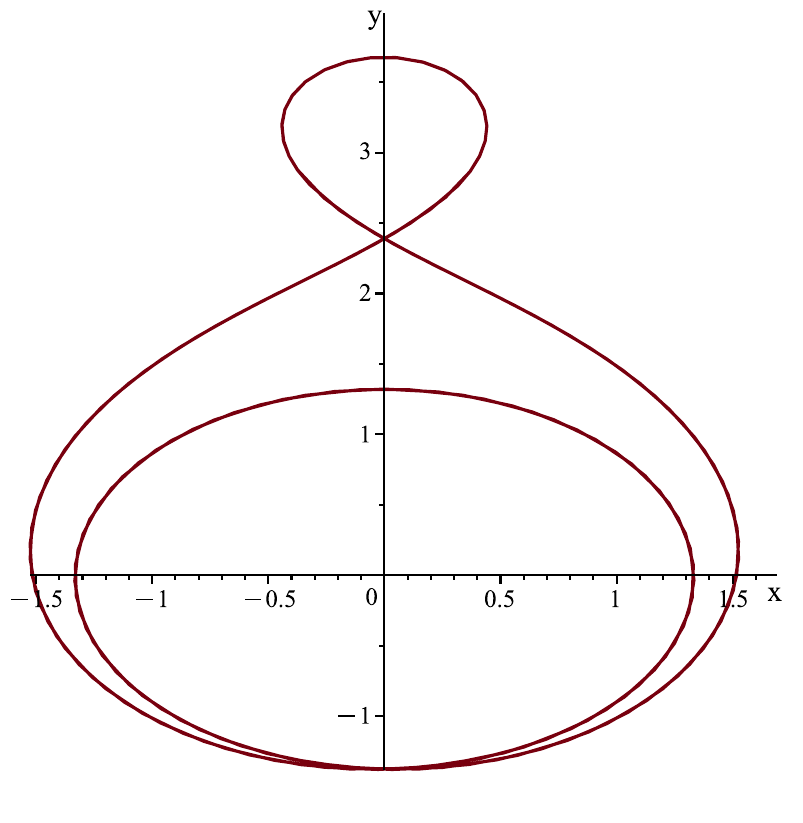}
\\
\includegraphics[width=0.3\textwidth,trim=2cm 12cm 6cm 2cm,clip]{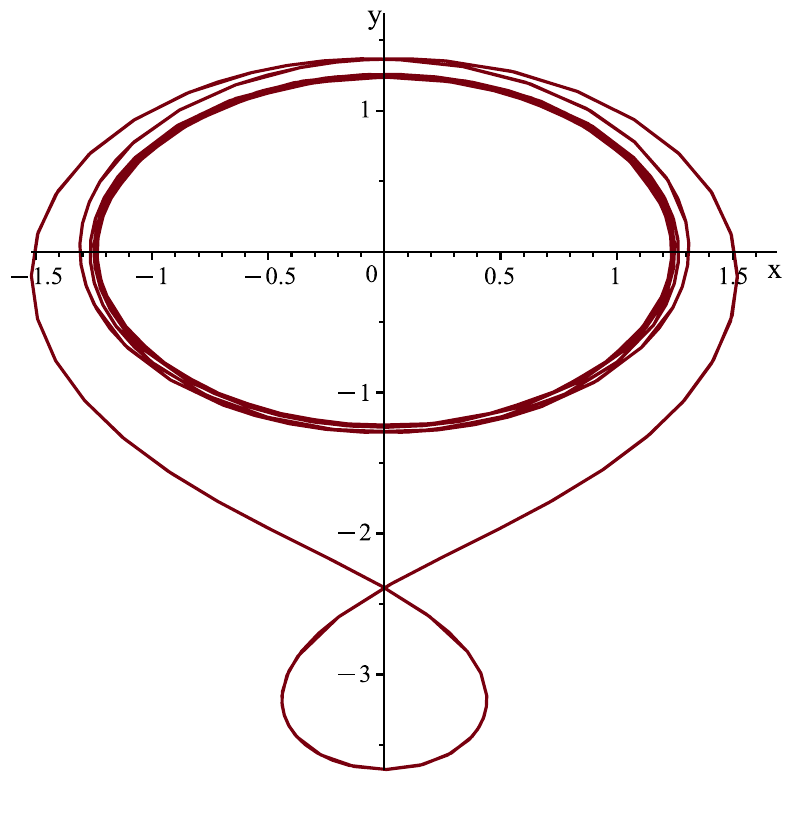}
\hfill
\includegraphics[width=0.3\textwidth,trim=2cm 12cm 6cm 2cm,clip]{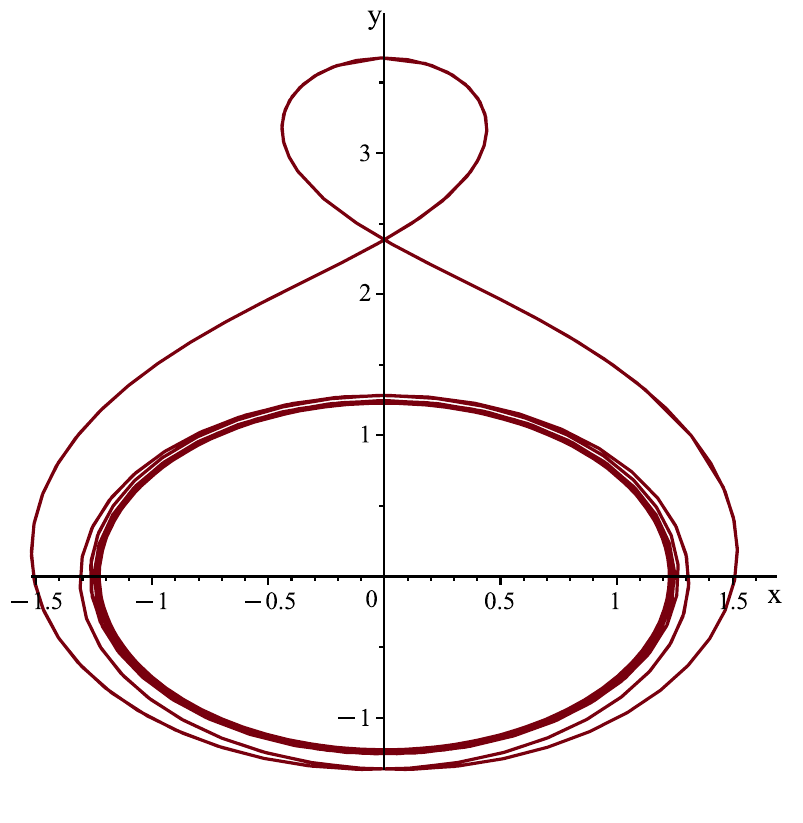}
\hfill
\caption{$C=0.886,m=1$. Top Left: $n=2$. Top Right: $n=3$. Bottom Left: $n=8$. Bottom Right: $n=11$.}
\end{subfigure}
\caption{rational cosine closed loops $c=1$}
\label{fig:ratcos_loop_row1}
\end{figure}

\begin{figure}[h]
\begin{subfigure}[t]{1\textwidth}
\centering
\includegraphics[width=0.3\textwidth,trim=2cm 12cm 6cm 2cm,clip]{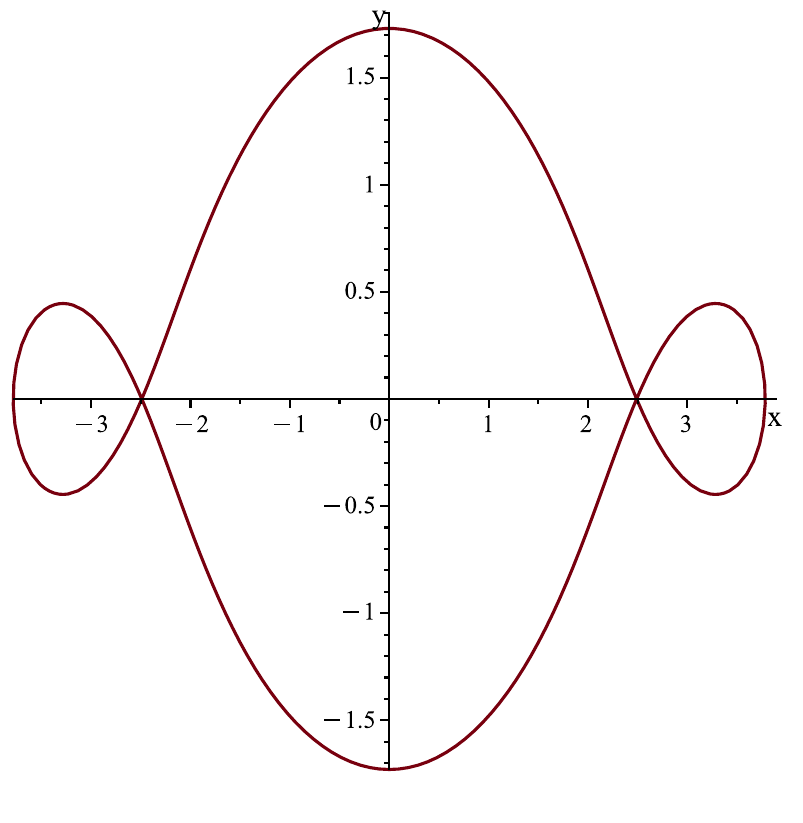}
\hfill
\includegraphics[width=0.3\textwidth,trim=2cm 12cm 6cm 2cm,clip]{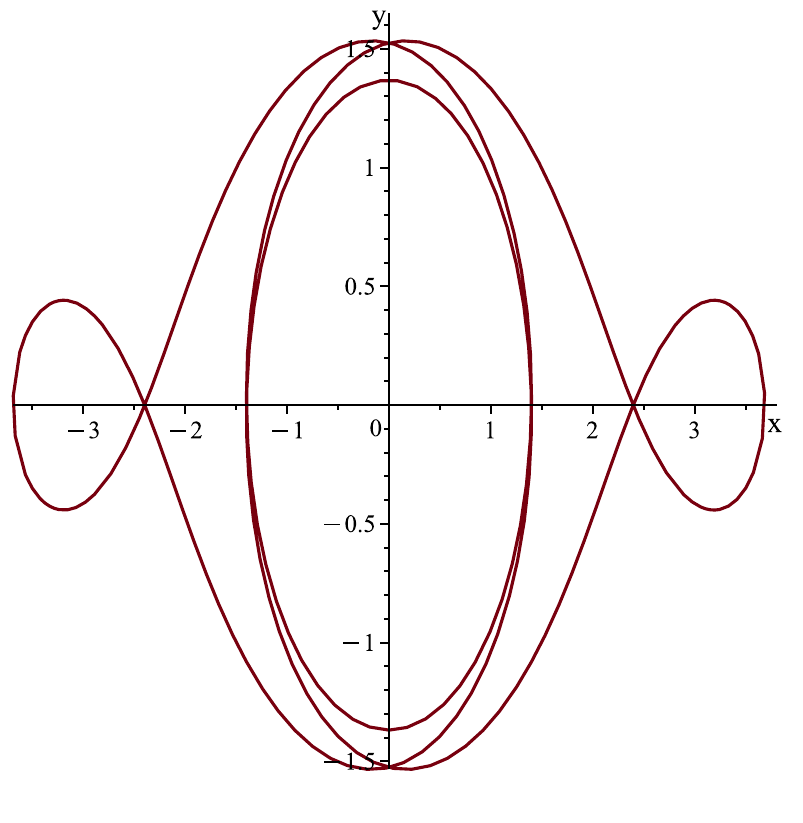}
\\
\includegraphics[width=0.3\textwidth,trim=2cm 12cm 6cm 2cm,clip]{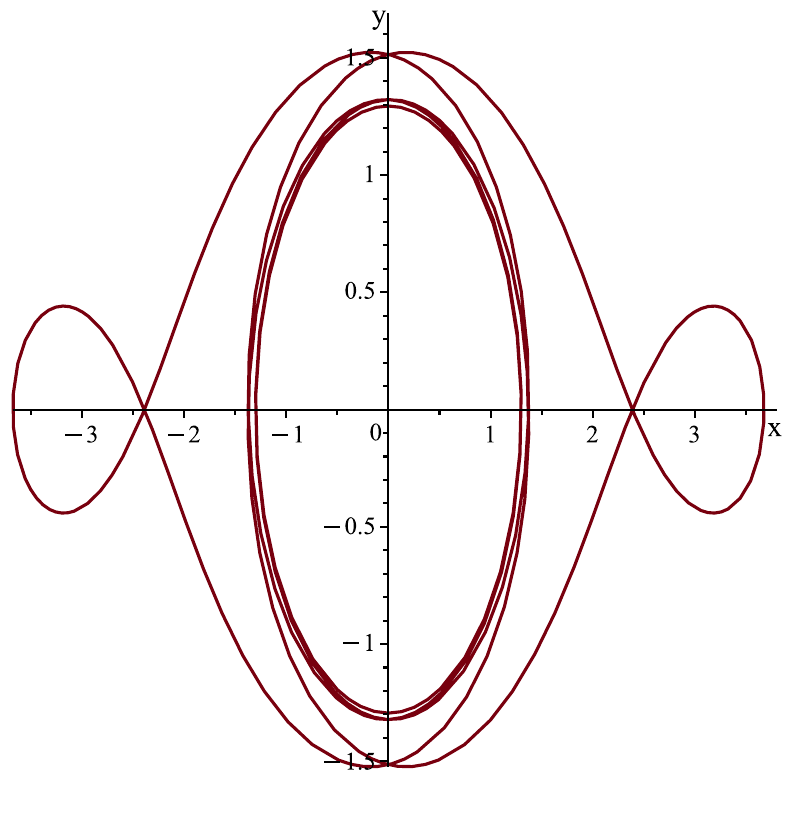}
\hfill
\includegraphics[width=0.3\textwidth,trim=2cm 12cm 6cm 2cm,clip]{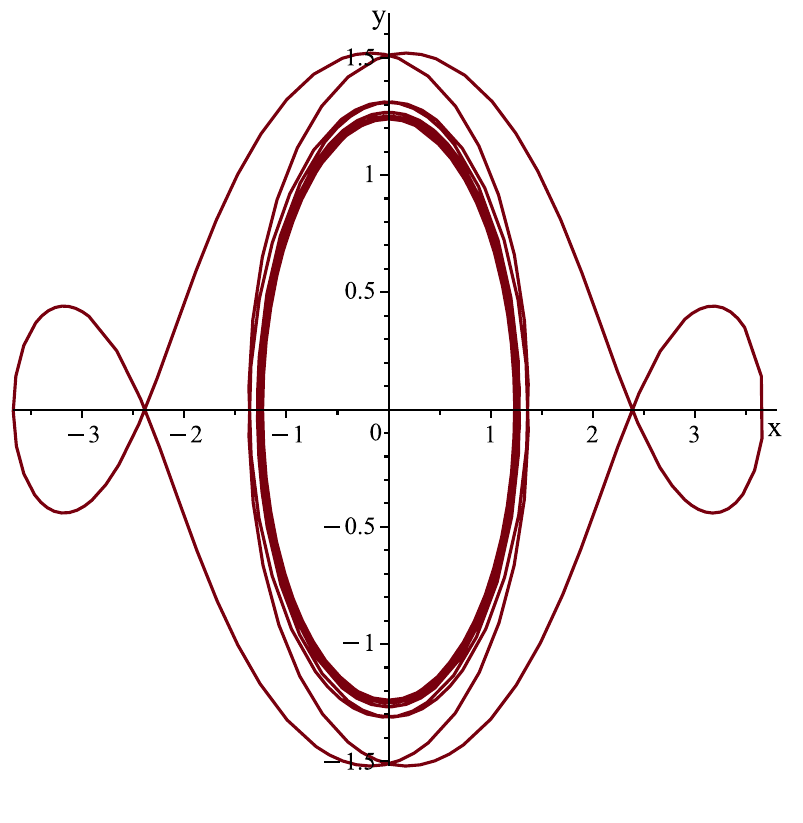}
\caption{$C=0.745,m=2$. Top Left: $n=3$. Top Right: $n=5$. Bottom Left: $n=7$. Bottom Right: $n=15$.}
\end{subfigure}
\caption{rational cosine closed loops $c=1$}
\label{fig:ratcos_loop_row2}
\end{figure}

\begin{figure}[h]
\begin{subfigure}[t]{1\textwidth}
\centering
\includegraphics[width=0.3\textwidth,trim=2cm 12cm 6cm 2cm,clip]{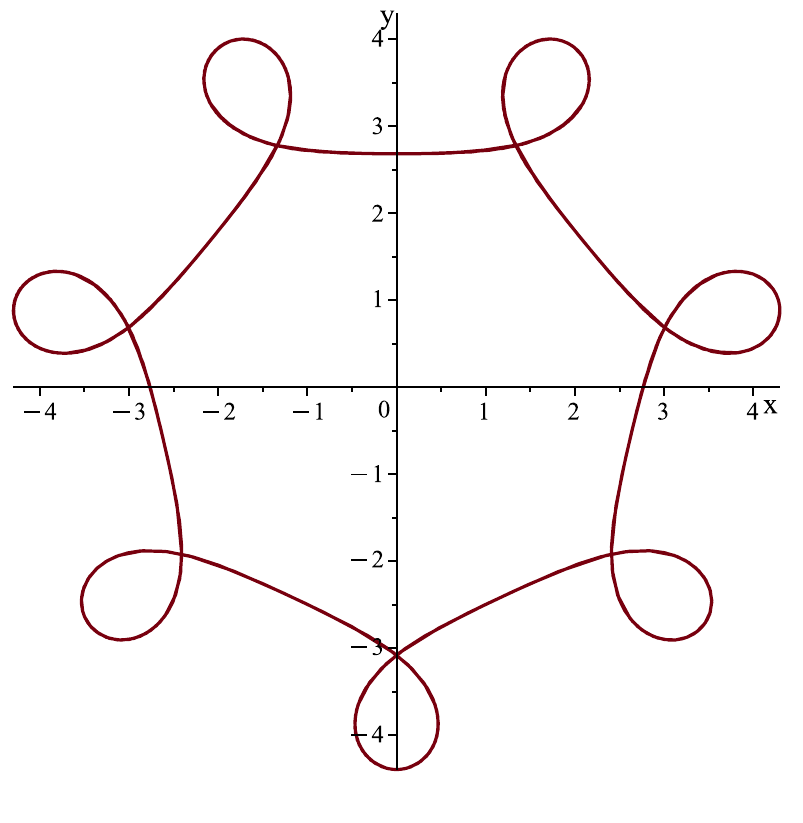}
\hfill
\includegraphics[width=0.3\textwidth,trim=2cm 12cm 6cm 2cm,clip]{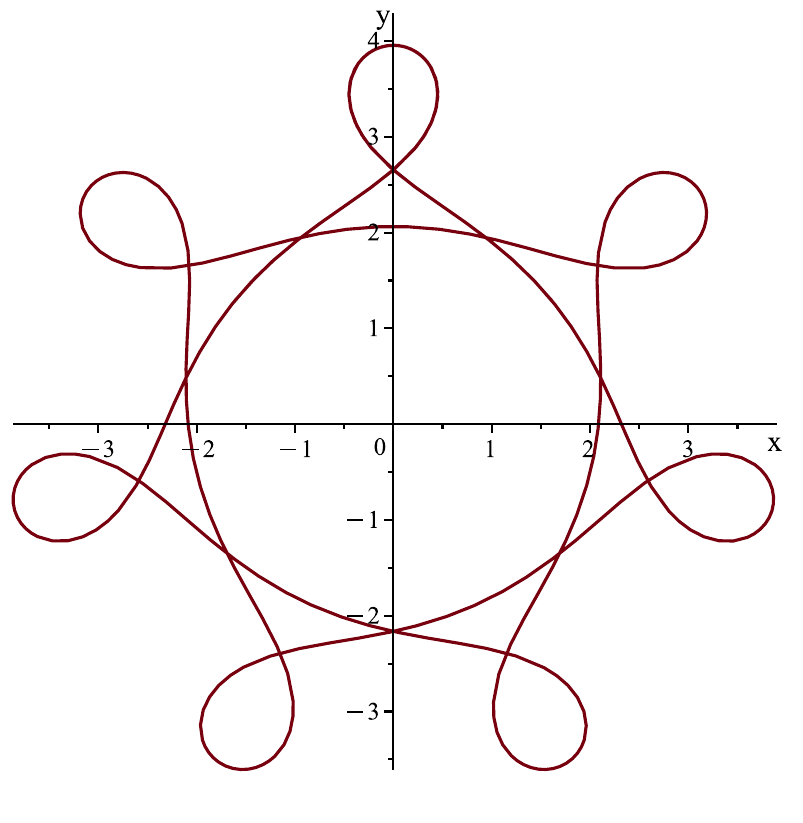}
\\
\includegraphics[width=0.3\textwidth,trim=2cm 12cm 6cm 2cm,clip]{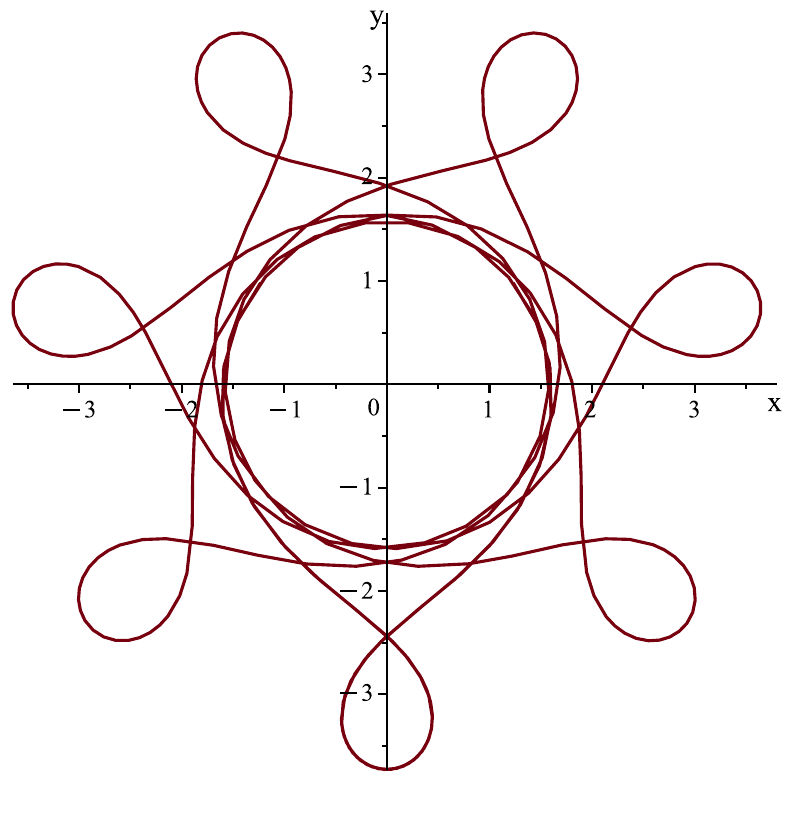}
\hfill
\includegraphics[width=0.3\textwidth,trim=2cm 12cm 6cm 2cm,clip]{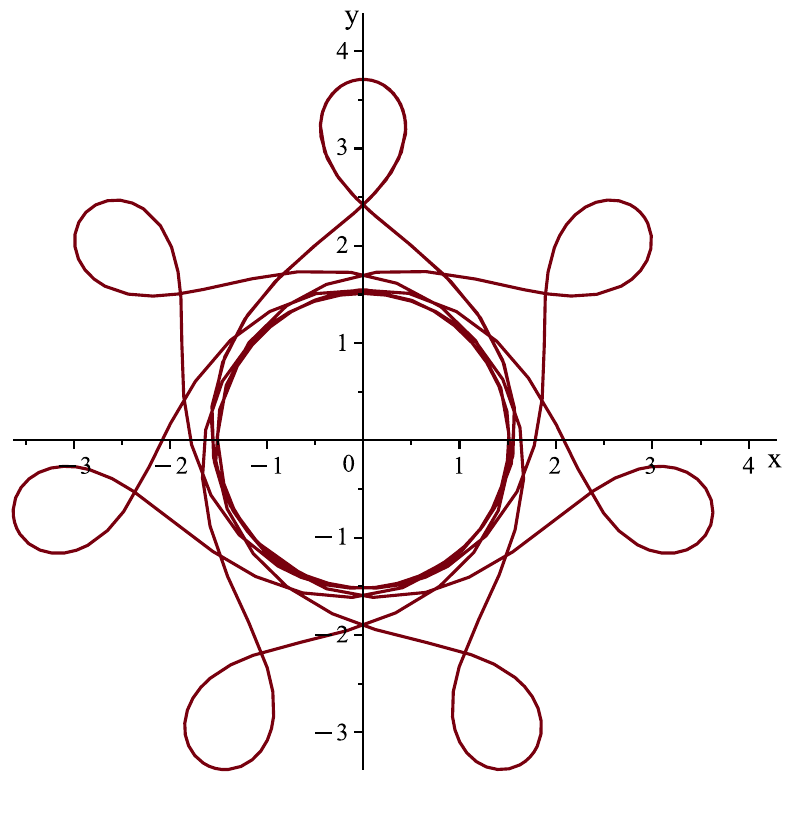}
\caption{$C=0.484,m=7$. Top Left: $n=8$. Top Right: $n=9$. Bottom Left: $n=12$. Bottom Right: $n=13$.}
\end{subfigure}
\caption{rational cosine closed loops $c=1$}
\label{fig:ratcos_loop_row3}
\end{figure}

Third, 
the secular period of the turning angle $\arg(\vec{r}(0,s)) = \vartheta(s) + \arg(J(s))$
is given by
\begin{equation}
  \pi\sqrt{2(1+2C^2)/c} . 
\end{equation}
Hence,
the curve $\vec{r}(s,0)$ is closed iff the ratio of this period to the period of the cosine function is commensurable.
This condition gives 
\begin{equation}
  \sqrt{1-C^2} = m/n
\end{equation}
for some non-zero integers $n$, $m$.
A closed curve has $m$ maxima points,
where $m \pi \sqrt{2(1+2C^2)}/\sqrt{c(1 -C^2)}$ is the period.
Plots are shown in Figs. ~\ref{fig:ratcos_loop_row1}, ~\ref{fig:ratcos_loop_row2}, and ~\ref{fig:ratcos_loop_row3}.

\section{Concluding remarks}
\label{sec:conclude}

The present work has derived the loop solutions corresponding to each type of mKdV travelling wave.
These loop solutions can be cleanly divided into two broad types:
travelling loops, and rotating loops.
Travelling loops arise from mKdV solitons and cnoidal (Jacobi cn) waves,
the latter being periodic.
Rotating loops can be subdivided into
asymptotically circular ones that come from mKdV solitary waves on a non-zero background and rational waves,
and periodic ones that are produced by mKdV rational elliptic period waves.
A specialization of periodic loops, both open and closed, yields rational cosine loops. 
An explicit description of each of these types of loop solutions has been derived
and used to characterize their main features. 

These features of the loop solutions
may prove useful in studying physical applications 
discussed in section~\ref{sec:intro}.
In particular,
travelling loops are relevant to the propagation of solitary waves in an elastic rod 
\cite{Nis.1997};
Rotating periodic loops are important in dynamics of boundaries
in vortex patches in incompressible fluid layers 
\cite{Gol.Pet.1991,Gol.Pet.1992}
and for electron cloud densities in thin-layered materials 
\cite{Wex.Dor.1999a,Wex.Dor.1999b}. 
Asymptotically circular rotating loops may be of interest in 
cylindrical membranes in fluid lipid bi-layers 
\cite{Vas.Djo.Mla.2008}. 

For future work,
the loop solutions corresponding to other (non-travelling wave) 
mKdV solutions can be pursued.
Specifically,
breathers and multi-solitons --- both on a zero background or a non-zero background ---
can be expected to produce interacting dynamical loops with interesting properties.

\appendix

\section{Appendix: Ellipitic function identities}\label{sec:identities}

All of the relevant algebraic identities that hold among Jacobi elliptic functions
are summarized as follows
(see e.g. \Ref{Abr.Ste-book}): 
\begin{gather}
\CN(x,k)^2 + \SN(x,k)^2 =1,
\quad
\DN(x,k)^2 = 1-k^2 \SN(x,k)^2 = 1-k^2 + k^2\CN(x,k)^2 ;
\label{squares}
\\
\CN(x,1/k) = \DN(x/k,k) ,
\quad
\SN(x,1/k) = k\,\SN(x/k,k) ;
\label{scaling}
\\
\CN(x,ik) = \dfrac{\CN(\sqrt{1+k^2}\,x,k/\sqrt{1+k^2})}{\DN(\sqrt{1+k^2}\,x,k/\sqrt{1+k^2})} ,
\quad
\SN(x,ik) = \dfrac{\SN(\sqrt{1+k^2}\,x,k/\sqrt{1+k^2})}{\sqrt{1+k^2}\,\DN(\sqrt{1+k^2}x,k/\sqrt{1+k^2})};
\label{imaginary.modulus}
\\
\CN(ix,k) = \CN(x,\sqrt{1-k^2}) ,
\quad
\SN(ix,k) = i\,\SN(x,\sqrt{1-k^2})/\CN(x,\sqrt{1-k^2}) ;
\label{imaginary.argument}
\\
\SN(x, e^{\phi})^2 =
e^{-\phi}\frac{1 - \CN(2 e^{\phi/2} x, \cosh(\phi/2))}{1+ \CN(2e^{\phi/2} x, \cosh(\phi/2))} . \label{SNsq.CN}
\end{gather}

Combining identities \eqref{squares} and \eqref{scaling} leads to the useful relation:
\begin{equation}\label{scaling.CNsq}
  \CN(x,1/k)^2 = k^2 \big( \CN(x/k,k)^2 -1 + 1/k^2 \big) . 
\end{equation}
Similarly, identity \eqref{SNsq.CN} can be expressed as
\begin{equation}\label{CN.SNsq.CNsq}
  \CN(x, \cosh\phi) 
  = \frac{ e^{2\phi} -1 +  \CN\big( \tfrac{1}{2}e^{\phi}\, x, e^{-2\phi} \big)^2}{ e^{2\phi} +1 - \CN\big( \tfrac{1}{2}e^{\phi} x, e^{-2\phi} \big)^2 }
 = \frac{ 1-e^{-2\phi} \SN\big( \tfrac{1}{2}e^{\phi}\, x, e^{-2\phi} \big)^2}{ 1 + e^{-2\phi} \SN\big( \tfrac{1}{2}e^{\phi} x, e^{-2\phi} \big)^2 } . 
\end{equation}

Note that, here and throughout the paper,
the Maple convention is used for the parameter in these functions, 
which is the square root of the standard modulus parameter.

\section{Curve flows as integrable systems}\label{sec:integrability}

The developments in section~\ref{sec:curveflows} show how
the structure equations of arclength-preserving curve flows in $\Rnum^2$
encode some of integrability structure of the mKdV hierarchy
independently of the actual motion of the curve.
There are two further aspects to this development,
as explained in \Refs{Mar-Bef.San.Wan.2002,Anc.2008}.

The first structure equation \eqref{eqn1} encodes a Hamiltonian operator: 
\begin{equation}
\kappa_t = \Hop(\omega)
\end{equation}
where
\begin{equation}\label{H.op}
  \Hop = \partial_s . 
\end{equation}
This operator trivially satisfies the Hamiltonian conditions \cite{Olv-book,Dor-book}
that it is skew-symmetric and has vanishing Schouten bracket.
Equivalently, it defines a Poisson bracket 
$\{\mathcal{F},\mathcal{G}\}_\Hop = \int_{-\infty}^{\infty} F \Hop(G)\,ds$
on arbitrary functionals $\mathcal{F}=\int_{-\infty}^{\infty}F\,ds$
and $\mathcal{G}=\int_{-\infty}^{\infty}G\,ds$, 
namely the bracket is skew and satisfies the Jacobi identity.

The other two structure equations \eqref{eqn2}--\eqref{eqn3}
encode a symplectic operator
\begin{equation}
  \omega = \Jop(h_\pe)
\end{equation}
where
\begin{equation}\label{J.op}
  \Jop = \partial_s + \kappa\partial_s^{-1}\kappa  . 
\end{equation}
This operator can be shown to have the property that it defines a bilinear 2-form
$(\mathrm{F},\mathrm{G})_\Jop
= \int_{-\infty}^{\infty} F \Jop(G)\,ds$
that is skew-symmetric and closed \cite{Olv-book,Anc.2008},
for arbitrary vector fields $\mathrm{F}= F\partial_s$ and $\mathrm{G}= G\partial_s$. 

Note that the recursion operator \eqref{R.op} is simply the composition of the Hamiltonian operator and the symplectic operator, $\Hop\Jop =\Rop$.
These operators exist for any non-stretching curve flow.
However, they do not imply themselves that the curvature evolution equation \eqref{curvature.evolution}
is an integrable system.
Integrability requires the curve flow to satisfy extra conditions.

To understand these conditions,
consider the circumstances under which a geometrical evolution for $\vec{r}(s,t)$
is obtained.
The normal component of the flow $h_\perp$ needs to be a function of the curvature $\kappa$ 
and its arclength derivatives $\kappa_s$, $\kappa_{ss}$, etc. up to any differential order. 
Any such function $h_\pe(\kappa,\kappa_s,\kappa_{ss},\ldots)$
will give rise, firstly, to a curve flow \eqref{r.flow.eqn}
in which both the tangential and normal components are functions of the geometrical Frenet frame and its arclength derivatives, 
\begin{equation}
\vec{r}_t = h_\pa(\N\cdot\T_s,\N\cdot\T_{ss},\ldots) \T + h_\pe(\N\cdot\T_s,\N\cdot\T_{ss},\ldots) \N . 
\end{equation}
Secondly, the resulting evolution of the curvature \eqref{curvature.evolution} is given by 
\begin{equation}\label{curvature.evolution.geom}
\kappa_t = h_\pe(\kappa,\kappa_s,\kappa_{ss},\ldots)_{ss} + \kappa^2 h_\pe(\kappa,\kappa_s,\kappa_{ss},\ldots) + \kappa_s\int \kappa h_\pe(\kappa,\kappa_s,\kappa_{ss},\ldots)\,ds 
\end{equation}
which is a (nonlocal) geometric partial differential equation
because it involves only $\kappa(s,t)$.

This equation \eqref{curvature.evolution.geom} will be integrable if it possesses
a bi-Hamiltonian structure \cite{Olv-book} or a Lax pair \cite{Abl.Cla-book}.
Existence of a bi-hamiltonian structure requires that 
$\Jop(h_\pe) = \delta H^{(1)}/\delta \kappa = \Rop^*(\delta H^{(0)}/\delta \kappa)$
holds for some functions $H^{(1)}$ and $H^{(0)}$ of  $\kappa,\kappa_s,\kappa_{ss},\ldots$,
and that the Hamiltonian operators $\Hop$ and $\Dop = \Rop\Hop$ are compatible.
Here $\delta/\delta \kappa$ is the variational derivative (Euler operator). 
In this situation, 
\begin{equation}
\kappa_t = \Hop(\delta H^{(1)}/\delta \kappa) = \Dop(\delta H^{(0)}/\delta \kappa) 
\end{equation}
is an evolution equation having two compatible Hamiltonian structures,
with $H^{(1)}$ and $H^{(0)}$ being the Hamiltonian functions.

The simple choice $h_\pe = \kappa_s$ always yields a bi-Hamiltonian curve flow.
This choice comes from the considering the symmetry of the pair of operators $\Hop$ and $\Jop$, in particular that they are invariant under translations
in the arclength parameter $s$.
The group of translations acts by
$\kappa(s) \to \kappa(s+\epsilon) = \kappa(s) +\epsilon \kappa_s(s) + O(\epsilon^2)$
where $\epsilon$ is the group parameter.
This transformation is generated infinitesimally by the operator
$\mathrm{X} = \partial_s$ on $\kappa(s)$. 

Remarkably,
the preceding discussion has a deep generalization to curve flows in any Riemannian symmetric space 
when a Frenet frame is generalized to a parallel frame \cite{Anc.2008}.
Any symmetry of the resulting frame structure equations can be used to generate
a bi-Hamiltonian structure and corresponding integrable geometric evolution equation.

\section*{Acknowledgements}
S.C.A. is supported by an NSERC Discovery grant.

\end{document}